\documentclass[twoside,11pt]{article}
\usepackage{standalone}
\usepackage[preprint]{jmlr2e}
\usepackage{amssymb, amsmath}
\usepackage{bm}
\usepackage[dvipsnames]{xcolor}
\usepackage{natbib}
\usepackage{microtype}
\usepackage{graphicx}
\usepackage{multirow}
\usepackage{setspace}
\usepackage{algorithm2e}
\RestyleAlgo{boxruled}
\usepackage{float}
\newcommand{\U}{\boldsymbol{U}}
\newcommand{\V}{{\boldsymbol{V}}}
\newcommand{\D}{\boldsymbol{D}}
\hypersetup{ hidelinks }

\newcommand{\mbepsilon}{\boldsymbol{\epsilon}}
\usepackage{booktabs}
\usepackage{siunitx}

\usepackage{algpseudocode}
\newcommand{\mbx}{\boldsymbol{x}}
\newcommand{\mby}{\boldsymbol{y}}
\newcommand{\mbX}{{\boldsymbol{X}}}
\newcommand{\mbY}{\boldsymbol{Y}}
\newcommand{\mbI}{\boldsymbol{I}}
\newcommand{\mbP}{\boldsymbol{\mathcal{P}}}

\newcommand{\mbv}{\boldsymbol{v}}
\newcommand{\mbu}{\boldsymbol{u}}
\newcommand{\mbb}{\boldsymbol{b}}

\newcommand{\mbTheta}{\boldsymbol{\Theta}}
\newcommand{\mbC}{\boldsymbol{C}}

\newcommand{\mbz}{\boldsymbol{z}}
\newcommand{\mbg}{\boldsymbol{g}}
\newcommand{\mba}{\boldsymbol{a}}

\newcommand{\mbt}{\boldsymbol{t}}

\newcommand{\mbS}{{\boldsymbol{S}}}

\newcommand{\mbM}{{\boldsymbol{M}}}
\newcommand{\mbU}{\boldsymbol{U}}
\newcommand{\mbV}{{\boldsymbol{V}}}
\newcommand{\mbK}{{\boldsymbol{K}}}
\newcommand{\mbD}{{\boldsymbol{D}}}
\newcommand{\mbB}{{\boldsymbol{B}}}
\newcommand{\mbO}{{\boldsymbol{O}}}
\newcommand{\mbZ}{{\boldsymbol{Z}}}
\newcommand{\mbQ}{{\boldsymbol{Q}}}
\newcommand{\mbR}{{\boldsymbol{R}}}
\newcommand{\mbG}{{\boldsymbol{G}}}

\usepackage{setspace}
\newcommand{\mbW}{{\boldsymbol{W}}}
\newcommand{\mbH}{{\boldsymbol{H}}}
\newcommand{\mbE}{{\boldsymbol{\mathcal{E}}}}
\newcommand{\mbA}{{\boldsymbol{A}}}

\newcommand{\mbbeta}{{\boldsymbol{\beta}}}
\newcommand{\mbGamma}{{\boldsymbol{\Gamma}}}
\newcommand{\mbPhi}{{\boldsymbol{\Phi}}}
\newcommand{\mbSigma}{{\boldsymbol{\Sigma}}}
\newcommand{\mbOmega}{{\boldsymbol{\Omega}}}
\newcommand{\mbDelta}{{\boldsymbol{\Delta}}}
\usepackage{lastpage}
\jmlrheading{23}{2022}{1-\pageref{LastPage}}{1/20; Revised
11/21}{4/22}{20-064}{Aaron J. Molstad}
\ShortHeadings{New Insights for the Multivariate Square-Root Lasso}{Molstad}
\newcommand{\argmin}{\operatorname*{arg \ min}}

\newcommand{\minim}{\operatorname*{minimize}}

\begin{document}

\title{New Insights for the Multivariate Square-Root Lasso}

\author{\name Aaron J. Molstad \email amolstad@ufl.edu \\
       \addr Department of Statistics and Genetics Institute\\
       University of Florida\\
       Gainesville, FL 32611, USA}

\editor{David Wipf}

\maketitle

\begin{abstract}
We study the multivariate square-root lasso, a method for fitting the multivariate response linear regression model with dependent errors. This estimator minimizes the nuclear norm of the residual matrix plus a convex penalty. Unlike existing methods that require explicit estimates of the error precision (inverse covariance) matrix, the multivariate square-root lasso implicitly accounts for error dependence and is the solution to a convex optimization problem. We establish error bounds which reveal that like the univariate square-root lasso, the multivariate square-root lasso is pivotal with respect to the unknown error covariance matrix. In addition, we propose a variation of the alternating direction method of multipliers algorithm to compute the estimator and discuss an accelerated first order algorithm that can be applied in certain cases. In both simulation studies and a genomic data application, we show that the multivariate square-root lasso can outperform more computationally intensive methods that require explicit estimation of the error precision matrix. \end{abstract}
\begin{keywords}
pivotal estimation, multivariate response linear regression, convex optimization, covariance matrix estimation
\end{keywords}


\section{Introduction}\label{sec:intro}
Modeling the linear relationship between a $p$-variate vector of predictors and a $q$-variate vector of responses is a central task in multivariate analysis. In this article, we will assume that the observed response vectors for the $n$ subjects in the study, $\mby_1, \dots, \mby_n$, are realizations of the random vectors
\begin{equation} \label{eq:MVR}
\mbbeta_{*0} + \mbbeta_*^\top\mbx_i + \mbepsilon_i
\end{equation}
for $i \in \{1, \dots, n\}$, where $\mbx_i \in \mathbb{R}^p$ is the predictor for the $i$th subject, $\mbbeta_{*0} \in \mathbb{R}^{q}$ is the unknown intercept vector, and $\mbbeta_* \in \mathbb{R}^{p \times q}$ is the unknown regression coefficient matrix. We assume that the $\mbepsilon_i$ are independent and identically distributed $q$-variate random vectors with mean zero and unknown error covariance matrix $\mbSigma_* \in \mathbb{S}^q_+$, where $\mathbb{S}_+^q$ is the set of $q \times q$ symmetric positive definite matrices. Let $\mbOmega_* = \mbSigma_*^{-1}$ be the unknown error precision matrix. For notational convenience, let $\mbY = (\mby_1 - \bar{\mby}, \dots, \mby_n - \bar{\mby})^\top \in \mathbb{R}^{n \times q}$ and $\mbX = (\mbx_1 - \bar{\mbx}, \dots, \mbx_n - \bar{\mbx})^\top \in \mathbb{R}^{n \times p}$, where $\bar{\mby} = n^{-1} \sum_{i=1}^n \mby_i$ and $\bar{\mbx} = n^{-1} \sum_{i=1}^n \mbx_i$. 

Many methods exist for fitting the multivariate response linear regression model in \eqref{eq:MVR}. When $n > p$ and the $\mbepsilon_i$ are multivariate normal, the maximum likelihood estimator (and equivalently, least squares estimator) of $\mbbeta_*$ does not require knowledge of nor an estimate of $\mbOmega_*$. When $p \geq n$ the least squares estimator is not unique, so a natural alternative is to estimate $\mbbeta_*$ by minimizing a penalized least squares criterion (i.e., penalized squared Frobenius norm of the residual matrix) using penalties that exploit the matrix structure of the unknown regression coefficients \citep{turlach2005simultaneous,yuan2007dimension,obozinski2011support,negahban2011}. However, the penalized least squares criterion implicitly assumes $\mbSigma_* \propto \mbI_q$: the penalized least squares estimator is equivalent to the penalized normal maximum likelihood estimator under the assumption that $\mbSigma_* \propto \mbI_q$. 

This limitation of penalized least squares has motivated numerous methods which incorporate an estimate of $\mbOmega_*$ into the estimation procedure for $\mbbeta_*$. One class of methods jointly estimates $\mbOmega_*$ and $\mbbeta_*$ by maximizing a penalized normal log-likelihood \citep{rothman2010sparse,yin2011sparse} using $L_1$-norm penalties---as defined in \eqref{eq:MSRLpen}---on the optimization variable corresponding to $\mbbeta_*$ and on off-diagonal entries of the optimization variable corresponding to $\mbOmega_*$. \citet{wang2015joint} proposed an alternative approach which performs estimation column-by-column, estimating the $k$th columns of $\mbbeta_*$ and $\mbOmega_*$ jointly for $k \in \{1,\dots, q\}$.  While these methods can perform well in certain settings, an estimate of $\mbOmega_*$ is often not needed by the practitioner. Regardless, the methods of \citet{rothman2010sparse}, \citet{yin2011sparse}, and \citet{wang2015joint} require estimating $O(q^2)$ precision matrix parameters, and in the case of \citet{rothman2010sparse} and \citet{yin2011sparse}, require solving a computationally burdensome nonconvex optimization problem. 

An ideal estimation criterion for $\mbbeta_*$ is convex and can account for error dependence without requiring an explicit estimate of $\mbOmega_*$ or $\mbSigma_*$. To this end, we study the class of estimators 
\begin{equation} \label{eqMSRL}
\argmin_{\mbbeta \in \mathbb{R}^{p \times q}} \left\{ \frac{1}{\sqrt{n}}\| \mbY - \mbX \mbbeta\|_{*} + \lambda g (\mbbeta) \right\},
\end{equation}
where $\|\mbA\|_* = {\rm tr}\{(\mbA^\top\mbA)^{1/2}\}$ denotes the nuclear norm of a matrix $\mbA$ (i.e., the norm which sums the singular values of its matrix-valued argument), $g$ is a nonnegative penalty function, and $\lambda > 0$ is a user-specified tuning parameter. When $g$ is a norm, which we will assume throughout, the objective function in \eqref{eqMSRL} is convex. 
The estimator in \eqref{eqMSRL} with $L_1$-norm penalty was originally proposed by \citet{van2016chi2}. Their focus was on using \eqref{eqMSRL} to construct confidence sets for high-dimensional regression coefficient vectors in univariate response linear regression. In this article, we study \eqref{eqMSRL} as a method for fitting \eqref{eq:MVR} in high-dimensional settings. 

Of course, the class of estimators defined by \eqref{eqMSRL} is applicable with penalties beyond the $L_1$-norm.  
We focus on three versions of \eqref{eqMSRL}, each defined by their choice of penalty $g$:
\begin{align}
&\text{Lasso (L)}  ~~~~~~~~~~~~~~~~ \|\mbbeta\|_{1\hspace{5pt}} = \sum_{j=1}^p \sum_{k=1}^q |\mbbeta_{j,k}|, \label{eq:MSRLpen}\\
&\text{Group lasso (GL)} ~~~~~\vspace{2pt} \|\mbbeta\|_{1,2} = \sum_{j=1}^p \left( \sum_{k=1}^q \mbbeta_{j,k}^2  \right)^{1/2}, \label{eq:MSRGLpen}\\
&\text{Nuclear norm (LR)}  ~~~ \|\mbbeta\|_{*\hspace{5pt}} = {\rm tr}\big\{(\mbbeta^\top \mbbeta)^{1/2}\big\}= \sum_{j=1}^{\min(p,q)}\hspace{-4pt} \sigma_j(\mbbeta) \label{eq:MSRLRpen},
\end{align}
where $\sigma_j(\mbA)$ and $\mbA_{j,k}$ denote the $j$th largest singular value and $(j,k)$th entry of the matrix $\mbA$, respectively. When referring to \eqref{eqMSRL} with the penalties \eqref{eq:MSRLpen}, \eqref{eq:MSRGLpen}, and \eqref{eq:MSRLRpen}, we use $\hat\mbbeta_{\rm L},$ $\hat\mbbeta_{\rm GL},$ and $\hat\mbbeta_{\rm LR},$ respectively. For simplicity, we refer to the class of estimators \eqref{eqMSRL} as the \textit{multivariate square-root lasso} regardless of the penalty $g$.

Relative to the $L_1$-norm penalty, which encourages estimates of $\mbbeta_*$ with unstructured sparsity, the group lasso and nuclear norm penalties are especially well-suited for multivariate response linear regression. The group lasso penalty exploits the assumption that many predictors are irrelevant for all $q$ responses by encouraging estimates of $\mbbeta_*$ with some rows entirely equal to zero \citep{yuan2006model,obozinski2011support,lounici2011oracle}. The nuclear norm penalty, in contrast, acts as a lasso penalty on the singular values of the optimization variable $\mbbeta$ and thus promotes estimates of $\mbbeta_*$ with low rank \citep{yuan2007dimension,negahban2011,chen2013reduced}, hence the shorthand LR. Low rankness of $\mbbeta_*$ is assumed in reduced rank regression \citep{velu2013multivariate}, a classical method for dimension reduction in \eqref{eq:MVR}. 

Computing \eqref{eqMSRL} is nontrivial because the nuclear norm of residuals, though convex, is nondifferentiable. To date, there are no specialized algorithms to compute \eqref{eqMSRL} with convergence guarantees. \citet{van2016chi2} suggested an iterative procedure for computing $\hat\mbbeta_{\rm L}$, but unfortunately, we found their algorithm cannot be used to solve the optimization in general.
In a later version of \citet{van2016chi2} appearing in a PhD thesis, \citet{stucky2017asymptotic} computed \eqref{eqMSRL} using the general purpose convex solver \texttt{CVX} \citep{cvx}, which can be slow in high-dimensional settings. 

In addition to the computational challenges, little is known about \eqref{eqMSRL} in terms of its statistical properties. While \citet{van2016chi2} and \citet{van2016estimation} pointed out the connection between \eqref{eqMSRL} and the univariate ($q=1$) square-root lasso \citep{belloni2011square, sun2012scaled, bunea2014group,derumigny2018improved}, their focus was on \eqref{eqMSRL} as a means for constructing confidence intervals. They did not establish any statistical properties of \eqref{eqMSRL} nor did they explore the empirical performance of \eqref{eqMSRL} in the context of fitting \eqref{eq:MVR}. 

In this article, we study \eqref{eqMSRL} from theoretical, computational, and empirical perspectives. We prove that like the univariate square-root lasso, \eqref{eqMSRL} is pivotal in the sense that the value of the tuning parameter $\lambda$ leading to near-oracle performance is determined by a random quantity whose distribution does not depend on the unknown error covariance $\mbSigma_*$. In so doing, we establish error bounds for \eqref{eqMSRL} with arbitrary $g$, then specialize these results to the penalties in \eqref{eq:MSRLpen}, \eqref{eq:MSRGLpen}, and \eqref{eq:MSRLRpen}. We also argue that \eqref{eqMSRL}, like the univariate square-root lasso, can be interpreted as implicitly incorporating an estimate of the error precision matrix into the criterion for estimating $\mbbeta_*$. 
Through simulation studies, we show that \eqref{eqMSRL} can perform as well or better than methods that estimate $\mbbeta_*$ and $\mbOmega_*$ jointly, both of which outperform penalized least squares estimators when $\mbOmega_*$ has many nonzero off-diagonals. Based on our theory, we also study a tuning procedure that requires computing \eqref{eqMSRL} for only a single value of the tuning parameter, i.e., does not require cross-validation. Finally, we propose two algorithms to compute \eqref{eqMSRL} efficiently: one algorithm that can be used in any setting and has convergence guarantees, and a second algorithm that can be applied when $n > q$ and the tuning parameter is sufficiently large. Our algorithms are often 100 or more times faster than \texttt{CVX} in the simulation settings we consider. An R package implementing our method is available for download at \url{https://github.com/ajmolstad/MSRL}. 

Before we proceed, we define notation which will be used throughout the article. Let $\mbI_s$ be the $s \times s$ identity matrix. When we write $(\mbU,\mbD,\mbV) = {\rm svd}(\mbA)$, we refer to the components of the singular value decomposition $\mbA = \mbU\mbD\mbV^\top \in \mathbb{R}^{a \times b}$, where letting $s = {\rm min}(a, b)$,  $\mbU \in \mathbb{R}^{a \times s}$ and $\mbV \in \mathbb{R}^{b \times s}$ with $\mbU^\top\mbU = \mbV^\top\mbV = \mbI_s$, and 
$\mbD \in \mathbb{R}^{s \times s}$ is a diagonal matrix with $\mbD_{k,k} = \sigma_k(\mbA) \geq 0$ for $k \in \{1, \dots, s\}$. Define the norms $\|\mbA\|_F = (\sum_{j,k} \mbA_{j,k}^2)^{1/2}$, $\|\mbA\|_{\infty} = \max_{j,k}|\mbA_{j,k}|$, $\|\mbA\| = \sigma_1(\mbA)$, and let $\varphi_j(\mbA)$ denote the $j$th largest eigenvalue of square matrix $\mbA$. For a symmetric matrix $\mbM$, define $\|\mbA\|_\mbM^2 = {\rm tr}(\mbA^\top\mbM \mbA)$. Let $\|\mbA\|_{\infty,2} = \max_{j}\|\mbA_{j,\cdot}\|_2$, where $\|\mba\|_2$ denotes the Euclidean norm of a vector $\mba$ and $\mbA_{j,\cdot}$ denotes the $j$th row of $\mbA.$ Similarly, let $\mbA_{\cdot,k}$ denote the $k$th column of $\mbA.$ For a subspace $\mathcal{R} \subseteq \mathbb{R}^d$, define the orthogonal complement of $\mathcal{R}$ as $\mathcal{R}^\perp = \{\mbv\in \mathbb{R}^d : \mbv^\top \mbu = 0~ \text{for all}~\mbu \in \mathcal{R}\}.$ For a set $\mathcal{T}$, let $|\mathcal{T}|$ be its cardinality. For sequences $a_n$ and $b_n$, we use the notation $a_n \lesssim b_n$ to mean that there exists a constant 
$K > 0$ such that $a_n \leq K b_n$ for all $n$ sufficiently large.
Finally, let $[n] =\{1, 2, \dots, n\}$ for all positive integers $n$.

\section{The Multivariate Square-root Lasso}

\subsection{Implicit Covariance Estimation}
If the $\mbepsilon_i$ were multivariate normal and the precision matrix $\mbOmega_*$ were known, the penalized maximum likelihood estimator of $\mbbeta_*$ would be 
\begin{equation} \label{eq:part_maxLik}
\argmin_{\mbbeta \in \mathbb{R}^{p \times q}} \left[ \frac{1}{n} {\rm tr}\big\{(\mbY - \mbX\mbbeta)\mbOmega_*(\mbY - \mbX\mbbeta)^\top \big\} + \lambda g(\mbbeta)\right],
\end{equation}
which can be interpreted as a penalized weighted least squares estimator. Based on the first order conditions for \eqref{eq:part_maxLik}, it can be verified that the solution depends on the error precision $\mbOmega_*$. Of course, \eqref{eq:part_maxLik} cannot be used in practice because $\mbOmega_*$ is generally unknown. Instead, a popular alternative proposed by \citet{rothman2010sparse} is the jointly penalized maximum likelihood estimator
\begin{equation} \label{eq:part_maxLik_2}
\argmin_{\mbbeta \in \mathbb{R}^{p \times q}, \mbOmega \in \mathbb{S}^q_+} \left[ \frac{1}{n} {\rm tr}\big\{(\mbY - \mbX\mbbeta)\mbOmega(\mbY - \mbX\mbbeta)^\top\big\}  - \log {\rm det}(\mbOmega) + \lambda g(\mbbeta) + \gamma \sum_{j\neq k}|\mbOmega_{j,k}| \right],
\end{equation}
where $\gamma > 0$ is a user-specified tuning parameter. Unlike \eqref{eq:part_maxLik}, the optimization problem in \eqref{eq:part_maxLik_2} is nonconvex. Solving \eqref{eq:part_maxLik_2} requires iteratively updating $\mbbeta$ with $\mbOmega$ held fixed and vice versa \citep{rothman2010sparse}, which can be time-consuming in high-dimensional settings.

As an alternative to the computationally intensive task of solving \eqref{eq:part_maxLik_2}, one could instead plug an estimate of $\mbOmega_* = \mbSigma_*^{-1}$ into \eqref{eq:part_maxLik}. However, standard estimators of $\mbSigma_*$ are themselves functions of $\mbbeta_*$, e.g., $n^{-1}(\mbY - \mbX\mbbeta_*)^\top(\mbY - \mbX\mbbeta_*)$. This naturally raises the question of whether one could construct a weighted least squares criterion like \eqref{eq:part_maxLik} wherein the weight is itself a function of the optimization variable $\mbbeta$. In fact, the nuclear norm can be interpreted in exactly this way because
$$ \frac{1}{\sqrt{n}}\| \mbY - \mbX \mbbeta\|_{*} =  \frac{1}{n}{\rm tr}\big\{(\mbY - \mbX\mbbeta)\tilde{\mbSigma}_{\mbbeta}^\dagger (\mbY - \mbX\mbbeta)^\top\big\}, $$
where the weight matrix $\tilde{\mbSigma}_{\mbbeta}$ is given by
$$
 \tilde{\mbSigma}_{\mbbeta} = \frac{1}{\sqrt{n}} \big\{(\mbY - \mbX\mbbeta)^\top(\mbY - \mbX\mbbeta)\big\}^{1/2},$$ and $\tilde{\mbSigma}_{\mbbeta}^{\dagger}$ is its Moore-Penrose pseudoinverse. That is, the nuclear norm of residuals can be expressed as a weighted least squares criterion where the weight matrix is an estimate of the square-root error precision matrix $\mbOmega_*^{1/2} = \mbSigma_*^{-1/2}$. Furthermore, the multivariate square-root lasso can, in some situations, be interpreted as jointly estimating the error covariance and regression coefficient matrix, like \eqref{eq:part_maxLik_2}.

\begin{lemma}\label{lemma:implicit}\cite[Lemma 1]{van2016chi2}
Define $\hat{\mbbeta}_g$ as the solution to \eqref{eqMSRL}, and define $(\bar{\mbbeta}_g, \bar{\mbSigma}_g^{1/2})$ as
\begin{equation}\label{eq:joint_interp}
 \argmin_{\mbbeta \in \mathbb{R}^{p \times q}, \mbSigma^{1/2} \in \mathbb{S}^{q}_+} \left[ \frac{1}{2n}{\rm tr} \big\{ (\mbY - \mbX\mbbeta)\mbSigma^{-1/2} (\mbY - \mbX\mbbeta)^\top\big\} + \frac{{\rm tr}(\mbSigma^{1/2})}{2}  + \lambda g(\mbbeta) \right],
\end{equation}
assuming the minimum is obtained for some $\mbSigma^{1/2} \in \mathbb{S}^q_+$. If $\mbY - \mbX\hat{\mbbeta}_g$ has $q$ nonzero singular values, then the estimator in \eqref{eq:joint_interp} satisfies $$\bar{\mbSigma}_g = \frac{1}{n} (\mbY-\mbX\hat{\mbbeta}_g)^\top(\mbY - \mbX \hat{\mbbeta}_g) ~~\text{  and  }~~\bar{\mbbeta}_g = \hat{\mbbeta}_g.$$
\end{lemma}
Lemma \ref{lemma:implicit} suggests that we can solve the joint optimization problem \eqref{eq:joint_interp} by solving \eqref{eqMSRL} directly: we need not explicitly estimate $\mbSigma_*$ or $\mbOmega_*$. It is in this sense that we argue \eqref{eqMSRL} implicitly estimates the error covariance. This is in contrast to \eqref{eq:part_maxLik_2}, which requires an explicit estimate of $\mbOmega_*$. In our simulation studies, we demonstrate that this implicit covariance estimation yields an estimator of $\mbbeta_*$ which performs similarly to methods that use $\mbOmega_*$, or an estimate thereof, in their estimation criterion. 

The relationship between \eqref{eqMSRL} and \eqref{eq:joint_interp} established in Lemma \ref{lemma:implicit} holds only when $\hat\mbbeta_g$ leads to a residual matrix with $q$ nonzero singular values. However, we emphasize that Lemma \ref{lemma:implicit} simply provides one way to characterize $\hat\mbbeta_g$ in this special setting. We do not require or assume that $\mbY - \mbX \hat\mbbeta_g$ has rank $q$. Our theory (Section \ref{sec:Theory}) and algorithm (Section \ref{sec:ADMM}) apply to \eqref{eqMSRL} even when $\mbY - \mbX \hat\mbbeta_g$ has fewer than $q$ nonzero singular values. In these settings, the interpretation of the nuclear norm of residuals as a weighted residual sum of squares with weight matrix $\tilde{\mbSigma}^\dagger_\mbbeta$ still applies, but $\hat\mbbeta_g$ cannot be interpreted as the solution to a joint optimization problem as neatly as in \eqref{eq:joint_interp}.

\subsection{Relationship to Existing Methods}
The univariate square-root lasso \citep{belloni2011square,sun2012scaled,bunea2014group,derumigny2018improved} is a special case of \eqref{eqMSRL} when $q=1$ and $g$ is the $L_1$-norm. However, there is an important difference between the univariate and multivariate square-root lasso estimators in terms of how they relate to their penalized least squares analogs. 

\begin{remark}\label{remark:solutionpath}
Suppose $g$ is the $L_1$-norm. When $q=1$, the univariate square-root lasso estimator has a solution path equivalent to that of the $L_1$-penalized least squares estimator \citep{tian2018selective}. When $q \geq 2$, \eqref{eqMSRL} and the $L_1$-penalized least squares estimator do not have equivalent solution paths in general.
\end{remark}

 Mathematically, Remark \ref{remark:solutionpath} follows from the fact that the unpenalized objective function for the multivariate square-root lasso (with $q \geq 2$), unlike the univariate square-root lasso, cannot be expressed as the square-root of its least squares analog, i.e., $\|\mbA\|_F \neq {\rm tr}\{(\mbA^\top\mbA)^{1/2}\}$ in general. Thus with $q \geq 2$, \eqref{eqMSRL} defines a class of estimators distinct from its least squares analog in the sense that their solution paths are distinct. Our simulation results show that the multivariate square-root lasso performs more like the normal penalized maximum likelihood estimator of \citet{rothman2010sparse}, which explicitly estimates the error precision matrix, than the penalized least squares estimator.

Our estimator \eqref{eqMSRL} is not the only multivariate generalization of the univariate square-root lasso.  \citet{liu2015calibrated} proposed an estimator which minimizes the sum of the Euclidean norm of residuals for each response plus a penalty on the optimization variable corresponding to $\mbbeta_*$. However, the method of \citet{liu2015calibrated} assumes that $\mbSigma_*$ is diagonal. In addition, when the penalty is separable across the columns of its matrix argument (e.g., when using \eqref{eq:MSRLpen}), the method of \citet{liu2015calibrated} is equivalent to performing $q$ separate univariate square-root lasso regressions with the same tuning parameter used for each response. In our simulation studies, the method of \citet{liu2015calibrated} outperforms penalized least squares estimators, but tends to be outperformed by \eqref{eqMSRL} when $\mbSigma_*$ is not diagonal. For more details, see our description of their method in Section \ref{sec:datagenmodels}.

\section{Statistical Properties}\label{sec:Theory}
\subsection{Overview}\label{subsec:theory_overview}
In this section, we establish Frobenius norm error bounds for \eqref{eqMSRL}. We first provide a general error bound, then specialize this result to penalties \eqref{eq:MSRLpen}, \eqref{eq:MSRGLpen}, and \eqref{eq:MSRLRpen}. 

For each of the following results, we assume that $\mbbeta_*$ belongs to a subspace $\mathcal{M}$, and choose the penalty $g$ according to $\mathcal{M}$. To make matters concrete, we define $\mathcal{M}$ as the \textit{model subspace} and assume throughout that $\mbbeta_* \in \mathcal{M}$. Let $\mathcal{M}^\perp$ denote the orthogonal complement of the model subspace $\mathcal{M}$. Define the subspace $\mathcal{N}^\perp$ as the \textit{perturbation subspace} and let $\mathcal{N}$ be its orthogonal complement. We will give concrete examples of $\mathcal{M}$ and $\mathcal{N}^\perp$ under the three different model assumptions momentarily.  For subspace pairs $(\mathcal{M},\mathcal{N}^\perp)$ for which $\mathcal{M} \subseteq \mathcal{N}$, a penalty function $g$ is said to be \textit{decomposable} with respect to the subspace pair $(\mathcal{M}, \mathcal{N}^\perp)$ if
$ g(\mbA + \mbB) = g(\mbA) + g(\mbB)$
for all $\mbA \in \mathcal{M}$ and $\mbB \in \mathcal{N}^\perp$. 
See \citet{negahban2012unified} for a further discussion of model subspaces, perturbation subspaces, and decomposability.

Throughout, let $\tilde{g}$ denote the dual norm of $g$, and define the \textit{subspace compatibility constant} \citep{negahban2012unified} with respect to $g$ as $$\Psi_g(\mathcal{N}) = \sup_{\mbA \in \mathcal{N}\setminus\{0\}} \frac{g(\mbA)}{\|\mbA\|_F}.$$

In the following, we consider three model subspaces for $\mbbeta_*$ (\textbf{M1}--\textbf{M3}): each corresponds to a distinct subspace pair and decomposable penalty function $g$.
\begin{itemize}
       \item[]\textbf{M1.} (Elementwise sparsity) We assume that many entries of $\mbbeta_*$ are zero. Letting $\mathcal{S} = \{(j,k): \mbbeta_{*j,k} \neq 0~,(j,k) \in [p] \times [q]\}$, define the subspace pair
       $$\mathcal{M}_{\rm L} = \{\mbbeta \in \mathbb{R}^{p \times q}: \mbbeta_{j,k} = 0, (j,k) \not\in \mathcal{S}\}, ~~ \mathcal{N}_{\rm L}^\perp = \{\mbbeta \in \mathbb{R}^{p \times q}: \mbbeta_{j,k} = 0, (j,k) \in \mathcal{S}\}.$$ 
       The penalty function $g(\cdot) = \|\cdot\|_1$ is decomposable with respect to $(\mathcal{M}_{\rm L}, \mathcal{N}_{\rm L}^\perp)$ and $\Psi_{\|\cdot\|_1}(\mathcal{N}_{\rm L}) \leq \sqrt{|\mathcal{S}|}$ \citep{negahban2012unified}.
       \item[]\textbf{M2.} (Row-wise sparsity) We assume that many rows of $\mbbeta_*$ are entirely zero. Letting $\mathcal{G} = \{j: \mbbeta_{*j,\cdot} \neq 0,~ j \in [p]\}$, define the subspace pair
       $$\mathcal{M}_{\rm GL} = \{\mbbeta \in \mathbb{R}^{p \times q}: \mbbeta_{j,\cdot} = 0, j \not\in \mathcal{G}\}, ~~ \mathcal{N}_{\rm GL}^\perp = \{\mbbeta \in \mathbb{R}^{p \times q}: \mbbeta_{j,\cdot} = 0, j \in \mathcal{G}\}.$$
       The penalty function $g(\cdot) = \|\cdot\|_{1,2}$ is decomposable with respect to $(\mathcal{M}_{\rm GL}, \mathcal{N}_{\rm GL}^\perp)$ and $\Psi_{\|\cdot\|_{1,2}}(\mathcal{N}_{\rm GL}) \leq \sqrt{|\mathcal{G}|}$ \citep{liu2015calibrated}.
       \item[]\textbf{M3.} (Low-rankness) We assume that ${\rm rank}(\mbbeta_*) = r$ where $r \ll \min(p,q)$. Letting $(\U_*, \D_*, \V_*) = {\rm svd}(\mbbeta_*)$, define $\mathcal{U} = {\rm span}(\mbu_{*1}, \dots, \mbu_{*{r}})$ and $\mathcal{V} = {\rm span}(\mbv_{*1}, \dots, \mbv_{*{r}})$ where $\mbu_{*k}$ and $\mbv_{*k}$ are the $k$th columns of $\mbU_*$ and $\mbV_*$, respectively, for $k \in [r].$ Let $\mathcal{U}^\perp$ and $\mathcal{V}^\perp$ denote the orthogonal complements of $\mathcal{U}$ and $\mathcal{V}$, respectively. Define the subspace pair
       $$ \mathcal{M}_{\rm LR} = \{ \mbbeta \in \mathbb{R}^{p \times q}: {\rm row}(\mbbeta) \subseteq \mathcal{V}, {\rm col}(\mbbeta)\subseteq \mathcal{U}\},$$ 
       $$\mathcal{N}_{\rm LR}^\perp = \{ \mbbeta \in \mathbb{R}^{p \times q}: {\rm row}(\mbbeta) \subseteq \mathcal{V}^\perp, {\rm col}(\mbbeta) \subseteq \mathcal{U}^\perp\},$$
       where ${\rm row}(\mbA)$ and ${\rm col}(\mbA)$ are the row and column spaces of a matrix $\mbA$, respectively.
       The penalty function $g(\cdot) = \|\cdot\|_{*}$ is decomposable with respect to $(\mathcal{M}_{\rm LR}, \mathcal{N}_{\rm LR}^\perp)$ and $\Psi_{\|\cdot\|_*}(\mathcal{N}_{\rm LR}) \leq \sqrt{2r}$ \citep{negahban2011}.
\end{itemize}

In the following subsection, we establish error bounds for $\hat\mbbeta_g$, the solution to \eqref{eqMSRL} for decomposable $g$. These bounds allow both $p$ and $q$ to grow with the sample size $n$. In Section \ref{subsec:normal_errors}, we then specialize our result to the case that the errors are multivariate normal.    

\subsection{Pivotal Estimation}\label{subsec:pivotal}
Throughout the remainder of this section, we treat $\mbX$ as nonrandom. For ease of display, we let $\check{c} = (c+1)/(c-1)$ and $\tilde{c} = c(c+1)/(c-1)$ for constant $c > 1$. We will require the following condition and assumptions. 
\begin{itemize}
       \item[] \textbf{C1.} The columns of $\mbX$ are scaled so that $\|\mbX_{\cdot,j}\|_2 = \sqrt{n}$ for $j \in [p]$.
 \item[] \textbf{A1.} The $n \times q$ error matrix $\mbE = (\mbepsilon_1, \dots, \mbepsilon_n)^\top$ has $q$ nonzero singular values almost surely. 
 \item[] \textbf{A2.} The distribution of the error matrix $\mbE$ is left-spherical, i.e., for any $n \times n$ orthogonal matrix $\mbO$, 
$\mbO\mbE$ has the same matrix-variate distribution as $\mbE$. 
\end{itemize}
Assumption \textbf{A1} requires that the sample size $n$ is at least as large as the number of responses $q$. Given $n \geq q$, assumptions \textbf{A1} and \textbf{A2} would hold if, for example, the rows of $\mbE$ were independent and each row followed a mean zero multivariate normal distribution with covariance $\mbSigma_* \in \mathbb{S}^q_+.$ Condition \textbf{C1} is simply a matter of rescaling the columns of $\mbX$.

In addition to Assumptions \textbf{A1} and \textbf{A2}, our bounds will depend on the quantity
$$\phi_{\mathcal{E},g}(\mathcal{M}, \mathcal{N}, c) = \inf_{\mbDelta \in \mathcal{C}_{g}(\mathcal{M},\mathcal{N}, c)} \left\{ \frac{\sup_{\|\boldsymbol{Q}\| \leq 1} {\rm tr}\left\{ (\boldsymbol{Q} - \U_\epsilon \V_\epsilon^\top)^\top(\mbE - \mbX \Delta) \right\} }{ \sqrt{n}\|\mbDelta\|_F^2}\right\},$$
$$\mathcal{C}_g(\mathcal{M}, \mathcal{N}, c) = \{\mbDelta \in \mathbb{R}^{p \times q}\hspace{-2pt}: \mbDelta  \neq 0, ~g(\mbDelta_{\mathcal{N}^\perp}) \leq \check{c} g(\mbDelta_{\mathcal{N}}) \}, ~~~ (\U_\epsilon, \D_\epsilon, \V_\epsilon) = {\rm svd}(\mathcal{\mbE}),$$
where $\mbDelta_{\mathcal{N}}$ denotes the projection of $\mbDelta$ onto $\mathcal{N}$, i.e., $\mbDelta_{\mathcal{N}} = \argmin_{\mbM \in \mathcal{N}}\|\mbDelta - \mbM\|_F^2$. Using the dual characterization of the nuclear norm, it is immediate that the $\mbQ$ which maximizes the numerator of $\phi_{\mathcal{E},g}(\mathcal{M}, \mathcal{N}, c)$ is $\mbQ = \tilde{\U}\tilde{\V}^\top$, where $(\tilde{\U},\tilde{\D},\tilde{\V}) = {\rm svd}(\mbE - \mbX\boldsymbol{\Delta})$.

The quantity $\phi_{\mathcal{E},g}(\mathcal{M},\mathcal{N}, c)$ is needed to ensure the restricted strong convexity \citep{negahban2012unified} of the nuclear norm of residuals. For this, we have another assumption. 
\begin{itemize}
       \item[] \textbf{A3.} There exists a constant $v$ such that $\phi_{\mathcal{E},g}(\mathcal{M},\mathcal{N}, c) \geq v > 0$ almost surely.
\end{itemize}
The quantity $\phi_{\mathcal{E},g}(\mathcal{M},\mathcal{N}, c)$ is closely related to the restricted eigenvalue of $\mbX$ \citep{raskutti2010restricted}, but also depends on $\mathcal{\mbE}$. As we will show in the next section, under some additional assumptions on the error matrix $\mathcal{\mbE}$ and the matrix $\mbX$, $\phi_{\mathcal{E},g}(\mathcal{M},\mathcal{N}, c)$ can be replaced with a restricted eigenvalue-type quantity which does not depend on $\mathcal{\mbE}$.




We are now ready to state our first error bound. The proof of this and all subsequent results can be found in Appendix \ref{appendixB}. 
\begin{theorem}\label{prop:main_prop}
For any fixed constant $c > 1$, define the event $\mathcal{A}_c = \{ \lambda \geq (c/\sqrt{n}) \tilde{g}(\mbX^\top\U_\epsilon \V_\epsilon^\top)\}$. If \textbf{C1}, \textbf{A1}, and \textbf{A3} hold, as long as $g$ is decomposable with respect to the subspace pair $(\mathcal{M}, \mathcal{N}^\perp)$, then $$\|\hat{\mbbeta}_g - \mbbeta_*\|_F \leq \frac{\check{c} \hspace{1pt}\Psi_g(\mathcal{N}) \hspace{1pt}\lambda}{\phi_{\mathcal{E},g}(\mathcal{M}, \mathcal{N},c)}$$
with probability at least $P(\mathcal{A}_c).$
If \textbf{A2} also holds, then the distribution of $\mbX^\top\U_\epsilon \V_\epsilon^\top$ does not depend on $\mbSigma_*$, i.e., $(c/\sqrt{n}) \tilde{g}(\mbX^\top\U_\epsilon \V_\epsilon^\top)$ is pivotal with respect to the unknown error covariance. 
\end{theorem}
Theorem \ref{prop:main_prop} reveals that optimal value of the tuning parameter $\lambda$ depends on the random quantity $\tilde{g}(\mbX^\top\U_\epsilon \V_\epsilon^\top)$. 
Under \textbf{A1} and \textbf{A2}, $\U_\epsilon \V_\epsilon^\top$ is a random matrix uniformly distributed on the set of matrices $V_q(n) = \{\mbS \in \mathbb{R}^{n \times q}: \mbS^\top\mbS = \mbI_q\}$ \citep{eaton,meckes2019random} regardless of $\mbSigma_*.$  
This result suggests that the tuning parameter $\lambda$ could be selected according to the quantiles of $\tilde{g}(\mbX^\top\mbS)$ where $\mbS$ is uniformly distributed on $V_q(n)$. For example, the result of Theorem \ref{prop:main_prop} would hold with probability $1 - \alpha$ if we selected $\lambda$ equal to the $(1-\alpha)$th quantile of $ (c/\sqrt{n}) \tilde{g}(\mbX^\top\mbS)$. Fortunately, we can easily sample from the distribution of $(c/\sqrt{n}) \tilde{g}(\mbX^\top\mbS)$, so we can approximate its quantiles using Monte Carlo. We study this tuning approach empirically in Section \ref{sec:sim_studies}.

We can use the distribution of $\mbX^\top\U_\epsilon \V_\epsilon^\top$ to establish explicit choices of $\lambda$ which yield more insightful error bounds under the three penalties discussed in Section \ref{sec:intro}.

\begin{corollary}\label{thm:asymp}
Suppose \textbf{C1} and \textbf{A1}--\textbf{A3} hold. 
\begin{itemize}
\item[] (i) Under \textbf{M1}, if $\lambda = c \{ 2\log(2pq^{k})/(n-1)\}^{1/2}$ and $ n > 2 \log(2 p q^{k}) + 1$ for fixed constants $c > 1$ and $k > 1$, then with probability at least $1 - q^{1-k},$
$$\|\hat{\mbbeta}_{\rm L} - \mbbeta_*\|_F \leq   \frac{\tilde{c}}{\phi_{\mathcal{E}, \|\cdot\|_1}(\mathcal{M}_{\rm L},\mathcal{N}_{\rm L},c)} \sqrt{\frac{2|\mathcal{S}| \log(2pq^{k})}{n-1}}.$$
\item[] (ii) Under \textbf{M2}, if $\lambda = c \{4 k \log p/(n-2)\}^{1/2} + c(q/n)^{1/2}$ and $k \log p > 4\pi$ for fixed constants $c > 1$ and $k >1$, then with probability at least $1 - p^{1 - k}$, 
$$\|\hat{\mbbeta}_{\rm GL} - \mbbeta_*\|_F \leq   \frac{ 2 \hspace{1pt} \tilde{c}}{\phi_{\mathcal{E},  \|\cdot\|_{1,2}}(\mathcal{M}_{\rm GL},\mathcal{N}_{\rm GL},c)}\left(\sqrt{\frac{ k |\mathcal{G}|\log p}{n-2}} + \sqrt{\frac{|\mathcal{G}|q}{4 n}}  \right).$$
\item[] (iii) Under \textbf{M3}, if $\lambda = 4 c \|\mbX\|[k_2 (p+q)/\{n(n-2)\}  ]^{1/2} + 4 c \|\mbX\|/n$ and $k_2\|\mbX\|^2(p + q) > 16 n \pi$ for fixed constants $c > 1$, $k_1 > 1$, and $k_2 = 4\log(7 + k_1)$, then with probability at least $1 - \{8/(7 + k_1)\}^{p+q},$
$$\|\hat{\mbbeta}_{\rm LR} - \mbbeta_*\|_F \leq   \frac{ 4 \hspace{1pt}\tilde{c}}{\phi_{\mathcal{E},  \|\cdot\|_*}(\mathcal{M}_{\rm LR},\mathcal{N}_{\rm LR},c)}\left(\frac{\|\mbX\|}{\sqrt{n}}\right)\left(\sqrt{\frac{2 k_2 r(p + q)}{n-2}} + \sqrt{\frac{2 r}{n}}  \right).$$
\end{itemize}
\end{corollary}

Corollary \ref{thm:asymp} demonstrates that we can set $\lambda$ equal to explicit quantities which will satisfy the condition of Theorem \ref{prop:main_prop} with high probability and do not depend on any unknown parameters. 

Before concluding this section, we emphasize that assumptions \textbf{A1} and \textbf{A2} are not assumptions on the residual matrix $\hat{\mbE}_g = \mbY - \mbX \hat{\mbbeta}_g$, but on the error matrix $\mbE$. After a version of this article had appeared on arXiv, \citet{massias2020support} derived bounds for $\|\hat{\mbbeta}_{\rm GL} - \mbbeta_*\|_{\infty,2}$. However, they required the assumption that $\hat{\mbE}_g$ was rank $q$. Of course, $\hat{\mbE}_g$ depends on both the random error matrix $\mbE$ and the estimator $\hat\mbbeta_g$ itself, so it not clear when their required choice of $\lambda$ would lead to a violation of this assumption.


\subsection{Asymptotics with Normal Errors}\label{subsec:normal_errors}
While Theorem \ref{prop:main_prop} and Corollary \ref{thm:asymp} verify that $\lambda$ can be chosen according to the quantiles of a pivotal quantity, we have not made any particular distributional assumptions on $\mbE$, which $\phi_{\mathcal{E},g}(\mathcal{M}, \mathcal{N}, c)$---itself a random quantity---depends upon. In this section, we establish asymptotic error bounds for \eqref{eqMSRL} under normality assumptions on $\mbE.$ To do so, we drop assumptions \textbf{A1} and \textbf{A2}, and add restricted eigenvalue-type conditions on the matrix $\mbX$ to replace \textbf{A3}. The assumptions we require are as follows.

\begin{itemize}
\item[] \textbf{A4.} The rows of $\mbE$ are independent and identically distributed from ${\rm N}_q(0, \mbSigma_*)$. Moreover, there exists a constant $v$ such that 
$$ 0 < v^{-1} \leq \varphi_q(\mbSigma_*) \leq \varphi_1(\mbSigma_*) \leq v < \infty.$$
\item[] \textbf{A5.}  
There exists a constant $\bar{v}$ such that
$$ \sup_{\mbDelta \in \mathcal{C}_g(\mathcal{M}, \mathcal{N}, c)} \frac{\|\mbX\mbDelta\|_F^2}{n \|\mbDelta\|_F^2}   \leq \bar{v} < \infty.$$
In addition, there exists a constant $\underline{v}$ such that $0 < \underline{v} \leq \nu_{g}(\mathcal{M},\mathcal{N}, c)$ where 
$$ {\nu}_{g}(\mathcal{M}, \mathcal{N}, c) = \inf_{\substack{\mbDelta \in \mathcal{C}_g(\mathcal{M},\mathcal{N},c)\\ \mbS \in V_q(n)}} \left\{ \frac{\sum\limits_{i=1}^q \sum\limits_{j=1}^q (\mbu_j^\top \mbX \mbDelta \mbv_i - \mbu_i^\top\mbX \mbDelta \mbv_j)^2 + 4 \sum\limits_{k = q+1}^{n} \sum\limits_{j=1}^q(\mbu_k^\top \mbX \mbDelta \mbv_j)^2}{4 n \|\mbDelta\|_F^2} \right\} 
$$
with $(\mbU, \mbI_q, \mbV) = {\rm svd}(\mbS)$, where  $\mbu_j$ denotes the $j$th column of $\mbU \in \mathbb{R}^{n \times q}$ for $j \in [q]$, $\mbv_l$ denotes the $l$th column of $\V \in \mathbb{R}^{q \times q}$ for $l \in [q]$, and $\mbu_k$ denotes the $(k - q)$th column of $\U_0 \in \mathbb{R}^{n \times (n-q)}$ where $\U_0^\top\U = 0$ and $\U_0^\top\U_0 = \mbI_{n-q}$ for $k \in \{q + 1, q + 2, \dots, n\}.$
\item[] \textbf{A6.} As $n \to \infty$, $q/n \to t$ for some $t \in (0,1).$
\end{itemize}
Assumption \textbf{A4} is standard in the multivariate response linear regression and precision matrix estimation literature. Assumptions \textbf{A4} and \textbf{A6} together would imply assumptions \textbf{A1} and \textbf{A2} asymptotically. Assumption \textbf{A5} consists of restricted eigenvalue-type conditions. The latter assumption made in \textbf{A5}, while tailored specifically to apply to our problem, can be seen as analogous to the standard restricted eigenvalue condition in penalized least squares \citep{raskutti2010restricted}. For example, we can write the spectral norm of $\mbX \mbDelta$ in variational form as 
$ \|\mbX\mbDelta\| = \sup_{\mbu \in S^{n-1}} \sup_{\mbv \in S^{q-1}} \mbu^\top\mbX\mbDelta \mbv,$ where $S^{n-1} = \{\mbu \in \mathbb{R}^{n}: \|\mbu\|_2 = 1\}.$ Both parts of \textbf{A5} are needed to establish the restricted strong convexity of the nuclear norm of residuals. 

With \textbf{A4}--\textbf{A6}, we are now ready to state a version of Theorem \ref{prop:main_prop} which applies to normally distributed error matrix  $\mbE$.  
\begin{theorem}\label{main_normal}
For fixed constants $c > 1$ and $d>1$, define the events $\mathcal{A}_c = \{ \lambda \geq (c/\sqrt{n})\tilde{g}(\mbX^\top\U_\epsilon \V_\epsilon^\top)\}$ and $\mathcal{B}_d = \{ \sigma^2_1(\mbE)/n \geq (t+d) \hspace{1pt} \varphi_1(\mbSigma_*)\}$. If \textbf{C1} and \textbf{A4}--\textbf{A6} hold, $g$ is decomposable with respect to the subspace pair $(\mathcal{M}, \mathcal{N}^\perp)$, and $\Psi_g(\mathcal{N})\lambda \to 0$ as $n \to \infty$, then  $$\|\hat{\mbbeta}_g - \mbbeta_*\|_F \lesssim \frac{(t+d) \hspace{1pt} \check{c} \hspace{1pt}\varphi_1^{1/2}(\mbSigma_*) \Psi_g(\mathcal{N})\lambda}{\nu_{g}(\mathcal{M},\mathcal{N},c)}$$
with probability at least $P(\mathcal{A}_c \cap \mathcal{B}_d)$ for $n$ sufficiently large.
\end{theorem}

In Theorem \ref{main_normal}, we have essentially replaced the random quantity $\phi_{\mathcal{E}, g}(\mathcal{M}, \mathcal{N}, c)$ from Theorem \ref{prop:main_prop} with a constant times $\nu_g(\mathcal{M}, \mathcal{N}, c)/\{(t+d)\varphi_1\textsuperscript{1/2}(\mbSigma_*)\}$, which is nonrandom. 

Applying the same concentration inequalities used to obtain the bounds in Corollary \ref{thm:asymp}---along with a concentration inequality on the largest singular value of the matrix $\mbE$---we arrive at the following set of asymptotic results concerning \eqref{eqMSRL} with penalties \eqref{eq:MSRLpen}, \eqref{eq:MSRGLpen}, and \eqref{eq:MSRLRpen}. 

\begin{corollary}\label{cor6}
Suppose \textbf{C1} and \textbf{A4}--\textbf{A6} hold. \begin{itemize}
\item[] (i) Under \textbf{M1}, if $\lambda = c \{2\log(2 pq^k)/(n-1)\}^{1/2}$,  $n > 2 \log(2pq^k) + 1$, and $|\mathcal{S}|\log(pq^k) = o(n)$ for fixed constants $c > 1$, $k>1$, and $d > 1$, then
\begin{equation}\label{eq:lasso}
\|\hat{\mbbeta}_{\rm L} - \mbbeta_*\|_F \lesssim  \frac{(t+d) \hspace{1pt}\tilde{c}  \hspace{1pt}\varphi_1^{1/2}(\mbSigma_*)}{\nu_{\|\cdot\|_1}(\mathcal{M}_{\rm L},\mathcal{N}_{\rm L},c)} \sqrt{\frac{|\mathcal{S}| \log(2pq^k)}{n-1}}
\end{equation}
with probability at least $1- q^{1-k} - 2e^{-(d-1)^2n/4}$ for $n$ sufficiently large.
\item[] (ii) Under \textbf{M2}, if $\lambda = c \{4k\log p/(n-2)\}^{1/2} + c(q/n)^{1/2}$, $k \log p > 4\pi,$ and $|\mathcal{G}|\max(\log p, q) = o(n)$ for fixed constants $c > 1$, $k>1$, and $d > 1$, then
\begin{equation}\label{eq:grouplasso}
\|\hat{\mbbeta}_{\rm GL} - \mbbeta_*\|_F \lesssim    \frac{(t+d)\hspace{1pt}\tilde{c}\hspace{1pt}\hspace{1pt}\varphi_1^{1/2}(\mbSigma_*) }{\nu_{\|\cdot\|_{1,2}}(\mathcal{M}_{\rm GL},\mathcal{N}_{\rm GL},c)}\left(\sqrt{\frac{k |\mathcal{G}|\log p}{n-2}} + \sqrt{\frac{|\mathcal{G}|q}{4 n}}  \right)
\end{equation}
with probability at least $1 -  p^{1 - k} - 2e^{-(d-1)^2n/4}$  for $n$ sufficiently large. 
\item[] (iii) Under \textbf{M3}, if $\lambda = 4 c \|\mbX\|[k_2(p+q)/\{n(n-2)\}]^{1/2} + 4 c \|\mbX\|/n$, $k_2 \|\mbX\|^2 (p+q) > 16 n \pi$, and $\|\mbX\|^2 r(p+q)  = o(n^2)$ for fixed constants $c > 1$, $k_1 > 1$, $k_2 = 4 \log (7 + k_1)$, and $d > 1$, then 
\begin{equation}\label{eq:NN}
\|\hat{\mbbeta}_{\rm LR} - \mbbeta_*\|_F \lesssim    \frac{ (t + d)\hspace{1pt}\tilde{c} \hspace{1pt}\varphi_1^{1/2}(\mbSigma_*) }{\nu_{\|\cdot\|_*}(\mathcal{M}_{\rm LR},\mathcal{N}_{\rm LR},c)}\left(\frac{\|\mbX\|}{\sqrt{n}}\right) \left(\sqrt{\frac{k_2 r(p + q)}{n-2}} + \sqrt{\frac{r}{n}}\right)
\end{equation}
with probability at least $1 -  \{8/(7 + k_1)\}^{p+q} - 2e^{-(d-1)^2n/4}$ for $n$ sufficiently large.
\end{itemize}
\end{corollary}
The error bounds in Corollary \ref{cor6} agree with those in the existing literature on penalized least squares estimators. For example, the bound in \eqref{eq:lasso} is asymptotically equivalent to the bound of \citet{price2018cluster}, who studied a version of the $L_1$-penalized least squares estimator. Similarly, the bound in \eqref{eq:grouplasso} coheres with the bound for the group lasso-penalized least squares estimator from \citet{lounici2011oracle}. Finally, our bounds for the nuclear norm penalized version of \eqref{eqMSRL} asymptotically agree with their least squares analog from \citet{negahban2011}. 


In Section \ref{sec:discussion}, we discuss the challenges in establishing the conditions for \eqref{eqMSRL} to consistently estimate the support of $\mbbeta_*$ under \textbf{M1} or \textbf{M2}, or the true rank of $\mbbeta_*$ under \textbf{M3}. In brief, because the nuclear norm of residuals is nondifferentiable in general, applying standard proof techniques (e.g., see the proof of Theorem 3.4 of \citet{lee2015model}) is difficult.

To conclude this section, we discuss a potential limitation of our theory. Assumptions \textbf{A1} and \textbf{A6} require that the number of subjects, $n$, is at least as large as the number of responses, $q$. However, we emphasize that \eqref{eqMSRL} can still be applied and perform well in finite sample settings where $q > n$, as we show in Section \ref{subsec:qgreatern} of Appendix A. We discuss possible ways to relax this requirement in Section \ref{sec:discussion}.

\section{Computation}
\subsection{Properties of the Solution}
In the low-dimensional setting, the minimizer of the unpenalized nuclear norm of residuals is equivalent to the minimizer of the unpenalized squared Frobenius norm of residuals. That is, the least squares estimator $(\mbX^\top\mbX)^{-1}\mbX^\top\mbY$, when it exists, is a minimizer of \eqref{eqMSRL} when $\lambda = 0$. 
Of course, the penalized solution to \eqref{eqMSRL} does not coincide with the penalized least squares estimator. This can be seen by examining the first order conditions for \eqref{eqMSRL} which we characterize in the following remark.
\begin{remark} When $\mbY - \mbX\hat{\mbbeta}_g$ has $q$ nonzero singular values, the first order conditions for \eqref{eqMSRL}, which are necessary and sufficient for optimality, are
 \begin{equation}  \frac{1}{\sqrt{n}} \mbX^\top(\mbY - \mbX\hat{\mbbeta}_g)[(\mbY - \mbX\hat{\mbbeta}_g)^\top(\mbY - \mbX\hat{\mbbeta}_g)]^{-1/2} \in \lambda \partial g(\hat{\mbbeta}_g)\label{eq:first_order}
 \end{equation}
where $\partial g(\hat{\mbbeta}_g)$ is the subdifferential of $g$ at $\hat{\mbbeta}_g.$  
If $\mbY - \mbX\hat{\mbbeta}_g$ has fewer than $q$ nonzero singular values, the first order conditions for \eqref{eqMSRL} are $$ \frac{1}{\sqrt{n}} \mbX^\top(\mbU_{\hat{\epsilon}}\mbV_{\hat{\epsilon}}^\top + \mbZ_1)=  \lambda \mbZ_2 ,$$
for some $\mbZ_2 \in \partial g(\hat{\mbbeta}_g)$ and $\mbZ_1 \in \{\mbZ \in \mathbb{R}^{n \times q}: \|\mbZ\| \leq 1, \mbU_{\hat{\epsilon}}^\top\mbZ = 0,   \mbZ \mbV_{\hat{\epsilon}} = 0, (\mbU_{\hat{\epsilon}}, \mbD_{\hat{\epsilon}}, \mbV_{\hat{\epsilon}}) = {\rm svd}(\mbY - \mbX \hat{\mbbeta}_g)\}.$
\end{remark}
The residual matrix $\mbY - \mbX\hat{\mbbeta}_g$ can only have $q$ nonzero singular values when $n > q$ and when $\lambda$ is sufficiently large (see Section \ref{eq:alt_comp}). In these cases, we could use \eqref{eq:first_order} as a termination criterion. 



\subsection{Prox-linear ADMM Algorithm}\label{sec:ADMM}
To compute \eqref{eqMSRL}, we must address that the nuclear norm of residuals is nondifferentiable in general. To do so, we employ a variation of the alternating direction method of multipliers (ADMM) algorithm which decouples the nuclear norm of residuals and penalty $g$ \citep{boyd2011distributed}. Throughout this and the subsequent section, we will refer to a proximal operator of a function $f$, defined as
$$ {\rm Prox}_f(\mbB) = \argmin_{\mbA}\left\{ \frac{1}{2}\|\mbA - \mbB\|_F^2 + f(\mbA) \right\}.$$
When $f$ is a proper and lower semi-continuous convex function, its proximal operator is unique \citep{parikh2014proximal,polson2015proximal}.
The proximal operators corresponding to \eqref{eq:MSRLpen}, \eqref{eq:MSRGLpen}, and \eqref{eq:MSRLRpen} all have closed forms and can be computed efficiently (see Table 1 of the Supplementary Materials to \citet{molstad2018explicit}). 

To apply the ADMM algorithm, following \citet{boyd2011distributed}, we first introduce an additional variable $\mbPhi\in \mathbb{R}^{n \times q}$ so that we may rewrite the problem in \eqref{eqMSRL} as the constrained optimization problem
\begin{equation} \minim_{\mbbeta \in \mathbb{R}^{p \times q}, \mbPhi \in \mathbb{R}^{n \times q}}  \big\{ \|\mbPhi\|_{*}  +  \tilde{\lambda} g(\mbbeta) \big\}~~~ \text{subject to} ~~~\mbPhi = \mbY - \mbX\mbbeta \label{eq:constrained},
 \end{equation}
 where $\tilde{\lambda} = \sqrt{n}\lambda.$
Then, we define the augmented Lagrangian for the constrained problem in \eqref{eq:constrained} as
$$ \mathcal{F}_\rho(\mbbeta, \mbPhi, \mbGamma) = \|\mbPhi\|_{*} + \tilde{\lambda}g(\mbbeta) + {\rm tr}\{\mbGamma^\top(\mbY - \mbX\mbbeta - \mbPhi) \} + \frac{\rho}{2}\|\mbY - \mbX\mbbeta - \mbPhi\|_F^2,$$
where $\rho > 0$ is fixed and $\mbGamma \in \mathbb{R}^{n \times q}$ is the Lagrangian dual variable. 
The updating equations for the $(k+1)$th iterate of the standard ADMM algorithm are
\begin{align}
  \mbbeta^{(k+1)} &= \argmin_{\mbbeta \in \mathbb{R}^{p \times q}} \mathcal{F}_{\rho}(\mbbeta, \mbPhi^{(k)}, \mbGamma^{(k)})\label{eq:beta_update}\\
  \mbPhi^{(k+1)} &= \argmin_{\mbPhi \in \mathbb{R}^{n \times q}} \mathcal{F}_{\rho}(\mbbeta^{(k+1)}, \mbPhi, \mbGamma^{(k)}) \label{eq:phi_update}\\
  \mbGamma^{(k+1)} &= \mbGamma^{(k)} + \tau \rho(\mbY - \mbX\mbbeta^{(k+1)} - \mbPhi^{(k+1)}),
\end{align}
where $\tau > 0$ modifies the step size for the dual variable update. The $\mbPhi$ updating equation of the ADMM algorithm, \eqref{eq:phi_update}, can be expressed in terms of the proximal operator of the nuclear norm
\begin{align*} 
\mbPhi^{(k+1)} & = {\rm Prox}_{\rho^{-1}\|\cdot\|_*}\left(\mbY + \rho^{-1} \mbGamma^{(k)} - \mbX\mbbeta^{(k+1)}\right),
\end{align*}
which can be solved efficiently in closed form by computing the singular value decomposition of $\mbY + \rho^{-1} \mbGamma^{(k)} - \mbX\mbbeta^{(k+1)}$ and soft thresholding its singular values (e.g., see 3 and 4 of Algorithm 1).

When $p$ is large, the first step of the ADMM algorithm, \eqref{eq:beta_update}, is more computationally burdensome since it requires solving the penalized least squares optimization problem
\begin{equation}\label{eq:penalized_reg}
\argmin_{\mbbeta \in \mathbb{R}^{p \times q}} \mathcal{F}_{\rho}(\mbbeta, \mbPhi^{(k)}, \mbGamma^{(k)}) = \argmin_{\mbbeta \in \mathbb{R}^{p \times q}} \left\{ \frac{1}{2}\|\mbY + \rho^{-1} \mbGamma^{(k)} - \mbPhi^{(k)} - \mbX\mbbeta\|_F^2 +  \frac{\tilde{\lambda}}{\rho}g(\mbbeta)\right\}.
\end{equation}
To avoid solving \eqref{eq:penalized_reg} at every iteration, we instead approximate \eqref{eq:beta_update} by minimizing a majorizing function of $\mathcal{F}_{\rho}(\mbbeta, \mbPhi^{(k+1)}, \mbGamma^{(k)})$ constructed at the previous iterate $\mbbeta^{(k)}.$ Specifically, we majorize $\mathcal{F}_{\rho}(\mbbeta, \mbPhi^{(k)}, \mbGamma^{(k)})$ in \eqref{eq:beta_update} with 
\begin{equation}
\mathcal{M}_{\rho, \eta}(\mbbeta, \mbPhi^{(k)}, \mbGamma^{(k)}; \mbbeta^{(k)}) = \mathcal{F}_\rho(\mbbeta, \mbPhi^{(k)}, \mbGamma^{(k)}) + \frac{\rho}{2} {\rm tr}\big\{ (\mbbeta - \mbbeta^{(k)})^\top\mbQ_\eta(\mbbeta - \mbbeta^{(k)})\big\},\notag
\end{equation}
where $\mbQ_\eta = \eta \mbI_p - \mbX^\top\mbX$ with $\eta > 0$ fixed and chosen so that $\mbQ_\eta$ is nonnegative definite. Thus, we replace \eqref{eq:beta_update} with
\begin{align}
\mbbeta^{(k+1)} & =  \argmin_{\mbbeta \in \mathbb{R}^{p \times q}} \mathcal{M}_{\rho, \eta}(\mbbeta, \mbPhi^{(k)}, \mbGamma^{(k)}; \mbbeta^{(k)})\notag \\
& = {\rm Prox}_{(\rho \eta)^{-1}\tilde{\lambda} g} \left\{ \mbbeta^{(k)} + \eta^{-1} \mbX^\top\big(\mbY + \rho^{-1} \mbGamma^{(k)} - \mbPhi^{(k)} - \mbX\mbbeta^{(k)}\big)\right\} \label{eq:prox_beta}.
\end{align}
It follows that using \eqref{eq:prox_beta}, $\mathcal{F}_{\rho}(\mbbeta^{(k+1)}, \mbPhi^{(k)}, \mbGamma^{(k)}) \leq \mathcal{F}_{\rho}(\mbbeta^{(k)}, \mbPhi^{(k)}, \mbGamma^{(k)})$ by the majorize-minimize principle \citep{lange2016mm}. This approximation can improve efficiency because for many $g$, \eqref{eq:prox_beta} can be computed efficiently in closed form. For example, in the case that $g$ is the $L_1$-norm, \eqref{eq:prox_beta} can be solved by soft thresholding $\mbbeta^{(k)} + \eta^{-1} \mbX^\top(\mbY + \rho^{-1} \mbGamma^{(k)} - \mbPhi^{(k)} - \mbX\mbbeta^{(k)})$. 

The complete prox-linear ADMM algorithm we implement is stated formally in Algorithm 1. In the algorithm statement, we use $(\cdot)_+$ to denote the elementwise positive part function, i.e., $(\mbA_{j,k})_+ = \max(\mbA_{j,k}, 0)$. 
This variation of the ADMM algorithm---which replaces the objective function in \eqref{eq:beta_update} with a quadratic majorization constructed at the previous iterate---was studied by \citet{deng2016global}, who called it the prox-linear ADMM algorithm. Fortunately, we can show that the iterates of our prox-linear ADMM algorithm converge to their optimal values.

  \begin{algorithm}[t]\caption{Prox-linear ADMM algorithm for \eqref{eqMSRL}}
  \onehalfspacing
  \noindent

  \begin{tabbing}
  \textit{1.} Given $\rho > 0$, $\eta \geq \|\mbX^\top\mbX\|$, $\tilde{\lambda} = \sqrt{n}\lambda$, $\tau \in (0, \frac{1 + \sqrt{5}}{2})$, and $s = \min(p,q)$, initialize\\ ~~~~$(\mbbeta^{(0)}, \mbPhi^{(0)},\mbGamma^{(0)}) \in \mathbb{R}^{p \times q} \times \mathbb{R}^{n \times q} \times \mathbb{R}^{n \times q}$ and set $k=0$ \\
  \textit{2.} $\mbbeta^{(k+1)} \leftarrow {\rm Prox}_{(\rho \eta)^{-1}\tilde{\lambda}g} \big\{ \mbbeta^{(k)} + \eta^{-1} \mbX^\top\left(\mbY + \rho^{-1} \mbGamma^{(k)} - \mbPhi^{(k)} - \mbX\mbbeta^{(k)}\right)\big\}$\\
    \textit{3.} $(\mbU,\mbD,\mbV) \leftarrow {\rm svd}(\mbY + \rho^{-1}\mbGamma^{(k)} - \mbX\mbbeta^{(k+1)})$\smallskip\\
  \textit{3.} $\mbPhi^{(k+1)} \leftarrow \mbU (\mbD - \rho^{-1} \mbI_s)_+ \mbV^\top$\\\bigskip
  \textit{4.} $\mbGamma^{(k+1)} \leftarrow \mbGamma^{(k)} + \tau \rho(\mbY - \mbX\mbbeta^{(k+1)} - \mbPhi^{(k+1)})$\\\bigskip
  \textit{5.} If not converged, set $k \leftarrow k + 1$ and return to \textit{2}
\end{tabbing}
\vspace{-10pt}
  \end{algorithm}

\begin{lemma}\label{conv_prop}
Suppose $0 < 2 \tau < 1 + \sqrt{5}$, $\rho > 0$, and $\eta \geq \|\mbX^\top\mbX\|$ are fixed. Then, as $k \to \infty$, the sequence of iterates $(\mbPhi^{(k)}, \mbbeta^{(k)}, \mbGamma^{(k)})$ generated from Algorithm 1 converge to $(\mbPhi^{\star}, \mbbeta^{\star}, \mbGamma^{\star})$, where $(\mbPhi^{\star}, \mbbeta^{\star})$ are optimal solutions to \eqref{eq:constrained} and $\mbGamma^{\star}$ is an optimal solution to the dual of \eqref{eq:constrained}.  In addition, if $\tau = 1$,
then the sequence $\{\theta_k, k = 0, 1, 2, \dots\}$ defined by 
$ \theta_k = \rho\|\mbbeta^{(k)} - \mbbeta^\star\|_{\mbQ_\eta}^2 + \rho \|\mbPhi^{(k)} - \mbPhi^\star\|_F^2 + \rho^{-1}\|\mbGamma^{(k)} - \mbGamma^\star\|_F^2$ 
is nonincreasing and $\theta_k = O(k^{-1})$ as $k \to \infty$.
\end{lemma}

The arguments used to prove Lemma \ref{conv_prop} are essentially identical to those from \citet{gu2018admm}, who proposed a prox-linear ADMM algorithm to compute a penalized (univariate response) quantile regression estimator. In our implementation, we found that $\tau = 1$ generally worked well, although setting $\tau$ closer to  $(1 + \sqrt{5})/2$ could lead to faster convergence in certain scenarios. Similarly, we set $\eta = (1 + 10^{-5})\|\mbX^\top\mbX\|$, which we found was fastest among the values we considered. 

The convergence criteria we use are based on the primal and dual residuals suggested by \citet{boyd2011distributed}. At each iteration we compute
$$ r^{(k+1)} = \|\mbY - \mbX\mbbeta^{(k+1)} - \mbPhi^{(k+1)}\|_F^2, ~~ s^{(k+1)} = \rho^2\|\mbX^\top(\mbPhi^{(k+1)} - \mbPhi^{(k)})\|_F^2.$$
We also compute ${\rm e}^{(k+1)}_{\rm primal} = \epsilon_{\rm abs}\sqrt{n} + \epsilon_{\rm rel}\max\{ \|\mbX\mbbeta^{(k+1)}\|_F, \|\mbPhi^{(k+1)}\|_F, \|\mbY\|_F \}$  and ${\rm e}_{\rm dual}^{(k+1)} = \epsilon_{\rm abs}\sqrt{p} + \epsilon_{\rm rel}\|\mbX^\top\mbGamma^{(k+1)}\|_F$ where $\epsilon_{\rm abs}$ and  $\epsilon_{\rm rel}$ are the absolute and relative convergence tolerances, respectively. Then, we terminate Algorithm 1 when $r^{(k+1)} \leq {\rm e}^{(k+1)}_{\rm primal}$ and $s^{(k+1)} \leq  {\rm e}^{(k+1)}_{\rm dual}$. Our default implementation sets $\epsilon_{\rm rel} = 10^{-4}$ and $\epsilon_{\rm abs} = 10^{-10}.$ We also adaptively update the step size $\rho$. Unlike the scheme originally proposed in \citet{boyd2011distributed}, we update $\rho$ every $\kappa$th iteration using
$\rho \leftarrow \rho  \left\{ \mathbf{1}(r^{(k+1)}  > 10 s^{(k+1)}) - 0.5 \cdot \mathbf{1}(s^{(k+1)}  > 10 r^{(k+1)}) + 1\right\}$. In our default implementation, we use $\kappa = 10.$ 

An R package implementing Algorithm 1, Algorithm 2 (see Section \ref{eq:alt_comp} and Section \ref{sec:additional_computational_details} of Appendix A), and a number of auxiliary functions are available for download at \url{https://github.com/ajmolstad/MSRL}.



\subsection{Alternative Computational Approaches}\label{eq:alt_comp}
Numerous other computational approaches could be applied to solve \eqref{eqMSRL}. One class of methods are those that, like the prox-linear ADMM algorithm, are designed to handle optimization problems where the objective function is the sum of two nondifferentiable, convex, and proximable functions. These include, for example, the accelerated primal-dual algorithm of \citet{chambolle2011first} and the graph projection ADMM algorithm \citep{parikh2014block,fougner2018parameter}. 

Another (arguably simpler) class of algorithms can be applied only in special settings. In particular, when $n > q$ and $\lambda$ is sufficiently large, we can treat the nuclear norm of residual as differentiable. This is because the subdifferential of the nuclear norm of the residual matrix with respect to $\mbbeta$ is the set 
\begin{align*}
\left\{\mbW \in \mathbb{R}^{p \times q}: \mbW = -\mbX^\top \right. &(\mbU \mbV^\top + \mbQ), \|\mbQ\| \leq 1, \\
& \left.\mbU^\top\mbQ = \mbQ\mbV  = 0, (\U, \D, \V) = {\rm svd}(\mbY - \mbX \mbbeta) \right\},
\end{align*}
for example, see \citet{watson1992characterization}.
Thus, when $\mbY - \mbX\mbbeta$ has $q$ nonzero singular values, the subdifferential of $\mbbeta \mapsto \|\mbY - \mbX\mbbeta\|_*$ is the singleton
\begin{equation}\label{eq:gradient_NN}
- \mbX^\top(\mbY - \mbX\mbbeta)\{(\mbY - \mbX\mbbeta)^\top(\mbY - \mbX\mbbeta)\}^{-1/2}
 \end{equation}
so that $\|\mbY - \mbX\mbbeta\|_*$ can effectively be treated as differentiable over the set of $\mbbeta$ such that $\mbY - \mbX\mbbeta$ has $q$ nonzero singular values. 

\begin{figure}[t]
\begin{center}
\makebox[\textwidth][c]{\includegraphics[width=\textwidth]{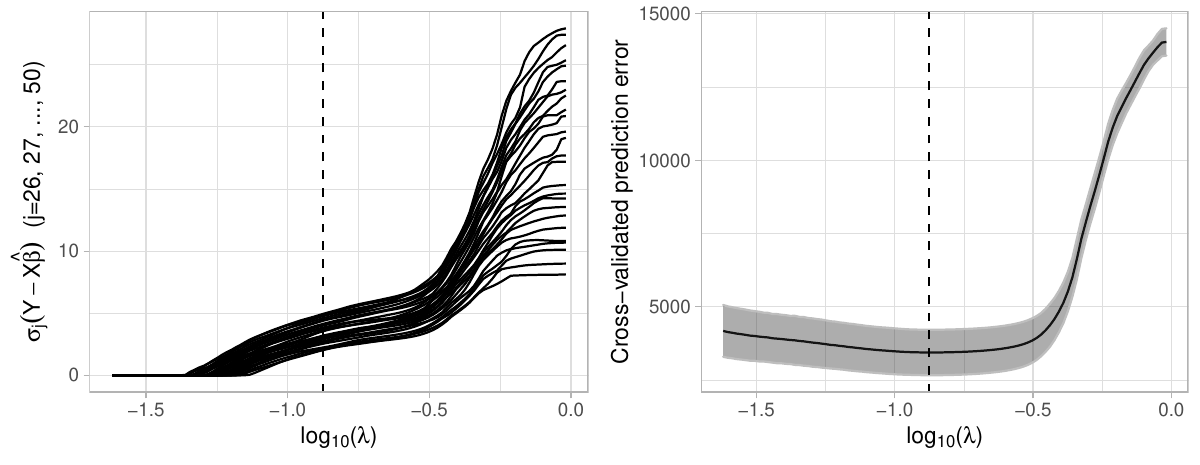}}
\end{center}
\caption{(Left) The solution path for the 25 smallest singular values of $\mbY - \mbX \hat\mbbeta_{\rm L}$ as a function of $\lambda$ for data generated under Model 1 and \textbf{M1} (see Section \ref{sec:datagenmodels} and \ref{subsec:Results_1_M1}) with $n = 200$, $p=500$, $q=50$, and normal errors with $\xi = 0.9$. (Right) Average five-fold cross-validation squared prediction error (and standard errors) for $\hat\mbbeta_{\rm L}$ on the same data set. In both panels the vertical dotted line denotes the tuning parameter value minimizing the average cross-validated squared prediction error. }\label{fig:SolutionPaths}
\end{figure}

This simple fact suggests that in these special settings, we can use first order algorithms to solve \eqref{eqMSRL}. To illustrate that this represents a range of interesting fitted models, we generated data from Model 1 of Section \ref{sec:datagenmodels} with $\mbbeta_*$ constructed according to \textbf{M1}, $g$ being the $L_1$-norm penalty, and $(n, p, q) = (200, 500, 50).$ In the left panel of Figure \ref{fig:SolutionPaths}, we display the path of the 25 smallest singular values of $\mbY - \mbX \hat\mbbeta_{\rm L}$ as a function of the tuning parameter $\lambda$; in the right panel, we display the cross-validated squared prediction errors. We see that for $\lambda$ sufficiently large, all $q$ singular values of the residual matrix are nonzero. In addition, we see that the cross-validated squared prediction error indicates that the best model fits are those occurring at points on the solution path where $\mbY - \mbX \hat\mbbeta_{\rm L}$ has $q$ nonzero singular values. As $\lambda$ approaches zero, we see that many singular values of $\mbY - \mbX \hat\mbbeta_{\rm L}$ become zero. This is because the nuclear norm acts like a lasso-type penalty on the singular values of its matrix argument, so reducing $\lambda$ is analogous to increasing the relative contribution of the nuclear norm of residuals to the overall objective function.

Hence, to solve \eqref{eqMSRL} when $n > q$ and $\lambda$ is sufficiently large, we consider using an accelerated proximal gradient descent algorithm \citep{beck2009fast,combettes2011proximal}. 
Letting $\mathcal{D}_{\underline{\kappa}} = \left\{\mbbeta \in \mathbb{R}^{p \times q}: \underline{\kappa} \leq \sigma_q(\mbY - \mbX\mbbeta) \leq \sigma_1(\mbY - \mbX\mbbeta) \leq \underline{\kappa}^{-1}\right\}$ and letting $(\mbU_{\epsilon^{(k)}},\mbV_{\epsilon^{(k)}})$ denote the left and right singular vectors of $\mbY - \mbX\mbbeta^{(k)}$ respectively, it follows from \eqref{eq:gradient_NN} that if we iteratively update $\mbbeta$ from $k$th to $(k+1)$th iterate using
\begin{align}
\mbbeta^{(k+1)} &= \argmin_{\mbbeta \in \mathbb{R}^{p \times q}} \left[  \frac{1}{2 \rho_k}\|\mbbeta - \mbbeta^{(k)}\|_F^2 - \frac{1}{\sqrt{n}} {\rm tr}\big\{ \mbV_{\epsilon^{(k)}}\mbU_{\epsilon^{(k)}}^\top\mbX(\mbbeta - \mbbeta^{(k)})\big\} + \lambda g(\mbbeta) \right] \notag \\
& = {\rm Prox}_{\rho_k \lambda g}\left( \mbbeta^{(k)} + \frac{\rho_k}{\sqrt{n}}\mbX^\top\mbU_{\epsilon^{(k)}}\mbV_{\epsilon^{(k)}}^\top \right)\notag
\end{align}
for step size $\rho_k$ sufficiently small, $\mbbeta^{(k+1)} \to \hat\mbbeta_g$ as $k \to \infty$ provided that $\hat\mbbeta_g$ and each $\mbbeta^{(k+1)}$ belong to $\mathcal{D}_{\underline{\kappa}}$ for some positive $\underline{\kappa}$ bounded away from zero. A similar computational approach was proposed and studied theoretically in \citet{li2020fast} for solving the univariate square-root lasso optimization problem. 

In contrast with Algorithm 1, accelerated versions of the proximal gradient descent algorithm are known to converge at a quadratic rate \citep{beck2009fast}, so this approach may be preferred in the settings where it can be applied. Of course, if the solution $\hat\mbbeta_g$ leads to residual matrix $\mbY - \mbX \hat\mbbeta_g$ with fewer than $q$ nonzero singular values, this algorithm cannot be used.  In practice, we use an accelerated proximal gradient descent algorithm to compute $\hat\mbbeta_g$ for large values of $\lambda$, but when an iterate of this algorithm leads to (nearly) rank deficient residuals, we then revert to using Algorithm 1 for that and all smaller values of $\lambda$. For example, in the setting displayed in Figure \ref{fig:SolutionPaths}, accelerated proximal gradient descent could be used to compute \eqref{eqMSRL} for all $\lambda$ such that $\log_{10}(\lambda) > -1.$ A formal statement of the accelerated proximal gradient descent algorithm we implement (Algorithm 2), along with details about our implementation, can be found in Section \ref{sec:additional_computational_details} of Appendix A.

\section{Simulation Studies}\label{sec:sim_studies}
\subsection{Overview}
In this section, we compare \eqref{eqMSRL} to alternative methods for fitting the multivariate response linear regression model in high-dimensional settings. We consider three data generating models under \textbf{M1}, \textbf{M2}, and \textbf{M3} as defined in Section \ref{subsec:theory_overview}.  In addition to comparing methods which use cross-validation for tuning parameter selection, we also consider versions of \eqref{eqMSRL} with tuning parameters chosen according to the theoretical results from Section \ref{subsec:pivotal}.

\subsection{Data Generating Models and Competing Methods}\label{sec:datagenmodels}
In each setting we consider, for one hundred independent replications, we generate $\mbX \in \mathbb{R}^{n \times p}$ to have rows being independent realizations of ${\rm N}_p(0, \mbSigma_{*\mbX})$ with $[\mbSigma_{*\mbX}]_{j,k} = 0.5^{|j-k|}$ for $(j,k) \in  [p] \times [p]$. Then, given $\mbX$, we generate $\mbY= \mbX \mbbeta_* + \mathcal{\mbE}$ where rows of $\mathcal{\mbE} \in \mathbb{R}^{n \times q}$ are independent and identically distributed with mean zero and covariance $\mbSigma_* \in \mathbb{S}^q_+$. We consider three distinct data generating models.
\begin{itemize}
\item[] \textbf{Model 1} (Compound symmetry):  $\mbSigma_* = 3 \hspace{1pt} \tilde{\mbSigma}_* $, where $\tilde{\mbSigma}_{*j,k} = \xi \mathbf{1}(j \neq k) + \mathbf{1}(j = k)$ for $(j,k) \in [q] \times [q]$ and $\mathbf{1}(\cdot)$ is the indicator function. 

\item[] \textbf{Model 2} (Varying condition number): $\mbSigma_* = 2 \hspace{1pt}\tilde{\mbSigma}_* $, where $\tilde{\mbSigma}_{*j,k} = \mbO \mbGamma\mbO^\top$,  $\mbO$ is a randomly generated ${q \times q}$ orthogonal matrix, and $\mbGamma$ is diagonal with equally spaced entries from $1$ to the inverse condition number.

\item[]  \textbf{Model 3} (Factor model): $\mbSigma_* = \mbR^\top\mbR + 0.05 \hspace{1pt} \mbI_q$, where $\mbR$ is obtained by first generating $\tilde{\mbR} \in \mathbb{R}^{m \times q}$ with $m \leq q$ to have independent standard normal entries and setting $\mbR = \tilde{\mbR} \mbK$ where $\mbK \in \mathbb{R}^{q \times q}$ is diagonal with entries chosen so that $\mbR^\top\mbR$ has diagonal entries equal to 1.45.  
\end{itemize}

Throughout our simulations, we set $n = 200$, $p = 500,$ $q=50,$ and let $\xi$, the condition number, and the number of factors $(m)$ vary, under Models 1, 2, and 3, respectively.  In addition to Models 1--3 with normally distributed errors, we also consider Models 1--3 with errors following a multivariate $t$-distribution with five degrees of freedom (henceforth, $t_5$).  

To select tuning parameters, we also generate a validation set of size $n$ from the same data generating model. For each method we consider, tuning parameters are chosen to minimize the squared prediction error averaged across all $q$ responses on the validation set. In a slight abuse of terminology, we refer to this as ``cross-validation'' for the remainder of this section. 

We will describe the construction of $\mbbeta_*$ separately in subsequent sections. Given a training data set, we estimate $\mbbeta_*$ using the following methods. 
\begin{itemize}
  \item[] \texttt{MSR-CV}: Our proposed estimator from \eqref{eqMSRL}. 
  \item[] \texttt{Calibrated}: A variation of the calibrated multivariate response linear regression method proposed by \citet{liu2015calibrated}: 
    $$\argmin_{\mbbeta \in \mathbb{R}^{p \times q}}  \left\{ \frac{1}{\sqrt{n}} \sum_{k=1}^{q}\|\mbY_{\cdot,k} - \mbX \mbbeta_{\cdot, k}\|_2   + \lambda g(\mbbeta) \right\} .$$
    Note that when $g$ is the $L_1$-norm, this estimator is equivalent to $q$ separate univariate square-root lasso estimators \citep{belloni2011square} with the same tuning parameter $\lambda$ used for each response. 
  \item[] \texttt{PLS}: The penalized least squares estimator of $\mbbeta_*$, i.e., 
  \begin{equation}\label{eq:PLS} 
  \argmin_{\mbbeta \in \mathbb{R}^{p \times q}}  \left\{\frac{1}{n}\|\mbY - \mbX \mbbeta\|_F^2   + \lambda g(\mbbeta)\right\}.
  \end{equation}
  \item[] \texttt{MRCE-Approx}: The approximate version of the multivariate regression with covariance estimation (MRCE) method proposed by \citet{rothman2010sparse}. This estimator is computed in three steps: 
  \begin{enumerate}
    \item Obtain $\mbbeta^{(0)}$, the \texttt{PLS} estimator. 
    \item Set $\hat\mbSigma =  n^{-1}(\mbY - \mbX \mbbeta^{(0)})^\top(\mbY - \mbX \mbbeta^{(0)})$ and compute 
    $$\mbOmega^{(1)}_\gamma = \argmin_{\mbOmega \in \mathbb{S}^q_+} \left\{  {\rm tr}(\hat\mbSigma\mbOmega) - \log {\rm det}(\mbOmega) + \gamma \sum_{j \neq k}|\mbOmega_{j,k}| \right\}.$$ 
    \item With $\mbOmega_\gamma^{(1)}$ fixed, compute the \texttt{MRCE-Approx} estimator of $\mbbeta_*$
    \begin{equation}\label{eq:MRCE}
    \argmin_{\mbbeta \in \mathbb{R}^{p \times q}} \left[  \frac{1}{n}{\rm tr}\big\{(\mbY - \mbX \mbbeta) \mbOmega^{(1)}_\gamma (\mbY - \mbX \mbbeta)^\top\big\} + \lambda g(\mbbeta) \right].
    \end{equation}
  \end{enumerate}
  \item[] \texttt{MRCE-Or}: The ``oracle'' penalized normal maximum likelihood estimator of $\mbbeta_*$ with $\mbOmega_*$ known, i.e., \eqref{eq:MRCE} with $\mbOmega^{(1)}_\gamma$ replaced with $\mbOmega_*$. 
\end{itemize}

We found that computing times for the exact version of the method proposed by \citet{rothman2010sparse} could be prohibitively long for our data generating models, so we only compare to the approximate version described above.

\subsection{Results under \textbf{M1} using Cross-Validation}\label{subsec:Results_1_M1}
In our first set of simulation studies, independently for each replication we generate the regression coefficient matrix $\mbbeta_* \in \mathbb{R}^{p \times q}$ such that $\mbbeta_* = \mbA \circ \mbG$, where $\mbA \in \mathbb{R}^{p \times q}$, $\mbG \in \mathbb{R}^{p \times q}$, and $\circ$ denotes the elementwise product. The matrix $\mbA$, which encodes the sparsity of $\mbbeta_*$, has five randomly selected entries equal to one per column and all other entries equal to zero. The matrix $\mbG$ has independent and identically distributed standard normal entries. Thus, the matrix $\mbbeta_*$ has proportion of nonzero entries equal to $(5/p)$. As this $\mbbeta_*$ corresponds to the model subspace under \textbf{M1}, for each method, we set $g$ to be the $L_1$-norm penalty.

\begin{figure}[t!]
\begin{center}
\makebox[\textwidth][c]{\includegraphics[width=\textwidth]{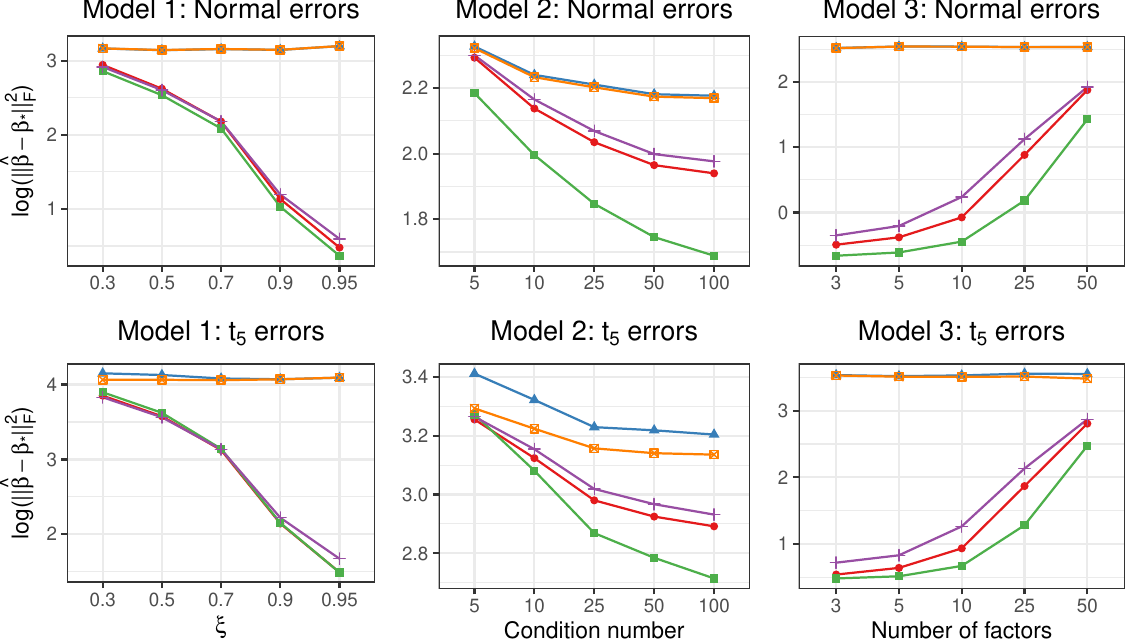}}
\makebox[\textwidth][c]{\includegraphics[width=14cm]{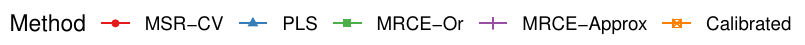}}
\vspace{-30pt}
\end{center}
\caption{Average log squared Frobenius norm error over one hundred independent replications under Model 1--3 with (top row) normal errors or (bottom row) $t_5$ errors and $\xi$, the condition number, and the number of factors varying. In these simulations, $\mbbeta_*$ is constructed according to \textbf{M1} and $g$ is the $L_1$-norm.}\label{fig:rho_Results}.
\end{figure}

In the top row of Figure \ref{fig:rho_Results}, we display the average log squared Frobenius norm errors, $\log(\|\hat\mbbeta - \mbbeta_*\|_F^2)$, for the five methods we considered under Models 1--3 with normally distributed errors. In every setting, \texttt{MRCE-Or}, which uses the true value of $\mbOmega_*$, performs best. Among the methods which could be used in practice, \texttt{MSR-CV} (our method) and \texttt{MRCE-Approx} tend to perform similarly. Under Model 1 with normal errors, when $\xi=0.3$, \texttt{MRCE-Approx} slightly outperforms the \texttt{MSR-CV}. As $\xi$ increases, \texttt{MSR-CV} only slightly outperforms \texttt{MRCE-Approx}.  The fact that \texttt{MRCE-Approx} performs well under Model 1 is not surprising: this method assumes that $\mbOmega_*$ is sparse and under Model 1, $\mbOmega_*$ is tri-diagonal. Under Models 2 and 3, however, \texttt{MSR-CV} outperforms \texttt{MRCE-Approx} in nearly every considered setting. Unlike Model 1, under Models 2 and 3, $\mbOmega_*$ is nonsparse. Notably, \texttt{MRCE-Or} still outperforms both estimators, which suggests that the relatively worse performance of \texttt{MRCE-Approx} is due to a poor estimate of the precision matrix being used in the criterion \eqref{eq:MRCE}.  

Similar results hold when errors are generated from the $t_5$-distribution, although the difference between \texttt{MRCE-Or} and \texttt{MSR-CV} is slightly less apparent than under normal errors. Overall, it appears that heavy tailed errors lead to worse estimation accuracy across all the methods.  Interestingly, when comparing \texttt{Calibrated} and \texttt{PLS}, we notice a difference in performance only under Model 2. This can be explained by the fact that the diagonals of $\mbSigma_*$ are different only under Model 2. \texttt{Calibrated} can exploit this fact, whereas \texttt{PLS} cannot.  In fact, with condition number equal to five under Model 2, the covariance is nearly diagonal, which corresponds to the modeling assumptions of \texttt{Calibrated}. This partly explains why it performs similarly to \texttt{MSR-CV} and \texttt{MRCE-Approx} in this setting. 

A reviewer suggested that it is counterintuitive that the performance of \texttt{MRCE-Or}, \texttt{MRCE-Approx}, and \texttt{MSR-CV} improves as errors become more correlated. To understand why this occurs,  consider that if the errors were perfectly correlated, observing $q$ responses for the $i$th subject would be like observing realizations of
$\mbbeta_{*0} + \mbbeta_*^\top\mbx_i + e_i \boldsymbol{1}_q$ for $i \in [n]$,
where $e_i \in \mathbb{R}$ is random and $\boldsymbol{1}_q = (1, 1, \dots, 1)^\top \in \mathbb{R}^q$ is a vector of ones. Of course, if we knew this were the case, we could estimate $\mbbeta_*$ much more efficiently than if we (incorrectly) assumed errors were independent (e.g., using least squares). The methods which improve as errors become more correlated (\texttt{MRCE-Or}, \texttt{MRCE-Approx}, and \texttt{MSR-CV}) are all able to exploit this situation through implicit or explicit covariance matrix estimation and thus estimate $\mbbeta_*$ more efficiently than the competitors. This phenomenon has been observed in numerous other works focused on multivariate response linear regression with correlated errors \citep{rothman2010sparse,molstad2020covariance}.

\begin{figure}[t]
\begin{center}
\makebox[\textwidth][c]{\includegraphics[width=\textwidth]{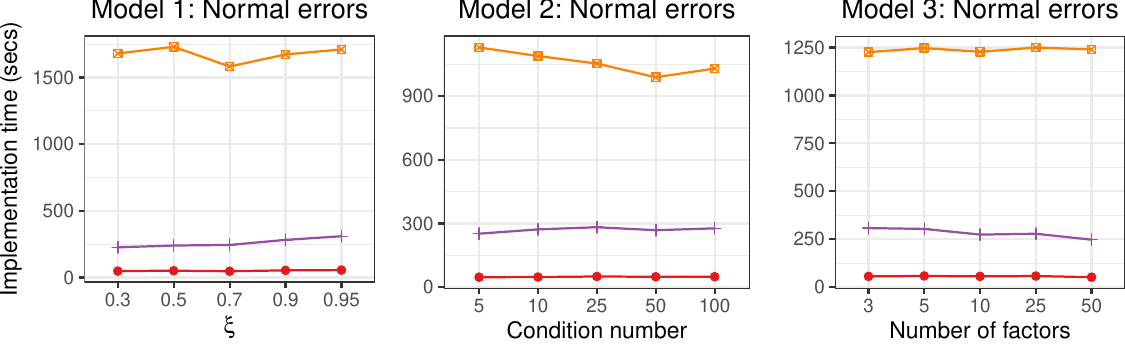}}
\makebox[\textwidth][c]{\includegraphics[width=9cm]{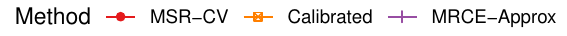}}
\vspace{-30pt}
\end{center}
\caption{Average implementation times over one hundred independent replications under Model 1--3 with normal errors, $\mbbeta_*$ constructed according to \textbf{M1}, and $g$ taken to be the $L_1$-norm. }\label{fig:implementation_time}
\end{figure}

In Figure \ref{fig:implementation_time}, we display the implementation times for \texttt{MSR-CV}, \texttt{Calibrated}, and \texttt{MRCE-Approx}. Focusing on Model 1 under normal errors, on average, \texttt{MSR-CV} never takes more than a minute to compute the entire solution path (for 100 candidate tuning parameter values). Average implementation times for \texttt{MRCE-Approx} in the same settings are all greater than 250 seconds. Note that \texttt{MRCE-Approx} requires the selection of two tuning parameters---and requires estimating $\mbOmega_*$---which explains the longer implementation times. Here, we consider 100 $\times$ 25 candidate tuning parameters for \texttt{MRCE-Approx}, but implement a rule wherein the solution path computation is terminated if the estimate of $\mbbeta_*$ leads to sufficiently poor prediction on the validation set. Thus, we generally compute the solution for less than half of the tuning parameter pairs under consideration. We implement no such rule for \texttt{MSR-CV} or \texttt{Calibrated}, so these results are somewhat biased in favor of \texttt{MRCE-Approx}. The estimator \texttt{Calibrated}, which we fit using the \texttt{flare} package in R, takes substantially longer than both other methods. However, it should be noted that the comparison to \texttt{Calibrated} is not entirely fair because the publicly available software we use requires fitting the solution path for each univariate square-root lasso estimator separately. Nonetheless, we see that \texttt{MSR-CV} is both the best performing method and can be obtained in the shortest amount of time given the existing software.

\begin{figure}[t]
\begin{center}
\makebox[\textwidth][c]{\includegraphics[width=\textwidth]{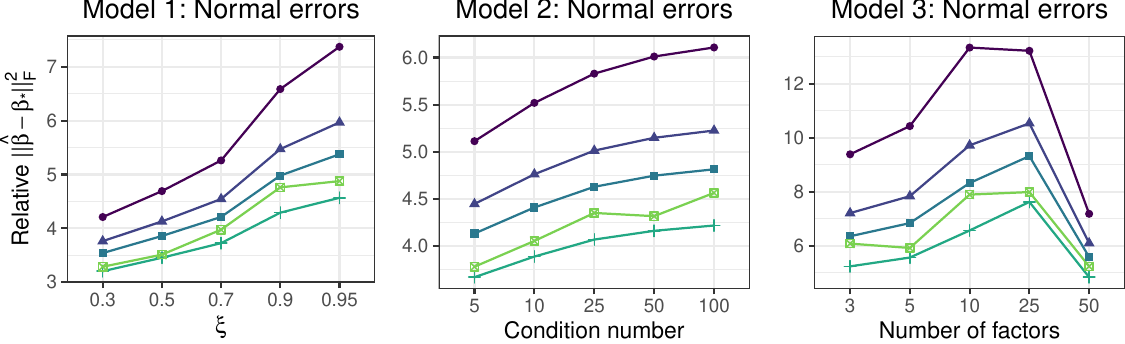}}
\makebox[\textwidth][c]{\includegraphics[width=13cm]{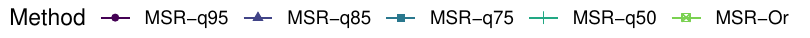}}
\end{center}
\begin{center}
\makebox[\textwidth][c]{\includegraphics[width=\textwidth]{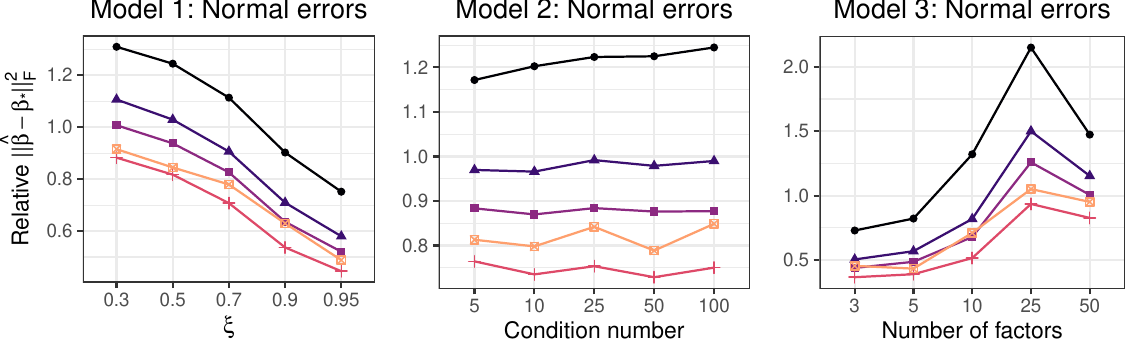}}
\makebox[\textwidth][c]{\includegraphics[width=16cm]{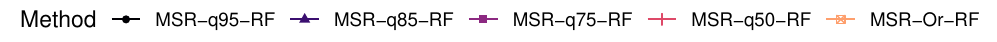}}
\vspace{-30pt}
\end{center}
\caption{Relative (to \texttt{MSR-CV}) average squared Frobenius norm errors over one hundred independent replications under Model 1--3 with normal errors, $\mbbeta_*$ constructed according to \textbf{M1}, and $g$ taken to be the $L_1$-norm both (top row) without refitting and (bottom row) with refitting.  }\label{fig:theory_tuning1}
\end{figure}
\subsection{Results under \textbf{M1} using Theoretical Tuning}\label{subsec:Results_1_M1_2}
In addition to the methods discussed in Section \ref{sec:datagenmodels}, we also consider multiple versions of \eqref{eqMSRL} using tuning parameters suggested by the theoretical results in Section \ref{subsec:pivotal}. Specifically, we also study selecting tuning parameters for \eqref{eqMSRL} based on quantiles of the distribution of the random variable $(c/\sqrt{n})\|\mbX^\top\mbS\|_\infty$ where $\mbS$ is uniformly distributed on $V_q(n)$ and $c > 1$. In our implementation, we set $c = 1.01$. We tried multiple quantiles: 0.95, 0.85, 0.75, and 0.50. We denote the corresponding estimators \texttt{MSR-q95, MSR-q85, MSR-q75,} and \texttt{MSR-q50}, respectively. For the sake of comparison, we also used the theoretically optimal tuning parameter $(c/\sqrt{n})\|\mbX^\top\mbU_{\mathcal{\epsilon}}\mbV_{\mathcal{\epsilon}}^\top\|_\infty$ where $(\mbU_{\mathcal{\epsilon}}, \mbD_{\mathcal{\epsilon}},\mbV_{\mathcal{\epsilon}}) = {\rm svd}(\mathcal{\mbE})$: we call this estimator \texttt{MSR-Or} since it uses oracle information. 

In Figure \ref{fig:theory_tuning1}, we display the average squared Frobenius norm errors of \texttt{MSR-q95, MSR-q85, MSR-q75, MSR-q50}, and \texttt{MRCE-Or} relative to \texttt{MSR-CV}. That is, an estimator with a relative error of 1.2 has a 20\% larger average squared Frobenius norm error than \texttt{MSR-CV}. Based on the results in the top row of Figure \ref{fig:theory_tuning1}, it seems that in general, all directly tuned estimators tend to perform substantially worse than \texttt{MSR-CV}, including \texttt{MSR-Or}---the estimator with theoretically optimal tuning parameter. In Table \ref{table:variableSelection_M1} of the Appendix, we display average true positive and false positive variable selection rates for each of the methods displayed in Figures \ref{fig:rho_Results} and \ref{fig:theory_tuning1}. In Table \ref{table:variableSelection_M1} we see why the directly tuned versions of \eqref{eqMSRL} tended to perform worse than \texttt{MSR-CV} in terms of average squared Frobenius norm error: the false positive rates for these estimators are extremely low, but true positive rates are often much lower than those of the estimators whose tuning parameters were chosen by cross-validation. A similar result was observed in \citet{belloni2011square}, who found that the direct choices of $\lambda$ based on theory often led to substantial bias. To alleviate this issue, we follow \citet{belloni2011square} who used a refitting procedure: we re-estimate the coefficients using a likelihood-based seemingly unrelated
regression estimator described in Section \ref{sec:refit} of Appendix A. We refer to all refitted estimators by appending \texttt{-RF} to their names (e.g., the refitted version of \texttt{MSR-q95} is \texttt{MSR-q95-RF}). Results for refitted estimators are displayed in the bottom row of Figure \ref{fig:theory_tuning1}.  In this figure, we see that the performance relative to \texttt{MSR-CV} (the non-refitted version) is much improved when using refitting. 

To conclude, it seems that when taking $g$ to be the $L_1$-norm, direct tuning may be most useful for obtaining very sparse models with few false positives, but cross-validation may be preferred for prediction accuracy. Refitting appears to alleviate some extra bias observed when using the theory-based tuning procedures. However, in a subsequent section, we will show that under \textbf{M2}, theory-based tuning can perform as well as cross-validation-based tuning in terms of squared Frobenius norm error even without refitting.

\subsection{Computing Time Comparisons under \textbf{M1}}\label{subsec:computing_time}
\begin{table}[t]
\begin{center}
\begin{tabular}{cccccc}
\toprule
   & \multicolumn{5}{c}{Model 1: $\xi$}\\
 & \multicolumn{1}{c}{$0.3$} &  \multicolumn{1}{c}{$0.5$} & \multicolumn{1}{c}{$0.7$}& \multicolumn{1}{c}{$0.9$} & \multicolumn{1}{c}{$0.95$} \\ 
\midrule
\texttt{ADMM}  & 0.61 & 0.62 & 0.63 & 0.68 & 0.76 \\ 
\texttt{AccPGD} & 0.49 & 0.64 & 0.68 & 0.94 & 1.06 \\ 
\texttt{CVX}  & 82.89 & 92.71 & 80.23 & 93.48 & 87.74 \\ 
\bottomrule
\end{tabular}

\end{center}
\caption{Average computing times (in seconds) for \eqref{eqMSRL} using Algorithm 1 (\texttt{ADMM}), and Algorithm 2 (\texttt{AccPGD}), and \texttt{CVXR} with $g$ taken to be the $L_1$-norm. Averages are taken over one hundred independent replications under Model 1 with normal errors and $\mbbeta_*$ constructed according to \textbf{M1}. In each replication, the tuning parameter $\lambda$ is that which minimizes average squared prediction error on the validation set. }\label{table:timing_results_comparison}
\end{table}
We also compare the computing time of our algorithms to the computing time using \texttt{CVX} \citep{cvx}, the off-the-shelf convex solver used to compute \eqref{eqMSRL} by \citet{stucky2017asymptotic}. In Table \ref{table:timing_results_comparison}, we display the average computing times for \eqref{eqMSRL} with the tuning parameter selected by minimizing the average squared prediction error on the validation set under Model 1 and \textbf{M1} with normal errors. Convergence tolerances for \texttt{ADMM} and \texttt{AccPGD} are discussed in Section \ref{sec:ADMM} and Section \ref{sec:additional_computational_details} of Appendix A, respectively. Convergence tolerances for \texttt{CVX} are left at their defaults in the \texttt{CVXR} R package. 

Briefly, the prox-linear ADMM algorithm takes less than one second on average, whereas \texttt{CVX} takes more than 80 seconds on average in every setting. In terms of solution accuracy, the objective function value at convergence of \texttt{CVX} is on average 
$1.000617$, $1.000688$, $1.000745$, $1.000917$, and $1.000988$ (for $\xi$ from 0.3 to 0.95) times larger than that obtained by \texttt{ADMM}. The solution using \texttt{AccPGD} is very similar to \texttt{ADMM}: on average their differences are even smaller than those between \texttt{CVX} and \texttt{ADMM}.

We attempted to compare the computing time of our algorithms to the iterative procedure suggested by \citet{van2016chi2}. In the settings we consider, however, we found that using their algorithm, the objective function value never converged to a value near that obtained by our algorithm or \texttt{CVX}. In personal communication with the authors, they suggested we use \texttt{CVX}, citing a lack of convergence guarantees for their approach. 

\begin{figure}[t!]
\begin{center}
\makebox[\textwidth][c]{\includegraphics[width=\textwidth]{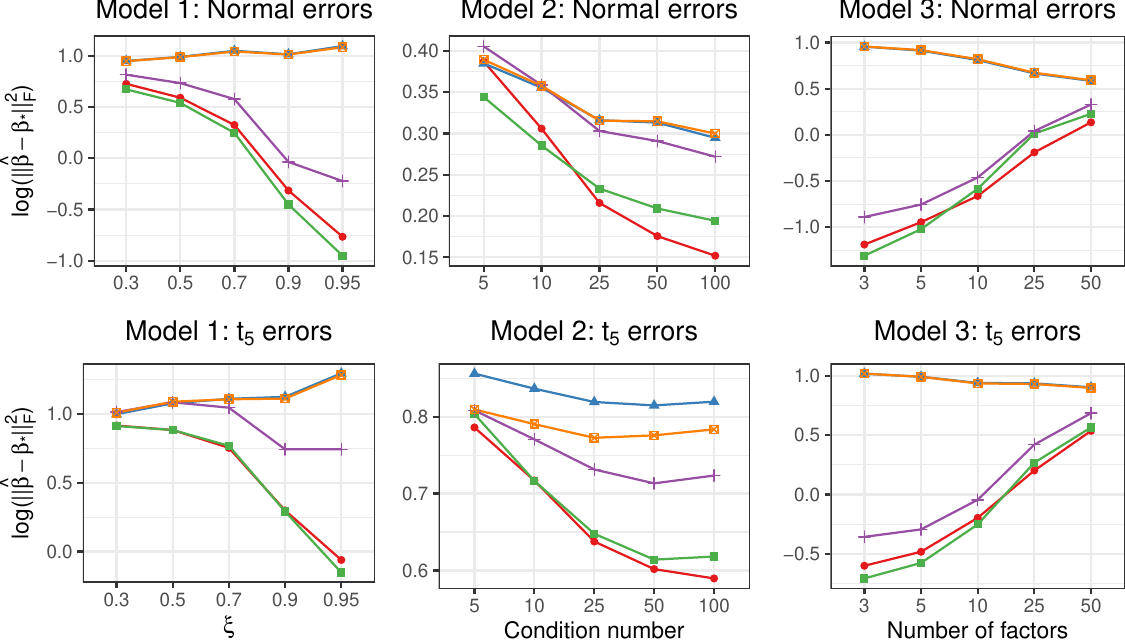}}
\makebox[\textwidth][c]{\includegraphics[width=14cm]{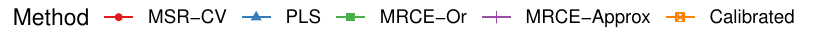}}
\vspace{-30pt}
\end{center}
\caption{Average log squared Frobenius norm errors over one hundred independent replications under Model 1--3 with (top row) normal errors or (bottom row) $t_5$ errors and $\xi$, the condition number, and number of factors varying. In these simulations, $\mbbeta_*$ is constructed according to \textbf{M2} and $g$ is the group lasso penalty.}\label{fig:Group_Results}.
\end{figure}
\subsection{Results under \textbf{M2} using Cross-Validation}
In this section, we consider the estimation of $\mbbeta_*$ under \textbf{M2} by setting $g$ to be the group lasso penalty for each of the methods discussed in Section \ref{sec:datagenmodels}. Specifically, for each replication under Models 1--3 as described in Section \ref{sec:datagenmodels}, we randomly generated $\mbbeta_*$ to be entirely zero except for five randomly chosen rows which have components drawn independently from a normal distribution with mean zero and standard deviation 0.1. Under this construction, only five predictors affect the $q$ responses and the same set of predictors is important for all $q$ responses.

We display average squared Frobenius norm error results in Figure \ref{fig:Group_Results}. We see that unlike under \textbf{M1}, \texttt{MSR-CV} outperforms \texttt{MRCE-Approx} in every setting we considered. We also see that \texttt{MSR-CV} performs similarly to \texttt{MRCE-Or}. This can be partly attributed to the fact that under \textbf{M2}, variable selection is a significantly easier task than under \textbf{M1}. Because predictors are either important for all $q$ responses or none, under \textbf{M2}, having a relatively large number of responses is helpful. Thus, since all estimators---\texttt{MSR-CV} included---more efficiently estimate the set of important predictors, the differences can more likely be attributed to the role of $\mbOmega_*.$ Evidently, using  $\mbOmega_*$ or an estimate thereof in \eqref{eq:MRCE} does not necessarily lead to better estimation than does using \eqref{eqMSRL}.

\begin{figure}[t!]
\begin{center}
\makebox[\textwidth][c]{\includegraphics[width=\textwidth]{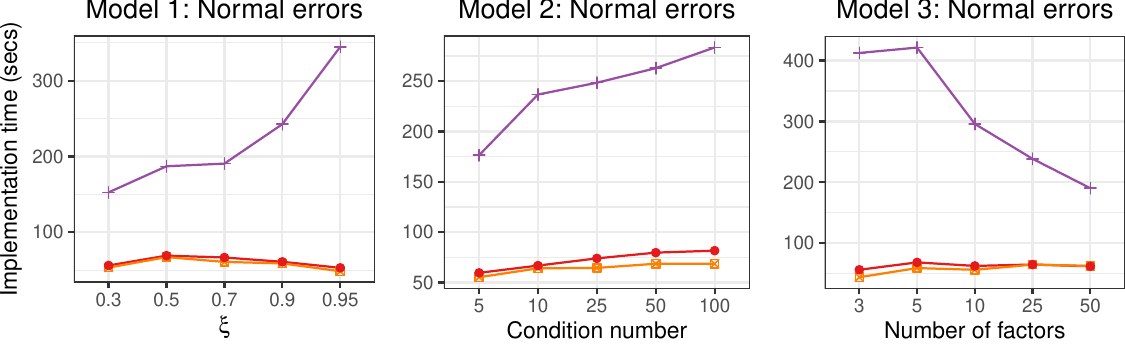}}
\makebox[\textwidth][c]{\includegraphics[width=9cm]{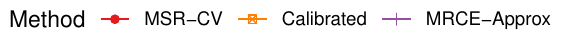}}
\vspace{-30pt}
\end{center}
\caption{Average implementation times for \texttt{MSR-CV}, \texttt{Calibrated}, and \texttt{MRCE-Approx} over one hundred independent replications under Model 1--3 with normal errors, $\mbbeta_*$ constructed according to \textbf{M2}, and $g$ taken to be the group lasso penalty. }\label{fig:implementation_time_1}\label{fig:Group_Results_Timing}.
\begin{center}
\makebox[\textwidth][c]{\includegraphics[width=\textwidth]{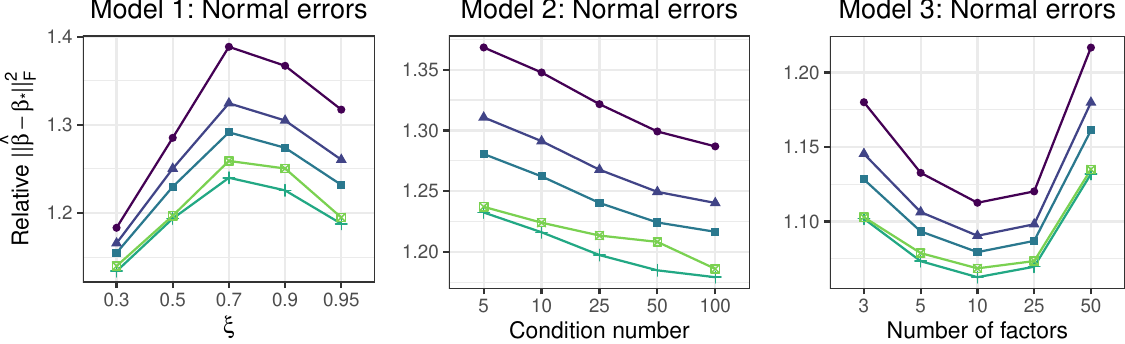}}
\makebox[\textwidth][c]{\includegraphics[width=13cm]{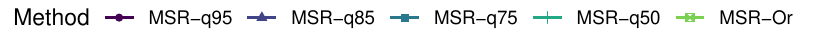}}
\vspace{-30pt}
\end{center}
\caption{Relative (to \texttt{MSR-CV}) average squared Frobenius norm errors over one hundred independent replications under Model 1--3 with normal errors, $\mbbeta_*$ constructed according to \textbf{M2}, and $g$ taken to be the group lasso penalty.}\label{fig:Group_Theory_Tuning}.
\end{figure}

In Figure \ref{fig:Group_Results_Timing}, we display implementation times of \texttt{MSR-CV}, \texttt{Calibrated}, and \texttt{MRCE-Approx}. To compute the solution path for \texttt{Calibrated}, we used the R package \texttt{camel}. To compute \eqref{eq:MRCE} with group lasso penalty, we wrote our own proximal gradient descent algorithm in R.  We see that both \texttt{MSR-CV} and \texttt{Calibrated} take around a minute or less to implement in every setting. \texttt{MRCE-Approx}, on the other hand, can take anywhere between two and seven minutes in the settings we considered. It is important to note that here, we are using a validation set to select tuning parameters. If instead one had to perform $K$-fold cross-validation, \texttt{MRCE-Approx} may become prohibitively time-consuming to implement. 

\subsection{Results under \textbf{M2} using Theoretical Tuning}
We again consider \eqref{eqMSRL} using tuning parameters chosen according to our results in Section \ref{subsec:pivotal}. As mentioned in the previous subsection, variable selection in this context is substantially easier than under \textbf{M1}, and as we will see, this leads to theoretically tuned versions of \eqref{eqMSRL} which perform nearly as well as those tuned using the validation set---even without refitting. 

Results for the same variations of \eqref{eqMSRL} (\texttt{MSR-q95}, \texttt{MSR-q85}, \texttt{MSR-q75}, \texttt{MSR-q50}, and \texttt{MRCE-Or}), except with $g$ as the group lasso penalty and quantiles based on the distribution of $(c/\sqrt{n})\|\mbX^\top\mbS\|_{\infty,2}$, are displayed in Figure \ref{fig:Group_Theory_Tuning}. In this context, we see that \texttt{MSR-q50} and \texttt{MRCE-Or} almost always have an average squared Frobenius norm error less than 1.25 that of \texttt{MSR-CV}. Examining the variable selection results displayed in Table \ref{table:variableSelection_M2} of the Appendix, we see that in general, the directly tuned estimators tend to have nearly perfect variable selection accuracy. The difference between the strong variable selection performance and the slight increase in squared Frobenius norm error (relative to \texttt{MSR-CV}) can be attributed to the bias induced from using the nuclear norm as a loss function. Refitting did slightly improve the Frobenius norm estimation error, but less so than under \textbf{M1}, so we omit these results. Finally, it is important to highlight that these estimators often take less than a single second to compute.

\subsection{Results under \textbf{M3} using Cross-Validation}
Lastly, we consider the estimation of $\mbbeta_*$ under \textbf{M3} by setting $g$ to be the nuclear norm penalty for a subset of the methods discussed in Section \ref{sec:datagenmodels}. The R package \texttt{camel} does not include an implementation of the nuclear norm penalized version of \texttt{Calibrated}, so this competitor is omitted from these comparisons.  We focus on the setting that $(n, p,q) = (200, 50, 40)$. We adjusted dimensions because when even when $\mbbeta^*$ is rank $r$, there are a large number of parameters, $r(p + q - r)$, to be estimated. 

For 100 independent replications under the data generating Models 1--3, we construct $\mbbeta_*$ by first computing $\mbU_*$ and $\mbV_*$, the left and right singular vectors of a randomly generated $p \times q$ matrix with independent and identically distributed standard normal entries. Then, we set $\mbbeta_* = \sum_{k=1}^{5} \boldsymbol{d}_k \mbu_{*k} \mbv_{*k}^\top$ where the $\boldsymbol{d} = (3, 2.5, 2, 1.5, 1)^\top$ and $\mbu_{*k}$ is the $k$th column of $\mbU_*$ and $\mbv_{*k}$ is the $k$th column of $\mbV_*$. This way, ${\rm rank}(\mbbeta_*) = 5$ almost surely. As before, we first consider the performance of the various methods using the validation set to select tuning parameters. 

\begin{figure}[t!]
\begin{center}
\makebox[\textwidth][c]{\includegraphics[width=\textwidth]{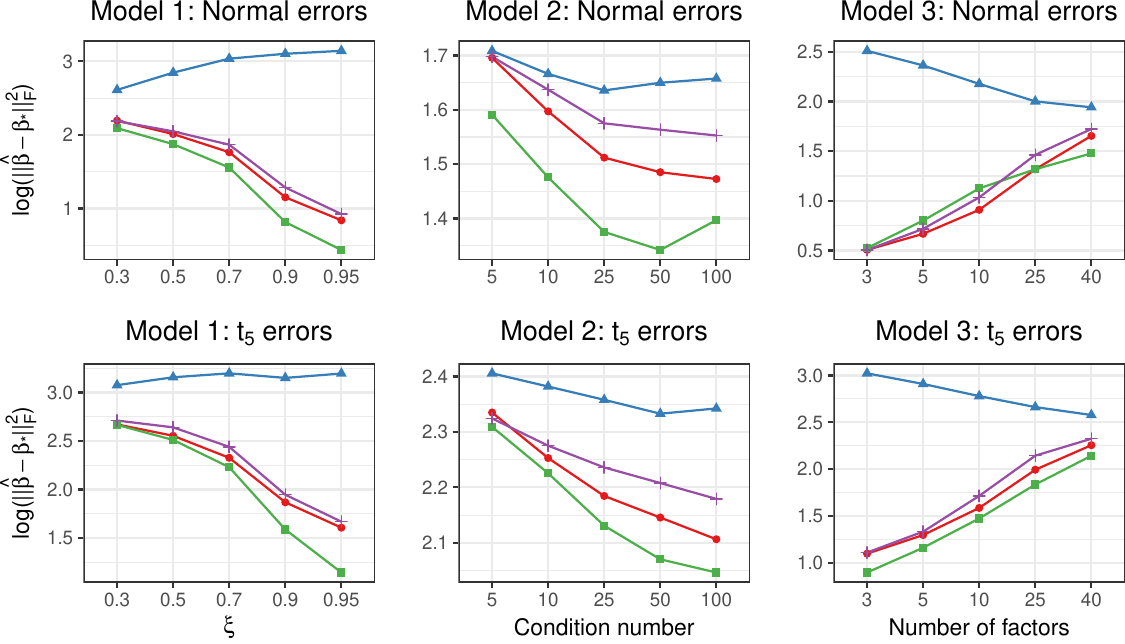}}
\makebox[\textwidth][c]{\includegraphics[width=11cm]{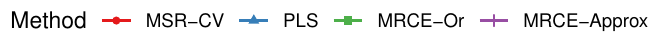}}
\vspace{-30pt}
 \end{center}
\caption{Average log squared Frobenius norm errors over one hundred independent replications under Model 1--3 with (top row) normal errors or (bottom row) $t_5$ errors and $\xi$, the condition number, and number of factors varying. In these simulations, $\mbbeta_*$ is constructed according to \textbf{M3} and $g$ is the nuclear norm.}\label{fig:NN_Results}.
\end{figure}

Results are displayed in Figure \ref{fig:NN_Results}. We see that like under \textbf{M1} and \textbf{M2}, in general, \texttt{MRCE-Or} performs best under \textbf{M3}. Interestingly, \texttt{MSR-CV} tends to outperform \texttt{MRCE-Approx} in the majority of settings considered. For example, under Model 1 and 2, when errors are more highly correlated, there is a more clear separation between \texttt{MSR-CV} and \texttt{MRCE-Approx} than under \textbf{M1}. Errors are larger overall for each method relative to \textbf{M1} or \textbf{M2} because in this setting $\mbbeta^*$ has $pq = 2000$ nonzero coefficients.  

\subsection{Results under \textbf{M3} using Theoretical Tuning}
Finally, we try selecting tuning parameters based on our theory. In general, however, these tuning parameters work about as poorly as under \textbf{M1} (e.g., performance was similar to that in the top row of Figure \ref{fig:theory_tuning1}). For this reason, we again consider refitted versions of these estimators. To refit $\hat\mbbeta_{\rm LR}$, we use the joint penalized maximum likelihood estimator from \eqref{eq:SUR} of the Appendix, except we constrain the optimization variable $\mbbeta$ to belong to the set of matrices which have rank less than or equal to that of $\hat\mbbeta_{\rm LR}.$ 

\begin{figure}[t]
\begin{center}
\includegraphics[width=\textwidth]{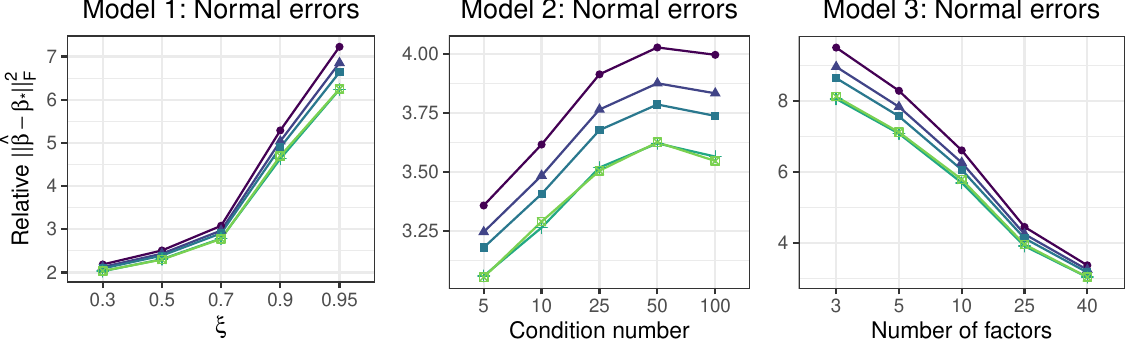}
\makebox[\textwidth][c]{\includegraphics[width=13cm]{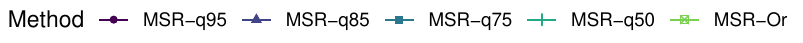}}\\
\end{center}
\begin{center}
\includegraphics[width=\textwidth]{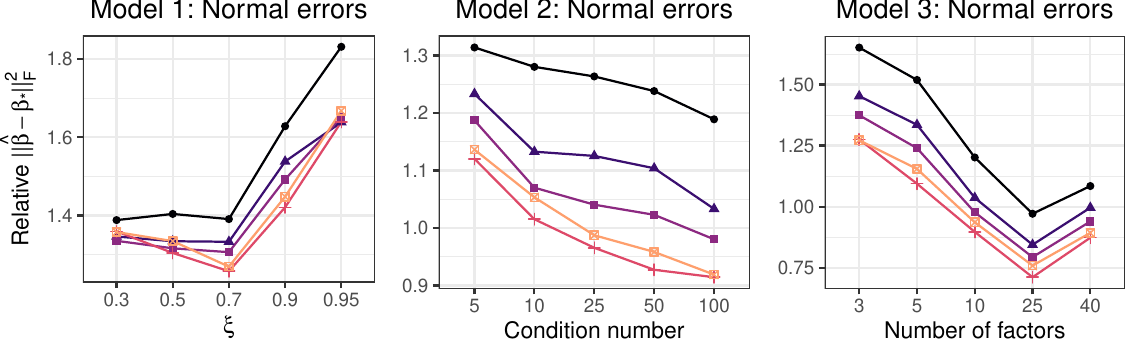}
\makebox[\textwidth][c]{\includegraphics[width=16cm]{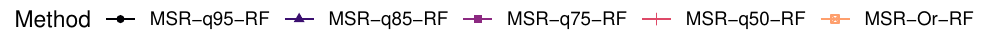}}
\end{center}
\vspace{-20pt}
\caption{Relative (to \texttt{MSR-CV}) average squared Frobenius norm errors under Model 1--3 with normal errors, $\mbbeta^*$ constructed according to \textbf{M3}, and $g$ taken to be the nuclear norm both (top row) without refitting and (bottom row) with refitting. }\label{fig:NN_refit}
\end{figure}
We display results relative to \texttt{MSR-CV} in Figure \ref{fig:NN_refit}. Here, we see that theoretical tuning combined with refitting can outperform \texttt{MSR-CV}. In Table \ref{table:rankEstimation_M3} of the Appendix, we see that the theoretically tuned versions tend to estimate the rank more accurately, but it seems that the combination of rank reduction and shrinkage of \texttt{MSR-CV} leads to improved performance in terms of squared Frobenius norm error compared to the refitted version which only imposes low-rankness.

\subsection{Conclusions}
In these simulation studies, we saw that \eqref{eqMSRL} can outperform \texttt{MRCE-Approx}, a method that requires an explicit estimate of the error precision matrix. In addition, in all of the settings we considered, \texttt{MSR-CV} required significantly less time to implement. While the tuning parameters suggested by our theory did not perform as well as those selected by cross-validation, under both \textbf{M1} and \textbf{M2}, these tuning parameters led to reasonable variable selection accuracy. Namely, the directly tuned versions of \eqref{eqMSRL} rarely included predictors which were not truly important, and could be computed in around one second on average. A similar result, although related to the rank of $\mbbeta_*$, was observed under \textbf{M3}. In practice, we advise practitioners to use cross-validation if computing time is not an issue. Otherwise, directly tuned versions of \eqref{eqMSRL} may be useful if short implementation times and model parsimony are of primary concern. 

The simulation settings considered here all have $n > q$. However, \eqref{eqMSRL} can be applied in settings where $q \geq n$. To demonstrate that \eqref{eqMSRL} can still perform well in these settings, we provide additional simulation results in Section \ref{subsec:qgreatern} of Appendix A in the case that $q = 60$, $n = 50$, and $p = 500$. To summarize briefly, with $\mbbeta_*$ constructed according to \textbf{M1} and data generated under Models 1--3 with normal errors, \eqref{eqMSRL} outperformed \texttt{MRCE-Approx} and other competitors (except \texttt{MRCE-Or}) under both Models 1 and 3, but both \texttt{MRCE-Approx} and \texttt{MSR-CV} performed very poorly under Model 2. This can be attributed to the difficulties in estimating (implicitly or explicitly) the $q \times q$ error covariance with such a small sample size.

\section{Glioblastoma Multiforme Application}\label{sec:GBM}
We used our method to model the linear relationship between microRNA expression and gene expression in patients with glioblastoma multiforme---an aggressive brain cancer---collected by The Cancer Genome Atlas program (TCGA, \citet{weinstein2013cancer}). Earlier versions of this data set were analyzed by \citet{wang2015joint} and \citet{lee2012simultaneous}, both of whom proposed new methods for multivariate response linear regression which explicitly estimate the error precision matrix. Following both \citet{wang2015joint} and \citet{lee2012simultaneous}, microRNA expression profiles were treated as the response and gene expression profiles were treated as predictors.

Similar to \citet{wang2015joint}, we reduce the dimension of both predictors and responses by retaining only the $p$ genes with largest median absolute deviation and the $q$ microRNAs with largest median absolute deviation. We then removed 93 subjects whose first two principal components for gene expression were substantially different than the majority of subjects. After removing these patients, there were 397 subjects in our complete data set. 

\begin{table}[t!]
\begin{center}
\scalebox{0.95}{
\centering
\begin{tabular}{ccc|cc|cc|cc}
  \toprule
  & \multicolumn{4}{c}{Weighted prediction error} & \multicolumn{4}{c}{Nuclear norm prediction error} \\
  \midrule
$q$ & \multicolumn{2}{c|}{20} & \multicolumn{2}{c|}{40} & \multicolumn{2}{c|}{20} & \multicolumn{2}{c}{40} \\
$p$ &  500 & 1000 & 500 & 1000 & 500 & 1000 & 500 & 1000 \\ 
  \midrule
\texttt{MSR-CV} & 0.6411 & 0.6161 & 0.6694 & 0.6510 & 0.2126 & 0.2077 & 0.3385 & 0.3328 \\ 
\texttt{PLS} & 0.6506 & 0.6198 & 0.6740 & 0.6488 & 0.2145 & 0.2090 & 0.3399 & 0.3333 \\ 
\texttt{PLS-q} & 0.6511 & 0.6200 & 0.6754 & 0.6496 & 0.2147 & 0.2091 & 0.3414 & 0.3348 \\ 
\midrule
  \texttt{MSR}$^*$ & 0.6395 & 0.6117 & 0.6689 & 0.6460 & 0.2124 & 0.2071 & 0.3382 & 0.3319 \\ 
  \texttt{MRCE-Approx}$^*$  & 0.6386 & 0.6091 & 0.6656 & 0.6399 & 0.2123 & 0.2070 & 0.3380 & 0.3313 \\ 
  \bottomrule
\end{tabular}
}
\end{center}
\caption{Weighted prediction errors and nuclear norm prediction errors averaged over 100 training/testing splits for the five considered methods from Section \ref{sec:GBM} with $g$ taken to be the $L_1$-norm. The superscript $*$ denotes a method which uses best-case tuning. Methods without the $*$ uses tuning parameters chosen by five-fold cross-validation. }\label{table:TCGA}
\end{table}

For one hundred independent replications, we randomly split the data into training and testing sets of size 250 and 147, respectively. We fit the multivariate response linear regression model using multiple methods described in Section \ref{sec:datagenmodels} with $g$ taken to be the $L_1$-norm: \texttt{MSR-CV}, \texttt{PLS}, and a version of \texttt{PLS} with different tuning parameters $\lambda$ for each of the $q$ responses (\texttt{PLS-q}). For \texttt{MSR-CV} and \texttt{PLS}, tuning parameters are selected by five-fold cross-validation minimizing squared prediction error averaged over all responses. Unfortunately, computing times for \texttt{MRCE-Approx} could be extremely long, so we tried ``best-case'' tuning, i.e., we select the tuning parameters which gave the minimum squared prediction error averaged over all responses on the test set. This approach is not applicable in practice, but is included to demonstrate that \eqref{eqMSRL} performs similarly to the much more computationally intensive approach. For comparison, we also include the best-case tuning version of \eqref{eqMSRL}.  We denote both of these versions with a superscript $*$ in Table \ref{table:TCGA}.

We compared the five methods in terms of two prediction metrics: nuclear norm prediction error, $\|\mbY_{\rm test} - \hat{\mbY}\|_*/1000$, and weighted prediction error, $\|(\mbY_{\rm test} - \hat{\mbY}) \boldsymbol{\Lambda}^{-1}\|_F^2/147q$, where $\boldsymbol{\Lambda}$ is a diagonal matrix with the complete data response standard deviations along its diagonal. 

Among the methods which could be used in practice, \texttt{MSR-CV} substantially outperformed both versions of \texttt{PLS} in terms of weighted prediction error when $p = 500$. When $p = 1000$, \texttt{MSR-CV} performed only similarly to \texttt{PLS}. Both best-case methods performed slightly better than \texttt{MSR-CV}, with the more computationally intensive method of \citet{rothman2010sparse}, \texttt{MRCE-Approx}, slightly outperforming \eqref{eqMSRL} with best-case tuning in the higher-dimensional settings. In terms of nuclear norm prediction error, \texttt{MSR-CV} outperformed both versions of \texttt{PLS} in every setting, and performed almost identically to the best-case version of \texttt{MRCE-Approx} in most settings. 

\section{Discussion}\label{sec:discussion}
In this article, we studied multiple versions of \eqref{eqMSRL}, the multivariate square-root lasso. There are numerous interesting directions for future research. First, the extension of \eqref{eqMSRL} to settings with matrix or tensor-valued responses may be of particular interest. In these situations, there is often a high degree of dependence across entries in the tensor-valued error (e.g., when the data are spatial and/or temporal). Implicit covariance estimation may be helpful as the dimension of the response often makes explicit covariance estimation computational infeasible. Second, it is also of interest to establish conditions under which \eqref{eqMSRL} estimates exactly the set of nonzero elements of $\mbbeta_*$ (for \textbf{M1} and \textbf{M2}) or consistently estimates the rank of $\mbbeta_*$ (for \textbf{M3}). However, the nondifferentiability of the nuclear norm of residuals makes the application of the standard proof techniques (e.g., the \textit{primal-dual witness} of \citet{wainwright2009sharp} and \citet{lee2015model}) nontrivial without requiring unreasonable assumptions.
For example, to establish a bound for $\|\hat\mbbeta_{\rm GL} - \mbbeta_*\|_{\infty,2}$, \citet{massias2020support} required that $\mbY - \mbX \hat\mbbeta_{\rm GL}$ was rank $q$, which as discussed in Section 3.2, is problematic.  Thus, we leave the conditions necessary for support recovery and rank estimation consistency---as well as the development of a proof technique for establishing such conditions---as future work.

A reviewer pointed out a connection between \eqref{eqMSRL} and a smoothed variation of \eqref{eq:joint_interp} proposed by \citet{massias2017generalized,massias2020support}. The method of \citet{massias2017generalized} assumes that columns of the error matrix are independent and identically distributed with covariance $\mbTheta_{*} \in \mathbb{S}^{n}_+$, which they estimate explicitly. However, their estimation criterion could be modified to accommodate our assumption that rows of $\mbE$ are independent and columns are correlated. The analog of their estimator conforming to our model assumptions in \eqref{eq:MVR} is 
\begin{equation}\label{eq:smoothed_analog}
\argmin_{\mbbeta \in \mathbb{R}^{p \times q}, \mbSigma^{1/2} \succeq \underline{\sigma} \mbI_q} \left[ \frac{1}{2n}{\rm tr} \big\{ (\mbY - \mbX\mbbeta)\mbSigma^{-1/2} (\mbY - \mbX\mbbeta)^\top\big\} + \frac{{\rm tr}(\mbSigma^{1/2})}{2}  + \lambda g(\mbbeta) \right],
\end{equation}
where the notation $\mbSigma^{1/2} \succeq \underline{\sigma} \mbI_q$ means $\mbSigma^{1/2} - \underline{\sigma} \mbI_q$ is positive semidefinite and $\underline{\sigma} > 0$ is a tuning parameter lower bounding the eigenvalues of $\mbSigma^{1/2} \in \mathbb{S}^{q}_+$.  Thus, we can view both the method of \citet{massias2017generalized} and  \eqref{eq:smoothed_analog} as smooth approximations to \eqref{eqMSRL}. As future work, it would be interesting to study whether the additional constraint on $\mbSigma^{1/2}$ in \eqref{eq:smoothed_analog} would allow one to relax the assumption that $n > q$. However, \eqref{eq:smoothed_analog} does have a potential drawback: \eqref{eq:smoothed_analog} can sometimes require explicit estimation of $\mbSigma_*^{1/2}$, so it is not clear when this estimator would be any easier to compute than the method of \citet{rothman2010sparse}.

\section*{Acknowledgments}
The author thanks three anonymous referees and the action editor for their many helpful comments. 
The author also thanks Benjamin Stucky and Sara van de Geer for sharing their code and their responses to inquiries; thanks Rohit K. Patra for a helpful conservation; and thanks Daniel J. Eck, Karl Oskar Ekvall, Keshav Motwani, Bradley S. Price, Adam J. Rothman, and Ben Sherwood for their feedback on earlier drafts of this article. This work was supported in part by National Science Foundation grant DMS-2113589.

\appendix
\section{Additional Details}
\subsection{Additional Computational Details}\label{sec:additional_computational_details}
In this section, we discuss our implementation of the accelerated proximal gradient descent algorithm in Algorithm 2. As mentioned in Section \ref{eq:alt_comp}, this algorithm can be used in situations where $\hat\mbbeta_g$ belongs to $\mathcal{D}_{\underline{\kappa}}$ for some positive $\underline{\kappa}$ bounded away from zero. Since we do not know, in general, whether $\mbY - \mbX \hat\mbbeta_g$ will be rank $q$ before computing $\hat\mbbeta_g$, we can attempt to use Algorithm 2, and if any iterates do not belong to $\mathcal{D}_{\underline{\kappa}}$, we may instead revert to using Algorithm 1. In our implementation, if $n > q$, we start computing the solution path for $\hat\mbbeta_g$ using Algorithm 2, but if at any iterate, the diagonal elements of $\bar{\mbD}$ or $\tilde{\mbD}$ (see 3 and 5 of Algorithm 2) are smaller than $10^{-3}$, we revert to Algorithm 1 and compute the rest of the solution path using Algorithm 1. 

To claim convergence, we check the first order conditions as described in Remark 7. For concreteness, we discuss the version we use with $g$ being the $L_1$-norm. Specifically, we let $(\mbU_{\epsilon^{(k+1)}},\mbD_{\epsilon^{(k+1)}},\mbV_{\epsilon^{(k+1)}}) = {\rm svd}(\mbY - \mbX \mbbeta^{(k+1)}).$ Then, we terminate the algorithm if (i) $\| \mbX^\top\mbU_{\epsilon^{(k+1)}}\mbV^\top_{\epsilon^{(k+1)}}\|_\infty \leq \sqrt{n}\lambda$, (ii)
$\max_{(l,m):[\hat\mbbeta^{(k+1)}]_{l,m} \neq 0}  |[\mbX^\top\mbU_{\epsilon^{(k+1)}}\mbV_{\epsilon^{(k+1)}}^\top - \sqrt{n} \lambda {\rm sign}(\hat\mbbeta^{(k+1)})]_{l,m}| < \tau$, and (iii) $ \mbY - \mbX \mbbeta^{(k+1)}$ is rank $q$. For the timing results in Table \ref{table:timing_results_comparison}, we set $\tau = 10^{-10}.$ We found that compared to the default implementation of the prox-linear ADMM (Algorithm 1), Algorithm 2 led to very slightly more accurate solutions.

  \begin{algorithm}[ht!]\caption{Accelerated proximal gradient descent algorithm for \eqref{eqMSRL}}
    \begin{tabbing}
 \textit{1.} 
  Given $\rho_0 > 0$ and $\gamma_{\rm decr} \in (0,1)$, initialize $\mbbeta^{(-1)} = \mbbeta^{(0)}\in \mathbb{R}^{p \times q}$, $\alpha^{(0)} = \alpha^{(-1)} = 1$, \\ ~~~~$(\dot{\mbU}, \dot{\mbD}, \dot{\mbV}) = {\rm svd}(\mbY - \mbX \mbbeta^{(0)})$ and set $k=0$\\
 \textit{2.} $\mbGamma^{(k)} \leftarrow \mbbeta^{(k)} + \left(\frac{\alpha^{(k-1)} - 1}{\alpha^{(k)}}\right) \left(\mbbeta^{(k)} - \mbbeta^{(k-1)}\right)$ \\
 \textit{3.} $(\tilde{\mbU}, \tilde{\mbD}, \tilde{\mbV}) \leftarrow {\rm svd}(\mbY - \mbX \mbGamma^{(k)})$\\
 \textit{4.} $\bar{\mbbeta} \leftarrow {\rm Prox}_{ \rho_k\lambda g} (\mbGamma^{(k)} + \frac{\rho_k}{\sqrt{n}}\mbX^\top\tilde{\mbU}\tilde{\mbV}^\top)$\\
 \textit{5.} $(\bar{\mbU}, \bar{\mbD}, \bar{\mbV}) \leftarrow {\rm svd}(\mbY - \mbX\bar{\mbbeta})$ \\
 \textit{6.} If ${\rm tr}(\bar{\mbD})  < {\rm tr}(\tilde{\mbD}) + 
  {\rm tr}\{\tilde{\mbV}\tilde{\mbU}^\top\mbX(\mbGamma^{(k)} - \bar{\mbbeta})\} + \frac{\sqrt{n}}{2\rho_k}\|\mbGamma^{(k)} - \bar{\mbbeta}\|_F^2$, go to \textit{7} \\
  \hspace{5pt}\quad \quad \quad Else, update $\rho_k \leftarrow \rho_k\gamma_{\rm decr}$ and return to \textit{4}\\
 \textit{7.} If ${\rm tr}(\bar{\mbD}) + \sqrt{n}\lambda g(\bar{\mbbeta}) \leq {\rm tr}(\dot{\mbD}) + \sqrt{n}\lambda g(\mbbeta^{(k)})$, set $\mbbeta^{(k+1)} \leftarrow \bar{\mbbeta}$ and $\dot{\mbD} \leftarrow \bar{\mbD}$ \\
 \hspace{5pt}\quad\quad\quad Else, set $\mbbeta^{(k+1)}\leftarrow \mbbeta^{(k)}$ \\
 \textit{8.} $\alpha^{(k+1)} \leftarrow (1 + \sqrt{1 + 4\{\alpha^{(k)} \}^2})/2$ \\
 \textit{8.} If not converged, set $\rho_{k+1} \leftarrow \rho_k$, update $k \leftarrow k + 1$, and return to \textit{2} 
\end{tabbing}
  \end{algorithm}

\subsection{Additional Simulation Results}\label{subsec:qgreatern}
\begin{figure}[t]
\begin{center}\makebox[\textwidth][c]{\includegraphics[width=\textwidth]{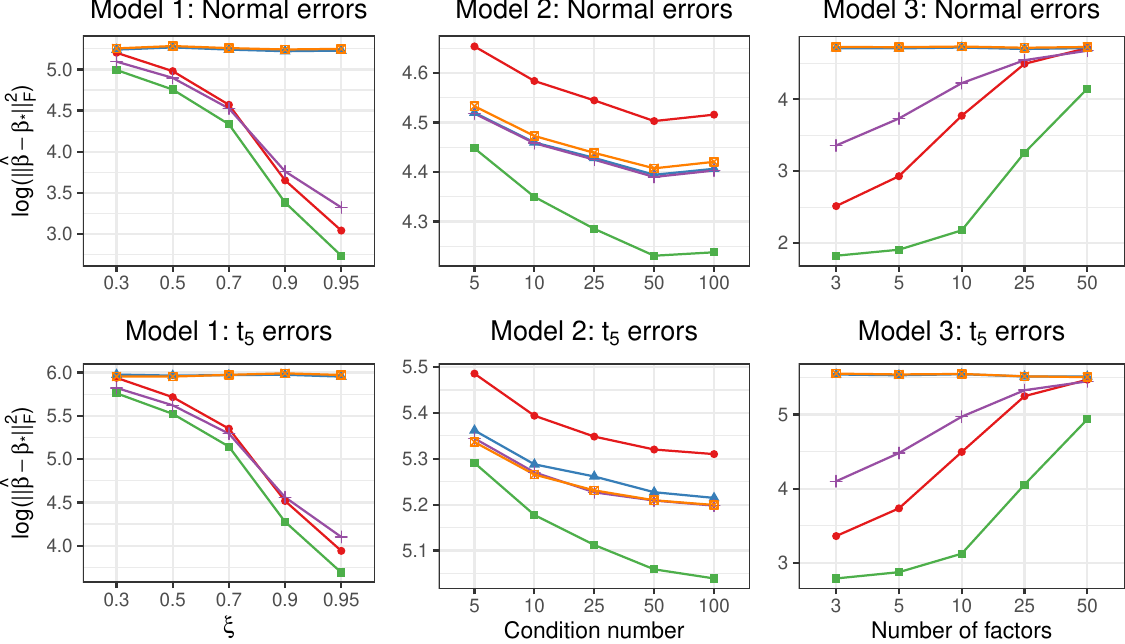}}
\makebox[\textwidth][c]{\includegraphics[width=12cm]{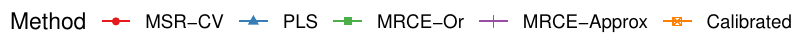}}
\vspace{-30pt}
\end{center}
\caption{Average log squared Frobenius norm errors over one hundred independent replications under Model 1--3 with $(n,p,q) = (50, 500, 60)$ and (top row) normal errors or (bottom row) $t_5$ errors and $\xi$, the condition number, and the number of factors varying. In these simulations, $\mbbeta_*$ is constructed according to \textbf{M1} and $g$ is the $L_1$-norm.}\label{fig:q60_Results}.
\end{figure}

In this section, we display additional simulation results with $\mbbeta_*$ constructed according to \textbf{M1} and $(n,p,q) = (50,60,500)$. The only difference between these data generating models and those from Section \ref{subsec:Results_1_M1} is that entries of $\mbG$ (from $\mbbeta_* = \mbA \circ \mbG$) are independent and identically distributed from a mean zero normal distribution with standard deviation two. Results from these simulations are displayed in Figure \ref{fig:q60_Results}. We observe that \texttt{MSR-CV} performs relatively well under both Model 1 and Model 3. Of course, compared to the results in Section \ref{subsec:Results_1_M1}, all estimators perform worse, which is expected given the smaller sample size and larger $q$. Notably, under Model 2, both \texttt{MSR-CV} and \texttt{MRCE-Approx} perform worse than \texttt{PLS} and \texttt{Calibrated}. However, we see that the oracle penalized maximum likelihood estimator, \texttt{MRCE-Or}, still performs well here. This suggests that the covariance structure under Model 2 is much more difficult to estimate than under Models 1 and 3 when the sample size is small relative to $q$. Together these results suggest that \texttt{MSR-CV} can work well in settings with $n > q$, although one may also consider \texttt{Calibrated} which makes the simplifying assumption that $\mbSigma_*$ is diagonal.

\begin{table}[ht!]
\centering
\scalebox{.85}{
\begin{tabular}{rcc|cc|cc|cc|cc}
   \toprule
  & \multicolumn{10}{c}{Model 1: $\xi$}\\
 & \multicolumn{2}{c}{$0.3$} &  \multicolumn{2}{c}{$0.5$} & \multicolumn{2}{c}{$0.7$}& \multicolumn{2}{c}{$0.9$} & \multicolumn{2}{c}{$0.95$} \\ 
   \toprule
\texttt{PLS} & 0.781 & 0.040 & 0.789 & 0.041 & 0.784 & 0.042 & 0.785 & 0.044 & 0.785 & 0.045 \\ 
  \texttt{MRCE-Or} & 0.820 & 0.042 & 0.848 & 0.044 & 0.881 & 0.045 & 0.931 & 0.047 & 0.952 & 0.049 \\ 
  \texttt{MRCE-Approx} & 0.819 & 0.046 & 0.848 & 0.049 & 0.884 & 0.052 & 0.932 & 0.056 & 0.954 & 0.062 \\ 
  \texttt{Calibrated} & 0.782 & 0.041 & 0.789 & 0.041 & 0.785 & 0.041 & 0.784 & 0.043 & 0.784 & 0.045 \\ 
  \midrule
  \texttt{MSR-CV} & 0.812 & 0.043 & 0.843 & 0.043 & 0.876 & 0.046 & 0.928 & 0.049 & 0.949 & 0.052 \\ 
  \texttt{MSR-q95} & 0.487 & 0.000 & 0.551 & 0.000 & 0.623 & 0.000 & 0.757 & 0.000 & 0.824 & 0.000 \\ 
  \texttt{MSR-q85} & 0.519 & 0.000 & 0.583 & 0.000 & 0.653 & 0.000 & 0.780 & 0.000 & 0.843 & 0.000 \\ 
  \texttt{MSR-q75} & 0.537 & 0.000 & 0.599 & 0.000 & 0.666 & 0.000 & 0.791 & 0.000 & 0.851 & 0.000 \\ 
  \texttt{MSR-q50} & 0.562 & 0.000 & 0.622 & 0.000 & 0.689 & 0.000 & 0.807 & 0.000 & 0.863 & 0.000 \\ 
  \texttt{MSR-Or} & 0.558 & 0.000 & 0.621 & 0.000 & 0.679 & 0.000 & 0.798 & 0.000 & 0.860 & 0.000 \\ 
   \toprule
  & \multicolumn{10}{c}{Model 2: Condition number}\\
   & \multicolumn{2}{c}{$5$} &  \multicolumn{2}{c}{$10$} & \multicolumn{2}{c}{$25$}& \multicolumn{2}{c}{$50$} & \multicolumn{2}{c}{$100$} \\ 
   \toprule
\texttt{PLS} & 0.847 & 0.045 & 0.846 & 0.045 & 0.847 & 0.044 & 0.847 & 0.043 & 0.848 & 0.045 \\ 
  \texttt{MRCE-Or} & 0.972 & 0.052 & 0.970 & 0.050 & 0.968 & 0.048 & 0.957 & 0.052 & 0.916 & 0.050 \\ 
  \texttt{MRCE-Approx} & 0.972 & 0.061 & 0.968 & 0.060 & 0.964 & 0.059 & 0.940 & 0.054 & 0.899 & 0.052 \\ 
  \texttt{Calibrated} & 0.847 & 0.044 & 0.847 & 0.045 & 0.849 & 0.044 & 0.847 & 0.043 & 0.848 & 0.044 \\ 
    \midrule
  \texttt{MSR-CV} & 0.970 & 0.051 & 0.966 & 0.050 & 0.962 & 0.048 & 0.936 & 0.045 & 0.892 & 0.046 \\ 
  \texttt{MSR-q95} & 0.879 & 0.000 & 0.864 & 0.000 & 0.821 & 0.000 & 0.696 & 0.000 & 0.626 & 0.000 \\ 
  \texttt{MSR-q85} & 0.895 & 0.000 & 0.882 & 0.000 & 0.849 & 0.000 & 0.732 & 0.000 & 0.658 & 0.000 \\ 
  \texttt{MSR-q75} & 0.903 & 0.000 & 0.890 & 0.000 & 0.860 & 0.000 & 0.748 & 0.000 & 0.675 & 0.000 \\ 
  \texttt{MSR-q50} & 0.911 & 0.000 & 0.901 & 0.000 & 0.876 & 0.000 & 0.775 & 0.000 & 0.700 & 0.000 \\ 
  \texttt{MSR-Or} & 0.907 & 0.000 & 0.899 & 0.000 & 0.866 & 0.000 & 0.772 & 0.000 & 0.688 & 0.000 \\ 
   \toprule
   & \multicolumn{10}{c}{Model 3: Number of factors}\\
   & \multicolumn{2}{c}{$2$} &  \multicolumn{2}{c}{$5$} & \multicolumn{2}{c}{$10$}& \multicolumn{2}{c}{$25$} & \multicolumn{2}{c}{$50$} \\ 
   \toprule
\texttt{PLS} & 0.862 & 0.044 & 0.870 & 0.044 & 0.875 & 0.045 & 0.876 & 0.044 & 0.874 & 0.044 \\ 
  \texttt{MRCE-Or} & 0.874 & 0.046 & 0.886 & 0.047 & 0.899 & 0.049 & 0.903 & 0.050 & 0.905 & 0.050 \\ 
  \texttt{MRCE-Approx} & 0.867 & 0.048 & 0.878 & 0.051 & 0.889 & 0.054 & 0.893 & 0.054 & 0.892 & 0.054 \\ 
  \texttt{Calibrated} & 0.864 & 0.045 & 0.871 & 0.045 & 0.876 & 0.046 & 0.877 & 0.045 & 0.875 & 0.045 \\ 
    \midrule
  \texttt{MSR-CV} & 0.866 & 0.046 & 0.876 & 0.045 & 0.886 & 0.046 & 0.891 & 0.046 & 0.890 & 0.046 \\ 
 \texttt{MSR-q95} & 0.601 & 0.000 & 0.622 & 0.000 & 0.637 & 0.000 & 0.638 & 0.000 & 0.639 & 0.000 \\ 
  \texttt{MSR-q85} & 0.631 & 0.000 & 0.652 & 0.000 & 0.665 & 0.000 & 0.668 & 0.000 & 0.669 & 0.000 \\ 
  \texttt{MSR-q75} & 0.645 & 0.000 & 0.666 & 0.000 & 0.681 & 0.000 & 0.683 & 0.000 & 0.684 & 0.000 \\ 
  \texttt{MSR-q50} & 0.668 & 0.000 & 0.690 & 0.000 & 0.703 & 0.000 & 0.706 & 0.000 & 0.704 & 0.000 \\ 
  \texttt{MSR-Or} & 0.663 & 0.000 & 0.684 & 0.000 & 0.693 & 0.000 & 0.701 & 0.000 & 0.695 & 0.000 \\ 
  \bottomrule
\end{tabular}
}
\caption{Average true positive and false positive variable selection rates for Models 1--3 under normal errors with $\mbbeta_*$ constructed according to \textbf{M1} and $g$ taken to be the $L_1$-norm.}\label{table:variableSelection_M1}
\end{table}

\begin{table}[th!]
\centering
\scalebox{.85}{
\begin{tabular}{r cc|cc|cc|cc|cc}
\toprule
  & \multicolumn{10}{c}{Model 1: $\xi$}\\
 & \multicolumn{2}{c}{$0.3$} &  \multicolumn{2}{c}{$0.5$} & \multicolumn{2}{c}{$0.7$}& \multicolumn{2}{c}{$0.9$} & \multicolumn{2}{c}{$0.95$} \\ 
   \toprule
\texttt{PLS} & 0.068 & 0.010 & 0.028 & 0.008 & 0.014 & 0.009 & 0.008 & 0.006 & 0.018 & 0.009 \\ 
  \texttt{MRCE-Or} & 0.910 & 0.093 & 0.976 & 0.110 & 1.000 & 0.120 & 1.000 & 0.149 & 1.000 & 0.156 \\ 
  \texttt{MRCE-Approx} & 0.778 & 0.116 & 0.888 & 0.148 & 0.922 & 0.165 & 0.984 & 0.222 & 0.992 & 0.246 \\ 
  \texttt{Calibrated} & 0.058 & 0.009 & 0.028 & 0.007 & 0.014 & 0.008 & 0.008 & 0.006 & 0.014 & 0.008 \\ 
  \midrule
  \texttt{MSR-CV} & 0.900 & 0.092 & 0.980 & 0.105 & 1.000 & 0.117 & 1.000 & 0.165 & 1.000 & 0.215 \\ 
  \texttt{MSR-q95} & 0.204 & 0.000 & 0.454 & 0.000 & 0.900 & 0.000 & 1.000 & 0.000 & 1.000 & 0.000 \\ 
  \texttt{MSR-q85} & 0.288 & 0.000 & 0.572 & 0.000 & 0.942 & 0.001 & 1.000 & 0.000 & 1.000 & 0.000 \\ 
  \texttt{MSR-q75} & 0.340 & 0.001 & 0.614 & 0.001 & 0.952 & 0.001 & 1.000 & 0.001 & 1.000 & 0.000 \\ 
  \texttt{MSR-q50} & 0.420 & 0.001 & 0.706 & 0.001 & 0.972 & 0.001 & 1.000 & 0.002 & 1.000 & 0.001 \\ 
  \texttt{MSR-Or} & 0.388 & 0.001 & 0.674 & 0.001 & 0.962 & 0.001 & 1.000 & 0.001 & 1.000 & 0.001 \\ 
   \toprule
  & \multicolumn{10}{c}{Model 2: Condition number}\\
   & \multicolumn{2}{c}{$5$} &  \multicolumn{2}{c}{$10$} & \multicolumn{2}{c}{$25$}& \multicolumn{2}{c}{$50$} & \multicolumn{2}{c}{$100$} \\ 
   \toprule
\texttt{PLS} & 0.152 & 0.018 & 0.242 & 0.020 & 0.676 & 0.050 & 0.910 & 0.071 & 0.972 & 0.080 \\ 
  \texttt{MRCE-Or} & 1.000 & 0.172 & 1.000 & 0.178 & 1.000 & 0.222 & 1.000 & 0.550 & 1.000 & 0.448 \\ 
  \texttt{MRCE-Approx} & 1.000 & 0.261 & 1.000 & 0.298 & 1.000 & 0.318 & 1.000 & 0.231 & 1.000 & 0.180 \\ 
  \texttt{Calibrated} & 0.144 & 0.017 & 0.228 & 0.020 & 0.672 & 0.050 & 0.904 & 0.070 & 0.966 & 0.071 \\ 
    \midrule
  \texttt{MSR-CV} & 1.000 & 0.192 & 1.000 & 0.152 & 1.000 & 0.154 & 1.000 & 0.144 & 1.000 & 0.100 \\ 
  \texttt{MSR-q95} & 1.000 & 0.000 & 1.000 & 0.000 & 1.000 & 0.000 & 1.000 & 0.000 & 0.998 & 0.000 \\ 
  \texttt{MSR-q85} & 1.000 & 0.000 & 1.000 & 0.000 & 1.000 & 0.000 & 1.000 & 0.000 & 1.000 & 0.000 \\ 
  \texttt{MSR-q75} & 1.000 & 0.000 & 1.000 & 0.001 & 1.000 & 0.000 & 1.000 & 0.000 & 1.000 & 0.000 \\ 
  \texttt{MSR-q50} & 1.000 & 0.001 & 1.000 & 0.001 & 1.000 & 0.001 & 1.000 & 0.001 & 1.000 & 0.001 \\ 
  \texttt{MSR-Or} & 1.000 & 0.001 & 1.000 & 0.001 & 1.000 & 0.001 & 1.000 & 0.001 & 1.000 & 0.001 \\ 
   \toprule
   & \multicolumn{10}{c}{Model 3: Number of factors}\\
   & \multicolumn{2}{c}{$2$} &  \multicolumn{2}{c}{$5$} & \multicolumn{2}{c}{$10$}& \multicolumn{2}{c}{$25$} & \multicolumn{2}{c}{$50$} \\ 
   \toprule
\texttt{PLS} & 1.000 & 0.100 & 0.998 & 0.103 & 1.000 & 0.102 & 1.000 & 0.098 & 1.000 & 0.102 \\ 
  \texttt{MRCE-Or} & 0.998 & 0.149 & 1.000 & 0.195 & 1.000 & 0.280 & 1.000 & 0.350 & 1.000 & 0.397 \\ 
  \texttt{MRCE-Approx} & 0.994 & 0.130 & 0.998 & 0.138 & 0.998 & 0.144 & 1.000 & 0.159 & 1.000 & 0.147 \\ 
  \texttt{Calibrated} & 1.000 & 0.103 & 0.998 & 0.102 & 1.000 & 0.100 & 1.000 & 0.096 & 1.000 & 0.107 \\ 
    \midrule
  \texttt{MSR-CV} & 1.000 & 0.117 & 1.000 & 0.114 & 1.000 & 0.117 & 1.000 & 0.120 & 1.000 & 0.118 \\ 
  \texttt{MSR-q95} & 0.850 & 0.000 & 0.928 & 0.000 & 0.982 & 0.000 & 0.988 & 0.000 & 0.992 & 0.000 \\ 
  \texttt{MSR-q85} & 0.894 & 0.000 & 0.960 & 0.000 & 0.990 & 0.000 & 0.994 & 0.000 & 0.996 & 0.000 \\ 
  \texttt{MSR-q75} & 0.920 & 0.000 & 0.970 & 0.000 & 0.992 & 0.000 & 0.994 & 0.001 & 0.998 & 0.001 \\ 
  \texttt{MSR-q50} & 0.942 & 0.001 & 0.982 & 0.001 & 0.998 & 0.002 & 0.996 & 0.001 & 1.000 & 0.001 \\ 
  \texttt{MSR-Or} & 0.928 & 0.001 & 0.972 & 0.001 & 0.996 & 0.001 & 0.990 & 0.001 & 0.998 & 0.001 \\ 
  \bottomrule
 \end{tabular}

}
\caption{Average true positive and false positive variable selection rates for Models 1--3 under normal errors with $\mbbeta_*$ constructed according to \textbf{M2} and $g$ taken to the be group lasso penalty.}\label{table:variableSelection_M2}
\end{table}

\begin{table}[th!]
\centering
\scalebox{.85}{
\begin{tabular}{r cc|cc|cc|cc|cc}
\toprule
  & \multicolumn{10}{c}{Model 1: $\xi$}\\
 & \multicolumn{2}{c}{$0.3$} &  \multicolumn{2}{c}{$0.5$} & \multicolumn{2}{c}{$0.7$}& \multicolumn{2}{c}{$0.9$} & \multicolumn{2}{c}{$0.95$} \\ 
  \toprule
\texttt{PLS} & 9.190 & 0.194 & 6.550 & 0.224 & 4.570 & 0.201 & 3.130 & 0.180 & 3.230 & 0.191 \\ 
  \texttt{MRCE-Or} & 15.060 & 0.134 & 16.360 & 0.131 & 17.960 & 0.141 & 20.610 & 0.138 & 21.990 & 0.165 \\ 
  \texttt{MRCE-Approx} & 16.830 & 0.132 & 19.590 & 0.174 & 23.010 & 0.190 & 27.730 & 0.212 & 28.920 & 0.366 \\
    \midrule 
  \texttt{MSR-CV} & 14.860 & 0.146 & 15.590 & 0.198 & 16.260 & 0.275 & 17.060 & 0.331 & 17.330 & 0.403 \\ 
  \texttt{MSR-q95} & 2.590 & 0.059 & 2.810 & 0.054 & 3.180 & 0.066 & 3.680 & 0.055 & 3.900 & 0.059 \\ 
  \texttt{MSR-q85} & 2.900 & 0.050 & 3.090 & 0.045 & 3.440 & 0.062 & 3.950 & 0.056 & 4.160 & 0.053 \\ 
  \texttt{MSR-q75} & 3.080 & 0.051 & 3.250 & 0.054 & 3.630 & 0.061 & 4.110 & 0.058 & 4.290 & 0.056 \\ 
  \texttt{MSR-q50} & 3.380 & 0.053 & 3.570 & 0.059 & 3.930 & 0.059 & 4.400 & 0.055 & 4.530 & 0.054 \\ 
  \texttt{MSR-Or} & 3.390 & 0.062 & 3.590 & 0.067 & 3.960 & 0.063 & 4.350 & 0.059 & 4.430 & 0.066 \\ 
   \toprule
  & \multicolumn{10}{c}{Model 2: Condition number}\\
   & \multicolumn{2}{c}{$5$} &  \multicolumn{2}{c}{$10$} & \multicolumn{2}{c}{$25$}& \multicolumn{2}{c}{$50$} & \multicolumn{2}{c}{$100$} \\ 
  \toprule
\texttt{PLS} & 7.470 & 0.070 & 8.980 & 0.080 & 10.810 & 0.101 & 12.900 & 0.117 & 14.130 & 0.110 \\ 
  \texttt{MRCE-Or} & 27.950 & 0.225 & 28.310 & 0.173 & 26.750 & 0.141 & 20.240 & 0.161 & 20.820 & 0.134 \\ 
  \texttt{MRCE-Approx} & 23.390 & 0.282 & 23.020 & 0.295 & 23.400 & 0.204 & 25.900 & 0.259 & 21.220 & 0.203 \\ 
    \midrule
  \texttt{MSR-CV} & 16.290 & 0.215 & 15.890 & 0.175 & 16.070 & 0.161 & 16.530 & 0.149 & 17.270 & 0.120 \\ 
  \texttt{MSR-q95} & 4.110 & 0.055 & 3.950 & 0.063 & 4.030 & 0.061 & 3.860 & 0.060 & 3.450 & 0.063 \\ 
  \texttt{MSR-q85} & 4.340 & 0.054 & 4.180 & 0.056 & 4.330 & 0.059 & 4.160 & 0.060 & 3.700 & 0.061 \\ 
  \texttt{MSR-q75} & 4.510 & 0.050 & 4.360 & 0.054 & 4.470 & 0.054 & 4.300 & 0.058 & 3.850 & 0.059 \\ 
  \texttt{MSR-q50} & 4.690 & 0.046 & 4.660 & 0.048 & 4.710 & 0.046 & 4.580 & 0.052 & 4.140 & 0.055 \\ 
  \texttt{MSR-Or} & 4.690 & 0.046 & 4.570 & 0.056 & 4.610 & 0.055 & 4.460 & 0.058 & 4.110 & 0.060 \\ 
   \toprule
   & \multicolumn{10}{c}{Model 3: Number of factors}\\
   & \multicolumn{2}{c}{$2$} &  \multicolumn{2}{c}{$5$} & \multicolumn{2}{c}{$10$}& \multicolumn{2}{c}{$25$} & \multicolumn{2}{c}{$40$} \\ 
  \toprule
\texttt{PLS} & 16.190 & 0.144 & 15.650 & 0.140 & 15.660 & 0.167 & 15.470 & 0.156 & 15.120 & 0.153 \\ 
  \texttt{MRCE-Or} & 18.000 & 0.133 & 18.650 & 0.124 & 19.370 & 0.141 & 19.630 & 0.143 & 17.500 & 0.326 \\ 
  \texttt{MRCE-Approx} & 17.110 & 0.131 & 17.600 & 0.169 & 18.610 & 0.239 & 19.630 & 0.260 & 19.970 & 0.242 \\ 
  \midrule
  \texttt{MSR-CV} & 18.260 & 0.143 & 18.410 & 0.126 & 18.580 & 0.146 & 18.730 & 0.134 & 18.310 & 0.126 \\ 
  \texttt{MSR-q95} & 3.040 & 0.060 & 3.220 & 0.060 & 3.300 & 0.059 & 3.380 & 0.055 & 3.480 & 0.059 \\ 
  \texttt{MSR-q85} & 3.260 & 0.063 & 3.560 & 0.056 & 3.610 & 0.060 & 3.650 & 0.054 & 3.790 & 0.057 \\ 
  \texttt{MSR-q75} & 3.440 & 0.062 & 3.740 & 0.050 & 3.780 & 0.054 & 3.820 & 0.048 & 3.950 & 0.059 \\ 
  \texttt{MSR-q50} & 3.700 & 0.058 & 4.010 & 0.054 & 4.040 & 0.057 & 4.130 & 0.054 & 4.200 & 0.064 \\ 
  \texttt{MSR-Or} & 3.700 & 0.067 & 3.960 & 0.065 & 4.010 & 0.064 & 4.100 & 0.063 & 4.210 & 0.064 \\ 
  \bottomrule
 \end{tabular}
}
\caption{Average estimated rank and standard errors for Models 1--3 under normal errors with $\mbbeta_*$ constructed according to \textbf{M3} and $g$ taken to be the nuclear norm.}\label{table:rankEstimation_M3}
\end{table}

\subsection{Method for Refitting}\label{sec:refit}
To refit the estimators as described in Section 5.4, we use a seemingly unrelated regressions-type \citep{zellner1962efficient} penalized normal maximum likelihood estimator. Suppose we are given $\hat{\mbbeta}_g$, an estimate of $\mbbeta_*$ from which we want to obtain a refitted version with, for example, an identical sparsity pattern as $\hat\mbbeta_{\rm L}$ (when $g$ is the $L_1$-norm), or a rank less than or equal to that of $\hat\mbbeta_{\rm LR}$ (when $g$ is the nuclear norm). Define the set $${\rm C}_{\rm L}(\hat{\mbbeta}_{\rm L}) = \big\{ \mbbeta \in \mathbb{R}^{p \times q}: \mbbeta_{j,k} = 0, \text{ for all } (j,k) \text{ such that } [\hat{\mbbeta}_{\rm L}]_{j,k} = 0  \big\},$$ 
and define 
$$ {\rm C}_{\rm LR}(\hat{\mbbeta}_{\rm LR}) =  \big\{ \mbbeta \in \mathbb{R}^{p \times q}:{\rm rank}(\mbbeta) \leq {\rm rank}(\hat{\mbbeta}_{\rm LR}) \big\}.$$
To obtain the refitted version of $\hat{\mbbeta}_g$, we solve 
\begin{equation}\label{eq:SUR} \argmin_{\mbbeta \in {\rm C}_g(\hat{\mbbeta}_g), \mbOmega \in \mathbb{S}^q_+} \left[ \frac{1}{n}{\rm tr}\big\{ (\mbY - \mbX\mbbeta)\mbOmega(\mbY - \mbX\mbbeta)^\top \big\} - \log{\rm det}(\mbOmega) + \frac{\alpha}{2}\|\mbOmega\|_F^2 \right],
\end{equation}
where we fix $\alpha = 10^{-4}$. To solve \eqref{eq:SUR}, we use blockwise coordinate descent. Specifically, for $k=1,2,3, \dots$, until convergence, we iterate between the following two steps:
\begin{enumerate}
\item $\mbOmega^{(k+1)} = 
\mbU \big\{ - \mbD + (\mbD^2 + 4\alpha \mbI_q)^{1/2}\big\} \mbU^\top/(2\alpha)$ where $(\mbU, \mbD, \mbU) = {\rm svd}\{(\mbY - \mbX\mbbeta^{(k)})^\top(\mbY - \mbX\mbbeta^{(k)})/n\}$. 
\item $\mbbeta^{(k+1)} = \argmin_{\mbbeta \in {\rm C}_g(\hat{\mbbeta}_g)} {\rm tr}\big\{(\mbY - \mbX \mbbeta) \mbOmega^{(k+1)}(\mbY - \mbX\mbbeta)^\top \big\}$
\end{enumerate}
Step 1 is the well-known solution for the ridge-penalized normal likelihood precision matrix estimation problem \citep{witten2009covariance}. When $g$ is the $L_1$-norm, Step 2 can be solved efficiently using an accelerated projected gradient descent algorithm. When $g$ is the nuclear norm, Step 2 has a closed form (e.g., see Chapter 2 of \citet{velu2013multivariate}). We terminate the algorithm when the objective function value converges. 

\section{Proofs}\label{appendixB}
\subsection{Notation and Preliminaries}
First, we clarify some of the notation that will be used in later sections. Recall that we assume that $\mbY = \mbX \mbbeta_* + \mbE$
where $\mbE \in \mathbb{R}^{n \times q}$ is a random matrix of errors which is assumed to have mean zero and be rank $q$ almost surely. For the remainder, define $(\mbU_\epsilon, \mbD_\epsilon, \mbV_\epsilon) = {\rm svd}(\mbE).$ 

For ease of display, we use $\phi$ and $\nu$ in place of $\phi_{\mathcal{E}, g}(\mathcal{M}, \mathcal{N}, c)$ and $\nu_g(\mathcal{M}, \mathcal{N}, c)$, respectively.
As before, define the quantities $\check{c} = (c+1)/(c-1)$ and $\tilde{c} = c (c+1)/(c-1)$ for constant $c > 1$. Also, for a symmetric matrix $\mbA$, let $\varphi_1(\mbA)$ denote the largest eigenvalue of $\mbA$, and for an arbitrary matrix $\mbA$, let $\sigma_j(\mbA)$ denote the $j$th largest singular value of $\mbA.$ For an $a \times b$ matrix $\mbA$, we let $\mbA_{j,\cdot} \in \mathbb{R}^b$ denote the $j$th row of $\mbA$, $\mbA_{\cdot, k} \in \mathbb{R}^{a}$ the $k$th column of $\mbA$, and $\mbA_{j,k}$ the $(j,k)$th entry of $\mbA.$  Let $\mbI_s$ be the $s \times s$ identity matrix. Finally, for sequences $a_n$ and $b_n$, let the notation $a_n \asymp b_n$ mean that $a_n = O(b_n)$ and $b_n  = O(a_n).$

Throughout, let $\tilde{g}$ denote the dual norm of $g$, i.e., $\tilde{g}(x) = \sup_z\{z^\top x: g(z) \leq 1\}$. For examples of penalty functions $g$ and their dual norms, see Table 1 of \citet{wainwright2014structured}. Following the notation of \citet{negahban2012unified}, let 
$\mathcal{M}$ and $\mathcal{N}^\perp$ denote the model subspace and the perturbation subspace, respectively. Under the various model assumptions, \textbf{M1}--\textbf{M3}, we assume that $\mbbeta_* \in \mathcal{M}$, where the specific form of $\mathcal{M}$ depends on the particular model. Stated in another way, we assume 
$\mbbeta_* = \mbbeta_{*\mathcal{M}} +  \mbbeta_{*\mathcal{M}^\perp} = \mbbeta_{*\mathcal{M}},$
where $\mbbeta_{*\mathcal{M}}$ denotes the projection of $\mbbeta_*$ onto $\mathcal{M}$, 
$ \mbbeta_{*\mathcal{M}} = \argmin_{\mbA \in \mathcal{M}}\|\mbA - \mbbeta_*\|_F^2,$
and $\mathcal{M}^\perp$ is the orthogonal complement of $\mathcal{M}.$
Under each of the model assumptions, \textbf{M1}--\textbf{M3}, we will assume the use of a penalty which is decomposable with respect to the pair $(\mathcal{M}, \mathcal{N}^\perp)$ in the sense that 
$ g(\mbA + \mbB) = g(\mbA) + g(\mbB)$   for all $\mbA \in \mathcal{M}$ and $\mbB \in \mathcal{N}^\perp.$
Under \textbf{M1}, the norm $\|\mbA\|_1 = \sum_{j,k}|\mbA_{j,k}|$ is decomposable (with respect to the subspace pair defined in \textbf{M1}); under \textbf{M2}, the norm $\|\mbA\|_{1,2} = \sum_{j}(\sum_{k} \mbA_{j,k}^2 )^{1/2}$ is decomposable; and under \textbf{M3}, the norm $\|\mbA\|_* =\sum \varphi_j(\mbA)$ is decomposable. Further, define the norms $\|\mbA\|_{\infty} = \max_{j,k}|\mbA_{j,k}|$, $\|\mbA\| = \sigma_1(\mbA)$, and $\|\mbA\|_{\infty,2} = \max_j \|\mbA_{j,\cdot}\|_2$, where $\|\cdot\|_2$ denotes the Euclidean norm of a vector.

For random quantities $\mbu$ and $\mbv$, we write $\mbu \sim \mbv$ to mean that $\mbu$ and $\mbv$ have the same distribution.  Throughout, let $O(n)$ denote the set of $n \times n$  matrices $\mbO$ such that $\mbO^\top\mbO = \mbI_n$ and let $V_q(n)$, a Stiefel manifold, denote the set of $n \times q$ matrices $\mbS$ such that $\mbS^\top\mbS = \mbI_q$. Let $S^{n-1} = \left\{\mbu \in \mathbb{R}^n: \|\mbu\|_2 = 1\right\}.$
In the following subsections, we refer to random matrices as having the uniform distribution on $O(n)$ and $V_q(n)$. Following \citet{eaton}, by uniform distribution we mean the unique translation-invariant probability measure. For a thorough treatment of the unique translation-invariant probability measure (called Haar measure) on $O(n)$, see Chapter 1 of \citet{meckes2019random}. For additional details on the uniform distribution on $V_q(n)$, see \citet{camano2006statistics} or Chapter 7 of \citet{eaton}.
\subsection{Preliminary Lemmas}
To begin, we first provide a preliminary lemma which will be used throughout our proofs. 
\begin{lemma} \label{lemma:Orthogonality_UniformDist}
(i) \citep[Section 3.5]{mattila1999geometry} If $\mbO$ is a random matrix having the uniform distribution on $O(n)$, then for any fixed vector $\mba \in S^{n-1}$, $\mbO \mba$ has a uniform distribution on $S^{n-1}.$
(ii)
If $\mbS$ is a random matrix having the uniform distribution on $V_q(n)$, then for any fixed unit vector $\mbb \in S^{q-1}$, $\mbS \mbb$ has the uniform distribution on $S^{n-1}$.
\end{lemma}
The second part of Lemma \ref{lemma:Orthogonality_UniformDist}, (ii), follows almost immediately from (i). For example, since $\mbS \sim \mbO \mbP_q$ where $\mbO$ is uniformly distributed on $O(n)$ and $\mbP_q \in \mathbb{R}^{n \times q}$ is the first $q$ columns of $\mbI_n$ (see (IV.1) of \citet{lyubarskii2010uncertainty}), it follows that $\mbS \mbv \sim \mbO \mbP_q \mbv$, so that because because $\mbP_q \mbv \in S^{n-1}$, an application of (i) yields (ii). 

The next lemma follows immediately from the proof of Proposition 7.1 in \citet{eaton}. 
\begin{lemma}\label{lemma:eaton}
Suppose $\mbR \in \mathbb{R}^{n \times q}$ is a random matrix which has $q$ non-zero singular values almost surely and suppose $\mbR$ is left-spherical, i.e., for any $\mbO \in O(n)$, $\mbO\mbR \sim \mbR$. Let $(\mbU_\mbR, \mbD_\mbR, \mbV_\mbR) = {\rm svd}(\mbR)$. Then, the random matrix $\mbR(\mbR^\top\mbR)^{-1/2}  = \mbU_\mbR \mbV_\mbR^\top$ follows a uniform distribution on $V_q(n)$.
\end{lemma}
The next lemma is a well-known result about the subdifferential of the nuclear norm. A proof sketch can be found in Section \ref{sec:Proofs}.
\begin{lemma}\label{lemma:grad}
Assume \textbf{A1} is true. Then, the subdifferential of $\mbbeta \mapsto \|\mbY - \mbX\mbbeta\|_*$ at $\mbbeta_*$ is the singleton
$$ - \mbX^\top\mbU_\epsilon \mbV_\epsilon^\top = - \mbX^\top(\mbY - \mbX\mbbeta_*)\{(\mbY - \mbX\mbbeta_*)^\top(\mbY- \mbX\mbbeta_*)\}^{-1/2}.$$

\end{lemma}

\subsection{Proof of Theorem \ref{prop:main_prop} and Corollary \ref{thm:asymp}}
We now focus our attention on the proofs of Theorem \ref{prop:main_prop} and Corollary \ref{thm:asymp}.  We begin with a preliminary lemma. 
\begin{lemma}\label{lemma:cone}
Assume \textbf{A1} is true. Define the event $\mathcal{A}_c = \{\lambda \geq (c/\sqrt{n}) \tilde{g}(\mbX^\top\mbU_\epsilon \mbV_\epsilon^\top)\}$ for a fixed constant $c > 1$. Then, on $\mathcal{A}_c$, $\hat{\mbDelta} = \hat{\mbbeta}_g - \mbbeta_*$ belongs to the set 
$$\mathcal{C}_{g}(\mathcal{M},\mathcal{N},c) = \left\{ \mbDelta \in \mathbb{R}^{p \times q}: g(\mbDelta_{\mathcal{N}^\perp}) \leq \check{c} \hspace{1pt} g(\mbDelta_{\mathcal{N}}) \right\}.$$
\end{lemma}
We omit the proof of Lemma \ref{lemma:cone} as it follows directly from the proof of Lemma 1 from \citet{negahban2012unified} using the fact that under \textbf{A1}, the gradient of nuclear norm of residuals with respect to $\mbbeta$ evaluated at $\mbbeta_*$ is $-\mbX^\top\mbU_\epsilon \mbV_\epsilon^\top$ (e.g., see Lemma \ref{lemma:grad}). Note that in the main text, we exclude the singleton $\mbDelta = 0$ from $\mathcal{C}_g(\mathcal{M}, \mathcal{N}, c)$. 

Next, we give a lower bound on the difference between the nuclear norm of residuals evaluated at $\mbbeta_* + \mbDelta$ and $\mbbeta_*$ for $\mbDelta \in \mathcal{C}_g(\mathcal{M},\mathcal{N}, c)$. The proof can be found in Section \ref{sec:Proofs}, but this follows straightforwardly from the convexity of the nuclear norm of residuals and definition of $\phi.$
\begin{lemma}\label{RSClemma}
Assume \textbf{A1} is true. Then, for all $\mbDelta \in \mathcal{C}_g(\mathcal{M}, \mathcal{N}, c)$, 
$$ \frac{1}{\sqrt{n}}\|\mbY - \mbX(\mbbeta_* + \mbDelta)\|_* - \frac{1}{\sqrt{n}}\|\mbY - \mbX\mbbeta_*\|_* \geq \phi \|\mbDelta\|_F^2 - \frac{1}{\sqrt{n}}|{\rm tr}( \mbDelta^\top\mbX^\top\mbU_\epsilon \mbV_\epsilon^\top)|.$$
\end{lemma}

We are now ready to prove Theorem \ref{prop:main_prop}. 
\smallskip

\noindent \textbf{Proof of Theorem \ref{prop:main_prop}.}
To prove Theorem \ref{prop:main_prop}, we follow the proof technique detailed in \citet{negahban2012unified}. For $\delta > 0$, define the set $\mathsf{B}_{\delta, c} = \left\{ \mbDelta \in \mathbb{R}^{p \times q}: \|\mbDelta\|_F = \delta, g(\mbDelta_{\mathcal{N}^\perp}) \leq  \check{c} \hspace{1pt} g(\mbDelta_{\mathcal{N}}) \right\}$ and let $\mathcal{L}(\mbbeta) = \|\mbY - \mbX \mbbeta\|_*/\sqrt{n} + \lambda g(\mbbeta)$. Because $\mathcal{L}$ is convex and $\hat{\mbbeta}_g$ is its minimizer, on $\mathcal{A}_c$,  $$\inf_{\mbDelta  \in \mathsf{B}_{\delta, c}} \{\mathcal{L}(\mbbeta_* + \mbDelta) - \mathcal{L}(\mbbeta_*) \} > 0 \implies \|\hat{\mbbeta}_g - \mbbeta_*\|_F \leq \delta.$$
For a proof of this fact, see the proof of Lemma 4 of the Supplementary Material to \citet{negahban2012unified}.  To simplify notation, let $D(\mbDelta) = \mathcal{L}(\mbbeta_*  + \mbDelta) - \mathcal{L}(\mbbeta_*)$ so that we need only show that $\inf_{\Delta \in\mathsf{B}_{\delta, c}} D(\mbDelta) > 0$ for $\delta = \lambda \check{c} \Psi_g(\mathcal{N})/\phi$. Notice first that
\begin{align}
D(\mbDelta) & =\frac{1}{\sqrt{n}}\|\mbY - \mbX(\mbbeta_* + \mbDelta)\|_* - \frac{1}{\sqrt{n}}\|\mbY - \mbX\mbbeta_*\|_* + \lambda g(\mbbeta_{*} + \mbDelta) - \lambda g(\mbbeta_{*})\notag \\
& \geq \phi \|\mbDelta\|_F^2 - \frac{1}{\sqrt{n}}|{\rm tr}(\mbDelta^\top\mbX^\top\mbU_\epsilon \mbV_\epsilon^\top)|+ \lambda g(\mbbeta_{*} + \mbDelta) - \lambda g(\mbbeta_{*})\label{eq:inequality1_Prop3}\\ 
& \geq \phi  \|\mbDelta\|_F^2 - \frac{1}{\sqrt{n}}|{\rm tr}(\mbDelta^\top\mbX^\top\mbU_\epsilon \mbV_\epsilon^\top)|+ \lambda g(\mbDelta_{\mathcal{N}^\perp}) -  \lambda g(\mbDelta_{\mathcal{N}})\label{eq:inequality2_Prop3}\\ 
 \intertext{where \eqref{eq:inequality1_Prop3} follows from the fact that $\mbDelta \in \mathsf{B}_{\delta, c}$ implies $\mbDelta \in \mathcal{C}_g(\mathcal{M}, \mathcal{N},c)$ and Lemma \ref{RSClemma}; and \eqref{eq:inequality2_Prop3} follows from the triangle inequality
 $$ g(\mbbeta_{*} + \mbDelta) -  g(\mbbeta_{*}) = g(\mbbeta_{*\mathcal{M}} + \mbDelta_{\mathcal{N}} + \mbDelta_{\mathcal{N}^\perp}) -g(\mbbeta_{*\mathcal{M}}) \geq g(\mbbeta_{*\mathcal{M}} + \mbDelta_{\mathcal{N}^\perp}) - g(\mbDelta_{\mathcal{N}}) -g(\mbbeta_{*\mathcal{M}}),$$
 and decomposability of the penalty function $g$ with respect to the pair $(\mathcal{M}, \mathcal{N}^\perp)$
 $$ g(\mbbeta_{*\mathcal{M}} + \mbDelta_{\mathcal{N}^\perp}) - g(\mbDelta_{\mathcal{N}}) -g(\mbbeta_{*\mathcal{M}}) = g(\mbbeta_{*\mathcal{M}}) + g(\mbDelta_{\mathcal{N}^\perp}) - g(\mbDelta_{\mathcal{N}}) -g(\mbbeta_{*\mathcal{M}}) = g(\mbDelta_{\mathcal{N}^\perp}) - g(\mbDelta_{\mathcal{N}}).$$ 
 Thus, applying H\"older's inequality to the second term in \eqref{eq:inequality2_Prop3}, we have }
 D(\mbDelta) & \geq \phi  \|\mbDelta\|_F^2 - \frac{1}{\sqrt{n}}g(\mbDelta) \tilde{g}(\mbX^\top\mbU_\epsilon \mbV_\epsilon^\top)+ \lambda  g(\mbDelta_{\mathcal{N}^\perp}) - \lambda g(\mbDelta_{\mathcal{N}}).\notag\\
\intertext{It then follows that on event $\mathcal{A}_c = \{\lambda \geq (c /\sqrt{n})\tilde{g}(\mbX^\top\mbU_\epsilon \mbV_\epsilon^\top)\}$,}
 D(\mbDelta)  & \geq \phi  \|\mbDelta\|_F^2 - \frac{\lambda}{c}g(\mbDelta) + \lambda  g(\mbDelta_{\mathcal{N}^\perp})  - \lambda g(\mbDelta_{\mathcal{N}}),\notag \\
& \geq \phi  \|\mbDelta\|_F^2 - \frac{\lambda}{c}g(\mbDelta_{\mathcal{N}}) - \frac{\lambda}{c}g(\mbDelta_{\mathcal{N}^\perp})+ \lambda  g(\mbDelta_{\mathcal{N}^\perp})  - \lambda g(\mbDelta_{\mathcal{N}}),\label{eq:tri_inequailty} \\
 &= \phi  \|\mbDelta\|_F^2 - \lambda \left(\frac{c + 1}{c}\right)g(\mbDelta_{\mathcal{N}}) + \lambda \left(\frac{c-1}{c}\right)g(\mbDelta_{\mathcal{N}^\perp}),\notag
\end{align}
where \eqref{eq:tri_inequailty} follows from the triangle equality $g(\mbDelta) \leq g(\mbDelta_{\mathcal{N}}) +g(\mbDelta_{\mathcal{N}^\perp})$.
Then, because $$ g(\mbDelta_{\mathcal{N}}) \leq \Psi_g(\mathcal{N})\|\mbDelta\|_F~~~~~\text{and}~~~~~\lambda\left(\frac{c-1}{c}\right)g(\mbDelta_{\mathcal{N}^\perp}) \geq 0,$$
it follows that 
$$D(\mbDelta) \geq \phi \|\mbDelta\|_F^2 - \lambda\left(\frac{c+1}{c}\right) \Psi_g(\mathcal{N}) \|\mbDelta\|_F.$$
Finally, since $\|\mbDelta\|_F = \delta$ for $\mbDelta \in \mathsf{B}_{\delta,c}$, setting $\lambda = \phi\delta/\{\check{c}\Psi_g(\mathcal{N})\}$, or equivalently $\delta = \check{c}\lambda \Psi_g(\mathcal{N})/\phi$, yields
$$D(\mbDelta) \geq \phi \delta^2 \left\{ 1 -  \lambda \left(\frac{c+1}{c}\right)\frac{\Psi_g(\mathcal{N})}{\phi \delta}\right\} =  \phi\delta^2 \left(1 - \frac{c-1}{c}\right) > 0, \quad \quad $$
from which the first conclusion follows since \textbf{A3} implies $\phi > 0$. 
The second claim follows immediately from Lemma \ref{lemma:eaton} under assumption \textbf{A2}.$\quad\quad \blacksquare$

We now turn our attention to the proof of Corollary \ref{thm:asymp}. By the result of Theorem \ref{prop:main_prop}, we need only select a $\lambda$ such $\mathcal{A}_c$ occurs with high probability.  To do so, we will need the following three concentration inequalities: proofs can be found in Section \ref{sec:Proofs}.

\begin{lemma}\label{lemma:max_norm_concentration}
Suppose $\mbS$ is a random matrix having the uniform distribution on $V_q(n)$. Let $k > 1$ be a fixed constant. If \textbf{C1}, \textbf{A1}, and \textbf{A2} hold, then 
$$P\left( \frac{1}{\sqrt{n}}\|\mbX^\top\mbS\|_{\infty} \geq \sqrt{\frac{2 \log(2pq^{k})}{n-1}} \right) \leq q^{1-k}$$
as long as $n > 2\log (2pq^k) + 1.$
Hence, under \textbf{M1}, setting $\lambda = c \{2\log(2pq^{k})/(n-1)\}^{1/2}$, 
$$ P(\mathcal{A}_c) \geq 1 - q^{1 - k}.$$
\end{lemma}

\begin{lemma}\label{lemma:max_2norm_concentration}
Suppose $\mbS$ is a random matrix having the uniform distribution on $V_q(n)$. Let $k > 1$ be a fixed constant such that $k \log p > 4\pi$. If \textbf{C1}, \textbf{A1}, and \textbf{A2} hold, then 
$$P\left( \frac{1}{\sqrt{n}}\|\mbX^\top\mbS\|_{\infty, 2} \geq \sqrt{\frac{4 k  \log p}{n-2}} + \sqrt{\frac{q}{n}}\right) \leq  p^{1 - k}.$$
Hence, under \textbf{M2}, setting $\lambda = c\{4 k\log p/(n-2)\}^{1/2} + c(q/n)^{1/2}$, 
$$ P(\mathcal{A}_c) \geq 1 - p^{1 - k}.$$
\end{lemma}
\begin{lemma}\label{lemma:spectral_norm_concentration}
Suppose $\mbS$ is a random matrix having the uniform distribution on $V_q(n)$. Let $k_1>1$ be a fixed constant such that $k_2 = 4\log(7 + k_1)$ and $ k_2 \|\mbX\|^2(p + q) > 16 n \pi$. If \textbf{C1}, \textbf{A1}, and \textbf{A2} hold, then 
$$P\left\{ \frac{1}{\sqrt{n}}\|\mbX^\top\mbS\| \geq \frac{4\|\mbX\|}{\sqrt{n}}\left(\sqrt{\frac{k_2(p + q)}{n-2}} + \sqrt{\frac{1}{n}}\right) \right\} \leq  \left(\frac{8}{7 + k_1}\right)^{p + q}.$$
Hence, under \textbf{M3}, setting $ \lambda = 4 c \|\mbX\| [k_2(p + q)/\{n(n-2)\}]^{1/2} +  4 c \|\mbX\|/n$, 
$$ P(\mathcal{A}_c) \geq 1 - \left(\frac{8}{7 + k_1}\right)^{p + q}.$$
\end{lemma}

\noindent \textbf{Proof of Corollary \ref{thm:asymp}.}
Under the conditions of Corollary \ref{thm:asymp}, \textbf{C1}, \textbf{A1}--\textbf{A3} hold so that we can apply the result of Theorem \ref{prop:main_prop}. 
\begin{itemize}
\item[] $(i)$ Under \textbf{M1} with $g(\cdot) = \|\cdot\|_1$, $\tilde{g}(\mbX^\top\mbU_\epsilon \mbV_\epsilon^\top) = \|\mbX^\top\mbU_\epsilon \mbV_\epsilon^\top\|_{\infty}$ and $\Psi_{\|\cdot\|_1}(\mathcal{N}) \leq \sqrt{|\mathcal{S}|}$ \citep{negahban2012unified}. If $\lambda = c \{ 2 \log(2 pq^k)/(n-1)\}^{1/2}$ and $n > 2 \log (2pq^k) + 1$ for fixed constants $c > 1$ and $k > 1$, an application of Lemma \ref{lemma:max_norm_concentration} implies $P(\mathcal{A}_c) \geq 1 - q^{1 - k}$. Hence, applying Theorem \ref{prop:main_prop}, it follows that
$$P\left( \|\hat\mbbeta_{\rm L} - \mbbeta_*\|_F  \leq \frac{\tilde{c}}{\phi} \sqrt{ \frac{2|\mathcal{S}| \log (2p q^k)}{(n-1)}} \right) \geq 1 - q^{1 - k}.$$
\item[]$(ii)$ Under \textbf{M2} with $g(\cdot) = \|\cdot\|_{1,2}$, $\tilde{g}(\mbX^\top\mbU_\epsilon \mbV_\epsilon^\top) = \|\mbX^\top\mbU_\epsilon \mbV_\epsilon^\top\|_{\infty, 2}$ and $\Psi_{\|\cdot\|_{1,2}}(\mathcal{N}) \leq \sqrt{|\mathcal{G}|}$ \citep{liu2015calibrated}. If $\lambda = c \{ 4 k \log p/(n-2)\}^{1/2} + c (q/n)^{1/2}$ for constants $c> 1$ and $k > 1$ such that $k \log p > 4\pi$, Lemma \ref{lemma:max_2norm_concentration} implies $P(\mathcal{A}_c) \geq 1 - p^{1 - k}$. Hence, applying Theorem \ref{prop:main_prop}, it follows that
$$P\left\{ \|\hat\mbbeta_{\rm GL} - \mbbeta_*\|_F  \leq \frac{2 \tilde{c}}{\phi} \left(\sqrt{\frac{ k|\mathcal{G}| \log p}{n-2}} + \sqrt{\frac{|\mathcal{G}|q}{4n}}\right)\right\}\geq 1 - p^{1 - k}.$$
\item[] $(iii)$ Under \textbf{M3} with $g(\cdot) = \|\cdot\|_*$, $\tilde{g}(\mbX^\top\mbU_\epsilon \mbV_\epsilon^\top) = \|\mbX^\top\mbU_\epsilon \mbV_\epsilon^\top\|$ and $\Psi_{\|\cdot\|_*}(\mathcal{N}) \leq  \sqrt{2r}$ \citep{negahban2011}. With $\lambda = 4 c \|\mbX\|[k_2(p + q)/\{n(n-2)\}]^{1/2} + 4 c \|\mbX\|/n$ for constants $c > 1$ and $k_1 > 1$ such that $k_2 = 4\log(7 + k_1)$ and $k_2\|\mbX\|^2 (p+q) > 16 n\pi$, Lemma \ref{lemma:spectral_norm_concentration} implies $P(\mathcal{A}_c) \geq 1 - \{8/(7+k_1)\}^{p+q}$. Hence,  applying Theorem \ref{prop:main_prop}, it follows that
$$P\left\{ \|\hat\mbbeta_{\rm LR} - \mbbeta_*\|_F  \leq \frac{4\tilde{c}}{\phi }\left(\frac{\|\mbX\|}{\sqrt{n}}\right)\left(\sqrt{\frac{2 k_2 r (p + q)}{n}} + \sqrt{\frac{2r}{n}}\right)\right\}\geq 1 - \left(\frac{8}{7+ k_1}\right)^{p+q}.~~\blacksquare$$ 
\end{itemize}

\subsection{Proof of Theorem \ref{main_normal} and Corollary \ref{cor6}}
We now focus on the proofs of Theorem \ref{main_normal} and Corollary \ref{cor6}. To prove Theorem \ref{main_normal}, we require the following lemma, which we prove in Section \ref{sec:Proofs}.

\begin{lemma}\label{eq:RSCII}
Under \textbf{C1} and \textbf{A4}--\textbf{A6}, if $n$ is sufficiently large, on the event $\mathcal{A}_c \cap \mathcal{B}_d$ 
$$ \frac{1}{\sqrt{n}} \|\mbE - \mbX \mbDelta\|_* - \frac{1}{\sqrt{n}}\|\mbE\|_* + \frac{1}{\sqrt{n}}{\rm tr}(\mbDelta^\top\mbX^\top\mbU_\epsilon \mbV_\epsilon^\top) \geq \frac{\nu}{4(t + d) \hspace{2pt} \varphi_1^{1/2}(\boldsymbol{\Sigma}_*)}\|\mbDelta\|_F^2,$$
provided $\|\mbDelta\|_F^2 \to 0$ as $n \to \infty.$
\end{lemma}

\noindent \textbf{Proof of Theorem \ref{main_normal}.}
By the same arguments used to prove Theorem \ref{prop:main_prop},  for sequence $\delta_n \to 0$ as $n \to \infty$, defining $\mathsf{B}_{\delta_n, c} = \{\mbDelta \in \mathbb{R}^{p \times q}:\|\mbDelta\|_F = \delta_n, g(\mbDelta_{\mathcal{N}^\perp}) \leq \check{c} \hspace{2pt} g(\mbDelta_{\mathcal{N}})\}$, we have that on $\mathcal{A}_c \cap \mathcal{B}_d,$ $$D(\mbDelta)  > 0 \text{ for all } \Delta \in \mathsf{B}_{\delta_n, c} \implies \|\hat\mbbeta_g - \mbbeta_*\|_F \leq \delta_n.$$
Recall that 
$$ D(\mbDelta)  = \underbrace{\frac{1}{\sqrt{n}}\|\mbY - \mbX\mbbeta_* - \mbX\mbDelta\|_* - \frac{1}{\sqrt{n}}\|\mbY - \mbX\mbbeta_*\|_*}_{T_1} + \underbrace{\lambda g(\mbbeta_{*} + \mbDelta) - \lambda g(\mbbeta_{*})}_{T_2}.$$
For $n$ sufficiently large, on $\mathcal{A}_c \cap \mathcal{B}_d$, we can bound $T_1$ using Lemma \ref{eq:RSCII} and H\"older's inequality, i.e., $|{\rm tr}(\mbDelta^\top\mbX^\top\mbU_\epsilon \mbV_\epsilon^\top)| \leq g(\mbDelta) \tilde{g}(\mbX^\top\mbU_\epsilon \mbV_\epsilon^\top)$, to see
$$ T_1 \geq \frac{\nu}{4(t + d) \varphi_1^{1/2}(\mbSigma_*)}\|\mbDelta\|_F^2 - \frac{1}{\sqrt{n}}g(\mbDelta) \tilde{g}(\mbX^\top\mbU_\epsilon \mbV_\epsilon^\top).$$ 
Similarly, we can bound $T_2$ by the same arguments as those used to obtain \eqref{eq:inequality2_Prop3}. Hence, 
\begin{align*}
 D(\mbDelta)
& \geq\frac{\nu}{4(t + d) \varphi_1^{1/2}(\mbSigma_*)} \|\mbDelta\|_F^2 - \frac{1}{\sqrt{n}}g(\mbDelta)\tilde{g}(\mbX^\top\mbU_\epsilon \mbV_\epsilon^\top) + \lambda g(\mbDelta_{\mathcal{N}^\perp}) - \lambda g(\mbDelta_{\mathcal{N}}),\notag
\intertext{so that on $\mathcal{A}_c = \{\lambda \geq (c/\sqrt{n}) \tilde{g}(\mbX^\top\mbU_\epsilon \mbV_\epsilon^\top)\}$, an application of the triangle inequality $g(\mbDelta) \leq  g(\mbDelta_{\mathcal{N}^\perp})+ g(\mbDelta_{\mathcal{N}})$ and the fact that $\lambda (c-1) g(\mbDelta_{\mathcal{N}^\perp})/c \geq 0$ yields 
$$D(\mbDelta)
\geq\frac{\nu}{4(t + d) \varphi_1^{1/2}(\mbSigma_*)} \|\mbDelta\|_F^2 - \lambda \left(\frac{c+1}{c}\right)g(\mbDelta_{\mathcal{N}}).$$
By definition of $\Psi_g(\mathcal{N})$, this implies 
\begin{align*}
D(\mbDelta) & \geq \frac{\nu}{4(t + d) \varphi_1^{1/2}(\mbSigma_*)} \|\mbDelta\|_F^2 -\lambda \left(\frac{c+1}{c}\right)\Psi_g(\mathcal{N})\|\mbDelta\|_F
\end{align*}
so that finally, since $\|\mbDelta\|_F = \delta_n$ for $\mbDelta \in \mathsf{B}_{\delta_n, c}$, we have 
$$ D(\mbDelta) \geq \nu\delta_n^2 \left(\frac{1}{4(t + d) \varphi_1^{1/2}(\mbSigma_*)} -  \lambda \left(\frac{c+1}{c}\right)\frac{\Psi_g(\mathcal{N})}{\nu \delta_n}\right)$$
which is positive if $\lambda = \delta_n \nu/\{\check{c}\Psi_g(\mathcal{N})4(t + d) \varphi_1^{1/2}(\mbSigma_*)\}$, or equivalently, $\delta_n = 4\lambda \check{c}\Psi_g(\mathcal{N})(t + d) \varphi_1^{1/2}(\mbSigma_*)/\nu$. Hence, because $\delta_n \asymp \lambda\Psi_g(\mathcal{N})$ under \textbf{A4}--\textbf{A6}, the result follows by requiring that $\lambda\Psi_g(\mathcal{N}) \to 0$ as $n \to \infty.~~\blacksquare$}
\end{align*}
Finally, to prove Corollary \ref{cor6}, in addition to Lemmas \ref{lemma:max_norm_concentration}--\ref{lemma:spectral_norm_concentration}, we need the following concentration inequality.

\begin{lemma}\label{lemma:E_singular}
Let $\mbE \in \mathbb{R}^{n \times q}$ be a matrix with rows independent and identically distributed from ${\rm N}_q(0, \boldsymbol{\Sigma}_*)$. Let $\varphi_1(\boldsymbol{\Sigma}_*)$ denote the the largest eigenvalue of $\boldsymbol{\Sigma}_*$, and let $\varphi_1^{1/2}(\boldsymbol{\Sigma}_*)$ be its square-root. Given fixed constant $d > 1$, if $q/n \to t \in (0,1)$ as $n \to \infty$, then for $n$ sufficiently large
$$P\left( \frac{1}{\sqrt{n}} \|\mbE\| \leq (t + d) \varphi^{1/2}_1(\boldsymbol{\Sigma}_*)\right) \geq 1 -  2 e^{-(d-1)^2n/4}.$$
\end{lemma}
The proof of Lemma \ref{lemma:E_singular} can be found in Section \ref{sec:Proofs}. With Lemma \ref{lemma:E_singular} in hand,  we can use Theorem \ref{main_normal} and Lemmas \ref{lemma:max_norm_concentration}--\ref{lemma:spectral_norm_concentration} to prove Corollary \ref{cor6}. \\

\noindent \textbf{Proof of Corollary \ref{cor6}.} Under the conditions of Theorem \ref{main_normal}, \textbf{C1} and \textbf{A4}--\textbf{A6} hold so that we can apply the result of Theorem \ref{main_normal}. 
\begin{itemize}
       \item[]$(i)$ Under \textbf{M1} with $g(\cdot) = \|\cdot\|_1$, $\tilde{g}(\mbX^\top\mbU_\epsilon \mbV_\epsilon^\top) = \|\mbX^\top\mbU_\epsilon \mbV_\epsilon^\top\|_{\infty}$ and $\Psi_{\|\cdot\|_1}(\mathcal{N}) \leq \sqrt{|\mathcal{S}|}$. If $n > 2 \log (2pq^k) + 1$ and $\lambda = c\{2 \log(2 pq^k)/(n-1)\}^{1/2}$ for fixed constants $c > 1$, $k > 1$, and $d > 1$, then applications of Lemma \ref{lemma:max_norm_concentration} and \ref{lemma:E_singular} imply that for $n$ sufficiently large,
       $P(\mathcal{A}_c \cap \mathcal{B}_d) \geq 1 - q^{1 - k} - 2 e^{-(d-1)^2 n/4}.$ Therefore, 
    Theorem \ref{main_normal} implies that under the same conditions,
$$P\left\{ \|\hat\mbbeta_{\rm L} - \mbbeta_*\|_F  \leq \varphi_1^{1/2}(\boldsymbol{\Sigma}_*)\left(\frac{4(t + d)\tilde{c}}{\nu}\right) \sqrt{ \frac{2  c_1  |\mathcal{S}| \log (2pq^k)}{(n-1)}} \right\}$$
with probability at least $1 - q^{1 - k}- 2 e^{-(d-1)^2n/4}$ as long as $ \lambda\sqrt{|\mathcal{S}|} \to 0$ as $n \to \infty$.

\item[]$(ii)$ Under \textbf{M2} with $g(\cdot) = \|\cdot\|_{1,2}$, $\tilde{g}(\mbX^\top \mbU_\epsilon \mbV_\epsilon^\top) = \|\mbX^\top \mbU_\epsilon \mbV_\epsilon^\top\|_{\infty, 2}$ and $\Psi_{\|\cdot\|_{1,2}}(\mathcal{N}) \leq \sqrt{|\mathcal{G}|}$. If $\lambda = c \{4 k \log p/(n-2)\}^{1/2} + c (q/n)^{1/2}$ and $k \log p > 4\pi$ for fixed constants $c > 1$, $k > 1$, and $d > 1$, then applications of Lemma \ref{lemma:max_2norm_concentration} and \ref{lemma:E_singular} imply that for $n$ sufficiently large,
 $ P(\mathcal{A}_c \cap \mathcal{B}_d) \geq 1 - p^{1-k} - 2 e^{-(d-1)^2n/4}.$ Therefore, 
    Theorem \ref{main_normal} implies that under the same conditions,
$$ \|\hat\mbbeta_{\rm GL} - \mbbeta_*\|_F  \leq \varphi_1^{1/2}(\boldsymbol{\Sigma}_*) \left(\frac{8(t + d)\tilde{c}}{\nu}\right) \left(\sqrt{\frac{k |\mathcal{G}|  \log p}{n-2}} + \sqrt{\frac{|\mathcal{G}|q}{4n}}\right)$$
with probability as least $1 - p^{1 - c_2} - 2 e^{-(d-1)^2 n/4}$  as long as $\lambda \sqrt{|\mathcal{G}|} \to 0$ as $n \to \infty.$

\item[]$(iii)$ Under \textbf{M3} with $g(\cdot) = \|\cdot\|_*$, $\tilde{g}(\mbX^\top \mbU_\epsilon \mbV_\epsilon^\top) = \|\mbX^\top\mbU_\epsilon \mbV_\epsilon^\top\|$ and $\Psi_{\|\cdot\|_*}(\mathcal{N}) \leq \sqrt{2r}$. If  $\lambda = 4 c \|\mbX\|[k_2 (p + q)/\{n(n-2)\}\}]^{1/2} + 4 c \|\mbX\|/n$  for fixed constants $c>1$, $k_1 > 1$, and $d > 1$ such that $k_2 = 4 \log(7 + k_1)$ and $k_2\|\mbX\|^2  (p + q) > 16 n\pi$, then applications of Lemma \ref{lemma:spectral_norm_concentration} and Lemma \ref{lemma:E_singular} imply that for $n$ sufficiently large, $P(\mathcal{A}_c \cap \mathcal{B}_d) \geq 1 - \{8/(7+k_1)\}^{p+q} - 2 e^{-(d-1)^2n/4}$. Therefore, 
    Theorem \ref{main_normal} implies that under the same conditions,
\begin{align*}
\|\hat\mbbeta_{\rm LR} - \mbbeta_*\|_F  \leq \varphi_1^{1/2}(\boldsymbol{\Sigma}_*)\left(\frac{16(t + d)\tilde{c} \|\mbX\| }{\nu \sqrt{n}}\right)\left(\sqrt{\frac{2k_2 r(p + q)}{n-2}} + \sqrt{\frac{2r}{n}}\right)
\end{align*}
with probability at least $1 - \{8/(7+k_1)\}^{p+q} - 2e^{-(d-1)^2 n/4}$ as long as $\lambda \sqrt{r} \to 0$ as $n \to \infty.~~~~ \blacksquare$
\end{itemize}

In our statement of Corollary \ref{cor6}(i), for example, we use that under $\textbf{C1}$ and \textbf{A4}--\textbf{A6}, $\lambda \sqrt{|\mathcal{S}|} \to 0$ as $n \to \infty$ is implied by $|\mathcal{S}|\log(pq^k) = o(n).$

\subsection{Proofs of Lemmas}\label{sec:Proofs}
\noindent \textbf{Proof of Lemma \ref{lemma:grad}.} We proceed with the following steps: first, we derive the subdifferential of $\mbbeta \mapsto \|\mbY - \mbX\mbbeta\|_*$, then we show that this set is a singleton at $\mbbeta$ such that $\mbY - \mbX\mbbeta$ has $q$ non-zero singular values. Let $\partial f(\mbA)$ denote the subdifferential of a function $f$ at $\mbA$.

To establish the subdifferential, we first apply the chain rule for subdifferentials:
\begin{equation}\label{eq:partial1}
\partial \|\mbY - \mbX\mbbeta\|_* = \left\{ - \mbX^\top\mbH: \mbH \in \partial\|\mbB\|_*\mid_{\mbB = \mbY - \mbX\mbbeta} \right\}.
\end{equation} 
From \citet{watson1992characterization}, letting $(\mbU_B,\mbD_B, \mbV_B) = {\rm svd}(\mbB)$, we have 
\begin{equation}\label{eq:partial2}
 \partial \|\mbB\|_* = \left\{\mbU_B\mbV_B^\top + \mbW_B: \mbW_B \in \mathbb{R}^{p \times q}, \|\mbW_B\|\leq 1, \mbU_B^\top\mbW_B = 0, \mbW_B \mbV_B = 0 \right\}.
 \end{equation}
Hence, combining \eqref{eq:partial1} and \eqref{eq:partial2}, with $(\mbU, \mbD, \mbV) = {\rm svd}(\mbY - \mbX \mbbeta),$
$$ \partial \|\mbY - \mbX\mbbeta\|_*  = \left\{ -\mbX^\top\mbU \mbV^\top - \mbX^\top\mbW:  \|\mbW\|\leq 1, \mbU^\top\mbW = 0, \mbW \mbV  = 0\right\}.$$
However, when $\mbY - \mbX\mbbeta$ has $q$ non-zero singular values, the only such $\mbW$ which can satisfy both $\mbU^\top\mbW = 0$ and $\mbW \mbV = 0$ is $\mbW = 0$. Thus, in this case, the subdifferential is the singleton $- \mbX^\top\mbU \mbV. ~~\blacksquare$

\bigskip

\noindent \textbf{Proof of Lemma \ref{RSClemma}.} First, recall that the nuclear norm can be expressed 
$$ \|\mbA\|_* = \sup_{\|\mbQ\| \leq 1} {\rm tr}(\mbQ^\top\mbA).$$
By Lemma \ref{lemma:grad}, under \textbf{A1} it suffices to show that 
\begin{equation}\label{eq:RSC_I}
\frac{1}{\sqrt{n}}\|\mbY - \mbX\mbbeta_* - \mbX\mbDelta\|_* - \frac{1}{\sqrt{n}}\|\mbY - \mbX\mbbeta_*\|_*  + \frac{1}{\sqrt{n}}{\rm tr}(\mbDelta^\top\mbX \mbU_\epsilon \mbV_\epsilon^\top)  \geq \phi \|\mbDelta\|_F^2.
\end{equation}
Clearly,
\begin{align*}
\frac{1}{\sqrt{n}}\|\mbY - \mbX\mbbeta_* &- \mbX\mbDelta\|_* - \frac{1}{\sqrt{n}}\|\mbY - \mbX\mbbeta_*\|_*  + \frac{1}{\sqrt{n}}{\rm tr}(\mbDelta^\top\mbX^\top \mbU_\epsilon \mbV_\epsilon^\top)\\
&=  \frac{1}{\sqrt{n}}\|\mbE - \mbX\mbDelta\|_* - \frac{1}{\sqrt{n}}\|\mbE\|_*  + \frac{1}{\sqrt{n}}{\rm tr}(\mbDelta^\top\mbX^\top \mbU_\epsilon \mbV_\epsilon^\top)  \\
&=  \frac{1}{\sqrt{n}}\|\mbU_\epsilon \mbD_\epsilon \mbV_\epsilon^\top - \mbX\mbDelta\|_* - \frac{1}{\sqrt{n}}\|\mbE\|_*  + \frac{1}{\sqrt{n}}{\rm tr}(\mbDelta^\top\mbX^\top \mbU_\epsilon \mbV_\epsilon^\top)  \\
&= \left[ \sup_{\|\mbQ\| \leq 1} \frac{1}{\sqrt{n}} {\rm tr}\left\{ \mbQ^\top(\mbU_\epsilon \mbD_\epsilon \mbV_\epsilon^\top - \mbX\mbDelta) \right\} \right] - \frac{1}{\sqrt{n}}\|\mbE\|_*  + \frac{1}{\sqrt{n}}{\rm tr}(\mbDelta^\top\mbX^\top \mbU_\epsilon \mbV_\epsilon^\top) 
\intertext{and because $\|\mbE\|_* = {\rm tr}(\mbD_\epsilon) = {\rm tr}(\mbU_\epsilon \mbV_\epsilon^\top \mbV_\epsilon \mbD_\epsilon \mbU_\epsilon^\top)$, the previous equality can be expressed}
& = \sup_{\|\mbQ\| \leq 1} \frac{1}{\sqrt{n}} {\rm tr}\left\{ (\mbQ - \mbU_\epsilon \mbV_\epsilon^\top)^\top(\underbrace{\mbU_\epsilon \mbD_\epsilon \mbV_\epsilon^\top}_{= \mbE} - \mbX\mbDelta) \right\}.
\end{align*}
Thus, the desired inequality must hold by the definition of $\phi$.  $~~ \blacksquare$
\bigskip

\noindent \textbf{Proof of Lemma \ref{lemma:max_norm_concentration}.} For any $\delta \geq 0$, by the union bound we have 
\begin{align}
P\left(\frac{1}{\sqrt{n}}\|\mbX^\top\mbS\|_{\infty} \geq 
\delta \right) & \leq  \sum_{l=1}^q P\left(\frac{1}{\sqrt{n}}\|\mbX^\top\mbS_{\cdot,l}\|_{\infty} \geq \delta\right) \notag
\intertext{where $\mbS_{\cdot,l} \in S^{n-1}$ is the $l$th column of $\mbS \in V_q(n)$. Under the conditions of Lemma \ref{lemma:max_norm_concentration}, Lemma \ref{lemma:Orthogonality_UniformDist}(ii) suggests $\mbS_{\cdot,l}$ is uniformly distributed on $S^{n-1}$. Thus, we know that $\mbS_{\cdot,l} \sim \mbg/\|\mbg\|_2$, where $\mbg \sim {\rm N}_n(0, \mbI_n)$, so the above inequality implies}
P\left(\frac{1}{\sqrt{n}}\|\mbX^\top\mbS\|_{\infty} \geq 
\delta \right) & \leq q \hspace{2pt} P\left(\frac{\|\mbX^\top\mbg\|_{\rm \infty}}{\sqrt{n}\|\mbg\|_2} \geq \delta \right ). \label{eq:union8}
\intertext{It remains only to bound the right hand side of \eqref{eq:union8}.
 By an application of Lemma \ref{lemma:max_norm}, 
$$P\left(\frac{\|\mbX^\top\mbg\|_{\rm \infty}}{\sqrt{n}\|\mbg\|_2} \geq \sqrt{\frac{2\log (2p/\alpha)}{n-1}}\right) \leq \alpha,$$
so that setting $\alpha = q^{-k}$ for fixed constant $k > 1$, as long as $2\log(2pq^k) < n-1$, we have
$$P\left(\frac{1}{\sqrt{n}}\|\mbX^\top\mbS\|_{\infty} \geq 
\sqrt{\frac{2\log (2pq^{k})}{n-1}} \right) \leq q^{1 - k},$$
from which our conclusion follows.$~~\blacksquare$} 
\end{align}

\noindent \textbf{Proof of Lemma \ref{lemma:max_2norm_concentration}.}
Recall that under \textbf{C1}, $\|\mbX_{\cdot, j}\|_2 = \sqrt{n}$, so that each  $\mbX_{\cdot,j}/\sqrt{n}$ is an element of $S^{n-1}$. By the union bound
$$ P\left( \frac{1}{\sqrt{n}}\|\mbX^\top\mbS\|_{\infty, 2} \geq \delta \right) \leq \sum_{j=1}^p P\left(\frac{1}{\sqrt{n}}\left\Vert \mbS^\top\mbX_{\cdot, j}\right\Vert_{2} \geq \delta\right)$$
for all $\delta \geq 0$. 
Hence, we need to establish a concentration inequality for the random quantity $\|\mbS^\top\mba \|_{2}$ where $\mba \in S^{n-1}$ is a fixed unit vector and $\mbS$ is a random matrix uniformly distributed on $V_q(n)$. Applying Lemma \ref{lemma:conc_euclidean}, as long as $\alpha > 4\{\pi/(n-2)\}^{1/2},$
$$ P\left(\frac{1}{\sqrt{n}}\left\Vert \mbS^\top\mbX_{\cdot, j}\right\Vert_{2} \geq  \alpha + \sqrt{\frac{q}{n}} \right) \leq {\rm exp}\left(- \frac{(n - 2) \alpha^2}{4}\right). $$ 
Thus, setting $ \alpha = \{4 k \log p/(n-2)\}^{1/2}$ for constant $k> 4 \pi/\log p$, we have 
$$P\left(\frac{1}{\sqrt{n}}\left\Vert \mbS^\top\mbX_{\cdot, j}\right\Vert_{2} \geq \sqrt{\frac{4 k \log p}{n-2}} + \sqrt{\frac{q}{n}}\right) \leq  {\rm exp}\left(- k \log p\right).$$
 Therefore, with $\delta = \{4 k \log p/(n-2)\}^{1/2} + (q/n)^{1/2}$, we can conclude
$$P\left( \frac{1}{\sqrt{n}}\|\mbX^\top\mbS\|_{\infty, 2} \geq \sqrt{\frac{4 k \log p}{n-2}} + \sqrt{\frac{q}{n}}\right) \leq  p \hspace{2pt}{\rm exp}\left(- k \log p \right) =  p^{1 - k}.~~~\blacksquare$$
\bigskip

\noindent \textbf{Proof of Lemma \ref{lemma:spectral_norm_concentration}.}
By identical arguments used to derive (F.3) of the Supplementary Material to \citet{negahban2011}, we have that for $\delta \geq 0$
\begin{equation}\label{eq:main_NN} P\left(\frac{1}{\sqrt{n}}\|\mbX^\top\mbS\| \geq 4 \delta \right) \leq 8^{p + q} \max_{\mbv_a, \mbt_b} P\left(\frac{1}{\sqrt{n}}|\mbt_b^\top\mbX^\top\mbS \mbv_a| \geq \delta\right)
\end{equation}
where $\{\mbv_1, \mbv_2, \dots, \mbv_A\}$ and $\{\mbt_1, \mbt_2, \dots, \mbt_B\}$ are 1/4 coverings of $S^{q-1}$ and $S^{p-1}$, respectively. Thus, following \citet{negahban2011}, we need to bound the random scalar $|\mbt^\top\mbX^\top\mbS \mbv|/\sqrt{n}$ for arbitrary (but fixed) vectors $\mbt \in S^{p-1}$, $\mbv \in S^{q-1}.$ First, note that the random vector $\mbz = \mbS \mbv$ has a uniform distribution on $S^{n-1}$ (see Lemma \ref{lemma:Orthogonality_UniformDist}). Hence, we need only concern ourselves with $|\mbt^\top\mbX^\top\boldsymbol{z}|$ where $\boldsymbol{z}$ is uniformly distributed on $S^{n-1}$. Because ${\rm E} |\mbt^\top\mbX^\top\boldsymbol{z}| \leq \|\mbX \mbt\|_2/\sqrt{n}$ (e.g., see Lemma 5.3.2(a) of \citet{vershynin2018high} and apply Jensen's inequality), and because $\boldsymbol{z} \mapsto \|\mbt^\top\mbX^\top\boldsymbol{z}\|_2/\sqrt{n}$ is Lipschitz with constant $\|\mbX \mbt\|_2/\sqrt{n}$, applying Lemma \ref{lemma:Levy2} we have
$$ P\left(\frac{1}{\sqrt{n}} |\mbt^\top\mbX^\top\mbz| \geq \alpha + \frac{1}{n}\|\mbX \mbt\|_2\right) \leq  {\rm exp}\left(-\frac{ (n-2)n \alpha^2}{4 \|\mbX \mbt\|_2^2}\right),$$
as long as $\alpha > 4 \{\pi/(n-2)\}^{1/2}.$ Since $\|\mbX\mbt\|_2 \leq \|\mbX\|$ for all vectors $\mbt \in S^{p-1}$, the above inequality implies
$$ P\left(\frac{1}{\sqrt{n}} |\mbt^\top\mbX^\top\mbz| \geq \alpha + \frac{1}{n}\|\mbX\|\right) \leq  {\rm exp}\left(-\frac{ (n-2)n \alpha^2}{4 \|\mbX\|^2}\right).$$
By setting $\alpha = \|\mbX\|[k_2(p + q)/\{n(n-2)\}]^{1/2}$ for constant $k_2 > 0$, we have 
\begin{align} 
P\left\{ \frac{1}{\sqrt{n}} |\mbt^\top\mbX^\top\mbz| \geq \frac{\|\mbX\|}{\sqrt{n}} \left(\sqrt{\frac{k_2(p + q)}{n-2}} + \sqrt{\frac{1}{n}}\right)  \right\}
& \leq {\rm exp}\left(-\frac{k_2(p+q)}{4}\right). \label{eq:NN_2}
\end{align}
Hence, using \eqref{eq:main_NN} and \eqref{eq:NN_2},
\begin{align*}
P\left\{\frac{1}{\sqrt{n}}\|\mbX^\top\mbS\| \geq 4  \frac{\|\mbX\|}{\sqrt{n}} \right. & \left. \left(\sqrt{\frac{k_2(p + q)}{n-2}} + \sqrt{\frac{1}{n}}\right) \right\}  \\
& \leq 8^{p+q} P\left\{ \frac{1}{\sqrt{n}} |\mbt^\top\mbX^\top\mbz| \geq \frac{\|\mbX\|}{\sqrt{n}} \left(\sqrt{\frac{k_2(p + q)}{n-2}} + \sqrt{\frac{1}{n}}\right)  \right\} \\
& \leq 8^{p+q}  \left\{  {\rm exp}\left(-\frac{k_2(p+q)}{4}\right)\right\}.
\intertext{ The conclusion follows by taking $k_2 = 4\log(7 + k_1)$ for $k_1 > 1$ large enough that $k_2 \|\mbX\|^2 (p + q) > 16 n \pi,$ in which case  }
P\left\{\frac{1}{\sqrt{n}}\|\mbX^\top\mbS\| \geq 4  \frac{\|\mbX\|}{\sqrt{n}} \right. & \left. \left(\sqrt{\frac{k_2(p + q)}{n-2}} + \sqrt{\frac{1}{n}}\right) \right\}  \\
&\leq 8^{p+q}  \left\{  {\rm exp}\left(-\frac{k_2(p+q)}{4}\right)\right\} = \left(\frac{8}{7 + k_1}\right)^{p+q}.~~\blacksquare
\end{align*}

\noindent \textbf{Proof of Lemma \ref{eq:RSCII}}.
To simplify notation, let 
$$\mathcal{H}(\mbDelta) = \frac{1}{\sqrt{n}}\|\mbE - \mbX \mbDelta\|_* - \frac{1}{\sqrt{n}}\|\mbE\|_* + \frac{1}{\sqrt{n}}{\rm tr}(\mbDelta^\top\mbX^\top\mbU_\epsilon \mbV_\epsilon^\top).$$
First, letting $\mbU$ and $\mbV$ denote the left and right singular vectors of $\mbE$ (momentarily omitting the subscript $\epsilon$ for ease of display), Lemma \ref{RSClemma2} and \textbf{A6} imply that for $n$ sufficiently large (so that $n \geq q$), 
\begin{align*}
\mathcal{H}(\mbDelta) &= \frac{1}{2}\sum_{i=1}^q \sum_{j=1}^q \frac{(\mbu_j^\top \mbX \mbDelta \mbv_i - \mbu_i^\top\mbX \mbDelta \mbv_j)^2}{2 \sqrt{n} \{\sigma_i(\mbE) + \sigma_j(\mbE)\}} + \frac{1}{2} \sum_{k = q+1}^n \sum_{j = 1}^q  \frac{(\mbu_k^\top \mbX \mbDelta \mbv_j)^2}{ \sqrt{n}\sigma_j(\mbE)}
 + o\left(\frac{\|\mbX\mbDelta\|_F^2}{n}\right)
 \intertext{where $\mbu_j$ denotes the $j$th column of $\mbU \in \mathbb{R}^{n \times q}$ for $j\in[q]$, $\mbv_k$ denotes the $k$th column of $\mbV \in \mathbb{R}^{q \times q}$ for $k \in [q]$ and $\mbu_{l}$ denotes the $(l-q)$th column of $\mbU_0 \in \mathbb{R}^{n \times (n-q)}$ for $l \in \{q+1,q+2, \dots, n\}$ where $\mbU_0^\top \mbU = 0$ and $\mbU_0^\top \mbU_0 = \mbI_{n-q}$. On $\mathcal{B}_d$, we have that for each $j \in[q]$,  $$\sigma_j(\mbE) \leq \sigma_1(\mbE) \leq \sqrt{n}(t + d) \varphi^{1/2}_1(\mbSigma_*),$$
from which it follows that}
\mathcal{H}(\mbDelta) \geq & \underbrace{\frac{1}{2}\sum_{i=1}^q \sum_{j=1}^q \frac{(\mbu_j^\top \mbX \mbDelta \mbv_i - \mbu_i^\top \mbX \mbDelta \mbv_j)^2}{4 n (t + d) \varphi^{1/2}_1(\mbSigma_*)}  + \frac{1}{2} \sum_{k = q+1}^n\sum_{j=1}^q  \frac{(\mbu_k^\top \mbX \mbDelta \mbv_j)^2}{n (t + d) \varphi^{1/2}_1(\mbSigma_*)}}_{T_1} + \underbrace{o\left(\frac{\|\mbX\mbDelta\|_F^2}{n}\right)}_{T_2}.
\intertext{
By assumption \textbf{A5} and Lemma \ref{lemma:cone},  $\|\mbX \mbDelta\|_F^2/n \leq  \bar{v} \|\mbDelta\|_F^2$, where $\bar{v}$ is a finite constant. Thus, because $\|\mbDelta\|_F^2 \to 0$ by assumption, there exists an $N$ such that for all $n > N$, $
T_2 \geq -\nu\|\mbDelta\|_F^2/\{4 (t + d) \varphi^{1/2}_1(\mbSigma_*) \}.$ Also by assumption \textbf{A5} and Lemma \ref{lemma:cone},  
$ T_1 \geq \nu \|\mbDelta\|_F^2/\{ 2(t + d)\varphi^{1/2}_1(\mbSigma_*)\}.$ Hence, for $n$ sufficiently large, 
$$\mathcal{H}(\mbDelta) \geq T_1 + T_2 \geq \|\mbDelta\|_F^2 \frac{\nu}{4(t + d) \varphi^{1/2}_1(\mbSigma_*)},$$
which completes the proof.$~~\blacksquare$}
\end{align*}

\noindent \textbf{Proof of Lemma \ref{lemma:E_singular}.}
Let $\mbG \in \mathbb{R}^{n \times q}$ be a matrix with independent and identically distributed standard normal entries. Hence, for $\delta \geq 0$, we can write
\begin{align}
P\left( \frac{1}{\sqrt{n}} \|\mbE\| \leq \delta\right) &= P\left( \frac{1}{\sqrt{n}} \|\mbG \boldsymbol{\Sigma}_*^{1/2}\| \leq\delta\right) \notag \\
& \geq P\left( \frac{1}{\sqrt{n}} \|\mbG\| \|\boldsymbol{\Sigma}_*^{1/2}\| \leq \delta\right) = P\left( \frac{1}{\sqrt{n}} \|\mbG\| \leq \frac{\delta}{\varphi^{1/2}_1(\boldsymbol{\Sigma}_*)}\right)\label{eq:gaussian_inequality_1}
\end{align}
Then, applying Lemma \ref{lemma:conc_of_gaussian_spec}, for $\alpha \geq 0$, 
$$ P\left( \frac{1}{\sqrt{n}} \|\mbG\| >  \sqrt{\frac{q}{n}} + 1 + \frac{\alpha}{\sqrt{n}}\right) \leq 2 e^{-\alpha^2/2}.$$
so that taking $\alpha = d\sqrt{\frac{n}{2}}$ for $d > 0$, and recalling 
that $q/n \to t$ for $t \in (0,1)$, for $n$ sufficiently large,
$$ P\left( \frac{1}{\sqrt{n}} \|\mbG\| > (t + d + 1) \right) \leq 2 e^{- d^2 n /4}.$$
Finally, setting $\delta = (t + d + 1) \varphi_1^{1/2}(\boldsymbol{\Sigma}_*),$ applying the previous inequailty to \eqref{eq:gaussian_inequality_1}, for $n$ sufficiently large
$$ P\left( \frac{1}{\sqrt{n}}\|\mbE\| \leq (t + d + 1) \varphi_1^{1/2}(\boldsymbol{\Sigma}_*)\right) \geq 1 - 2e^{-d^2 n/4}.$$
The result follows by replacing $d$ with $\tilde{d} - 1$ for $\tilde{d} > 1. ~~\blacksquare$
\bigskip

\subsection{Technical Lemmas}

In this section, we provide several technical lemmas which were used in the previous sections. 

\begin{lemma}[Lemma 8.2, \cite{van2016estimation}]\label{lemma:max_norm}
Let $\mbg \sim {\rm N}_n(0, \mbI_n)$ and suppose $\mbX \in \mathbb{R}^{n \times q}$ has columns $\mbX_{\cdot,j}$ such that $\|\mbX_{\cdot,j}\|_2= \sqrt{n}$ for $j \in [p].$ If $0 < \delta < 1$ and $2\log (2 p/\delta) < n-1$, then
$$ P\left(\frac{\|\mbX^\top\mbg\|_\infty}{\sqrt{n}\|\mbg\|_2} \geq \sqrt{\frac{2 \log (2 p/\delta)}{n-1}}\right) \leq \delta.$$

\end{lemma}

\begin{lemma}[Proposition 66, \cite{dubois2019fast}]\label{RSClemma2}
Let $n \geq q$ and let $\mbA \in \mathbb{R}^{n \times q}$ be a full rank matrix with $(\mbU, \mbD, \mbV) = {\rm svd}(\mbA)$. Let $\mbU_0 \in \mathbb{R}^{n \times (n-q)}$ such that $\mbU_0^\top\mbU_0 = \mbI_{n-q}$ and $\mbU^\top\mbU_0 = 0$. Let $\mbu_i$ denote the $i$th column of $\mbU$ for $i \in [q]$, $\mbv_j$ the $j$th column of $\mbV$ for $j\in[q]$, $\rho_j$ the $j$th diagonal entry of $\mbD$ (i.e., $j$th largest singular value of $\mbA$) for $j \in [q]$, and $\mbu_k$ the $(k - q)$th the column of $\mbU_0$ for $k \in \{q+ 1, q + 2, \dots, n\}$. Then, for any $\mbC \in \mathbb{R}^{n \times q}$, it follows that 
\begin{align*} 
\|\mbA - \mbC\|_* & =  \|\mbA\|_* - {\rm tr}(\mbC^\top\mbU \mbV^\top) + \frac{1}{2}\sum_{i = 1}^q \sum_{j=1}^q \frac{(\mbu_j^\top \mbC \mbv_i - \mbu_i^\top\mbC\mbv_j)^2}{2 (\rho_i + \rho_j)}\\&~~~~~~~~~~~~~~~~~~~~~~~~~~~~~~~~~~~~~~  + \frac{1}{2} \sum_{k=q+1}^n \sum_{j=1}^q  \frac{(\mbu_k^\top \mbC \mbv_j)^2}{\rho_j}
 + o(\|\mbC\|_F^2).
\end{align*}
\end{lemma}

\begin{lemma}[Corollary 5.35, \cite{vershynin2010introduction}]\label{lemma:conc_of_gaussian_spec}
Let $\mbG \in \mathbb{R}^{n \times q}$ be a matrix with independent and identically distributed standard normal entries. If $\delta \geq 0$, then
$$P\left( \|\mbG\| \leq \sqrt{q} + \sqrt{n} + \delta\right) \geq 1 - 2 {\rm exp}\left(-\frac{\delta^2}{2}\right).$$ 
\end{lemma}

Next, we provide a general result on the concentration of Lipschitz functions $f:S^{n-1} \to \mathbb{R}$. In order to establish this result, we need a preliminary lemma regarding the concentration of a function $f$ near its median on $S^{n-1}$. 
\begin{lemma}[Theorem 3.4.1, \cite{raginsky2012concentration}]\label{median_lemma}
 Let $f:S^{n-1} \to \mathbb{R}$ be an $\eta$-Lipschitz function and let $\mbz$ be a random vector having the uniform distribution on $S^{n-1}$. If $\delta \geq 0$, then
$${\rm (i)} P(f(\mbz) \geq M_f + \delta) \leq {\rm exp}\left(-\frac{(n-2)\delta^2}{2\eta^2}\right),$$
where $M_f$ is the median of $f$ with respect to the uniform probability measure on $S^{n-1}.$
Moreover, 
$${\rm (ii)} |M_f - {\rm E}f(\mbz)| \leq \eta \sqrt{\frac{\pi}{n-2}}.$$
\end{lemma}
For (ii), see the proof of Corollary 5.4 of \citet{meckes2019random}.
This leads to our main lemma, which we use throughout the remainder of this section.  

\begin{lemma}\label{lemma:Levy2}
Let $f:S^{n-1} \to \mathbb{R}$ be an $\eta$-Lipschitz function and let $\boldsymbol{z}$ be a random vector having the uniform distribution on $S^{n-1}$.  If $\delta > 4 \eta  \{\pi/(n-2)\}^{1/2}$, then
$$  P\left(f(\boldsymbol{z}) -  {\rm E}f(\boldsymbol{z}) \geq \delta  \right) \leq {\rm exp}\left(\frac{-(n-2) \delta^2}{4\eta^2}\right).$$

\end{lemma}
Note that Lemma \ref{lemma:Levy2} would hold if we had $\delta > \eta \left\{\sqrt{2}/(\sqrt{2} - 1)\right\} \{\pi/(n-2)\}^{1/2}$. We use $\delta > 4 \eta  \{\pi/(n-2)\}^{1/2}$ in Lemma \ref{lemma:Levy2} for ease of display. \\

\noindent \textbf{Proof of Lemma \ref{lemma:Levy2}.}
We combine the two results from Lemma \ref{median_lemma} to obtain a bound on $P(f(\mbz) - {\rm E}f(\mbz) \geq \delta).$ First, 
\begin{align*}
P(f(\mbz) - {\rm E}f(\mbz) \geq \delta) &= P(f(\mbz) -M_f   \geq \delta + {\rm E}f(\mbz) - M_f )\\
&\leq   P(f(\mbz) -M_f   \geq \delta -|{\rm E}f(\mbz) - M_f|)
\intertext{so that an application of Lemma \eqref{median_lemma}(ii), $\delta > \eta \left\{\sqrt{2}/(\sqrt{2} - 1)\right\} \{\pi/(n-2)\}^{1/2}$---which is implied by $\delta > 4 \eta \{\pi/(n-2)\}^{1/2}$---and Lemma \eqref{median_lemma}(i), respectively, yield}
P(f(\mbz) - {\rm E}f(\mbz) \geq \delta) &\leq   P\left(f(\mbz) -M_f   \geq \delta -  \eta \sqrt{\frac{\pi}{n-2}} \right)\\
&\leq   P\left(f(\mbz) - M_f   \geq \frac{\delta}{\sqrt{2}}\right) \\
& \leq {\rm exp}\left(-\frac{(n-2)\delta^2}{4\eta^2}\right). ~~~\blacksquare
\end{align*}

This result can be generalized to quantities of the form $\|\mbS^\top\mba\|_2$ where $\mbS$ is uniformly distributed on $V_q(n).$



\begin{lemma} \label{lemma:conc_euclidean}
Let $\mba \in S^{n-1}$ be fixed and let $\mbS$ be a random matrix having the uniform distribution on $V_q(n)$. If $\delta > 4 \{\pi/(n-2)\}^{1/2},$ then
$$P\left(\|\mbS^\top\mba\|_2 \geq \delta + \sqrt{\frac{q}{n}}\right) \leq {\rm exp}\left(-\frac{(n-2)\delta^2}{4}\right).$$
\end{lemma}

\noindent \textbf{Proof of Lemma \ref{lemma:conc_euclidean}.} We apply the same arguments as in the proof of Lemma 4.2 of \citet{lyubarskii2010uncertainty}. Specifically, let $\mbO$ be a random matrix uniformly distributed on $O(n)$. We know then that $\mbS \sim \mbO \mbP_q$ where $\mbP_q \in \mathbb{R}^{n \times q}$ is the first $q$ columns of $\mbI_n$. Thus, $\mbS^\top\mba \sim \mbP_q^\top\mbO^\top \mba$ and consequently, because $\mbO^\top \mba$ is uniformly distributed on $S^{n-1}$ (Lemma \ref{lemma:Orthogonality_UniformDist}), for all $\delta \geq 0$ it follows that
$$P(\|\mbS^\top\mba\|_2 \geq \delta) = P(\|\mbP_q^\top\mbz \|_2 \geq \delta)$$
for random vector $\mbz$ having the uniform distribution on $S^{n-1}.$
Hence, applying Lemma \ref{lemma:Levy2} and using that $\mbz \mapsto \|\mbP_q^\top \mbz \|_2$ is 1-Lipschitz,
$$  P\left\{ \|\mbP_q^\top\mbz\|_2 \geq \alpha + {\rm E}(\|\mbP_q^\top\mbz\|_2)\right\} \leq {\rm exp}\left(-\frac{(n-2)\alpha^2}{4}\right).$$
Then, again applying a result from the proof of Lemma 4.2 of \citet{lyubarskii2010uncertainty}, ${\rm E}(\|\mbP_q^\top\mbz \|_2) \leq (q/n)^{1/2}$, so that finally, applying Lemma \ref{lemma:Levy2}, we conclude
$$ P\left(\|\mbS^\top\mba\|_2 \geq \alpha + \sqrt{\frac{q}{n}}\right) \leq {\rm exp}\left(-\frac{(n-2)\alpha^2}{4}\right),$$
as long as $\alpha > 4 \{\pi/(n-2)\}^{1/2}.~~\blacksquare$

\bibliography{JMLR_Final}

\end{document}